\documentclass[twocolumn,prb,aps]{revtex4}
\usepackage{graphicx}
\usepackage{amsmath}
\usepackage{color}
\newcommand{\be}{\begin{equation}}
\newcommand{\ee}{\end{equation}}
\newcommand{\bea}{\begin{eqnarray}}
\newcommand{\eea}{\end{eqnarray}}
\newcommand{\bs}{\begin{split}}
\newcommand{\bse}{\begin{subequations}}
\newcommand{\ese}{\end{subequations}}

\begin{document}

\title{Unified Molecular Field Theory for Collinear and Noncollinear Heisenberg Antiferromagnets}
\author {David C. Johnston} 
\affiliation {Department of Physics and Astronomy and Ames Laboratory, Iowa State University, Ames, Iowa 50011}

\date{\today}

\begin{abstract}

A unified molecular field theory (MFT) is presented that applies to both collinear and planar noncollinear Heisenberg antiferromagnets (AFs) on the same footing.  The spins in the system are assumed to be identical and crystallographically equivalent.  This formulation allows calculations of the  anisotropic magnetic susceptibility~$\chi$ versus temperature~$T$ below the AF ordering temperature $T_{\rm N}$ to be carried out for arbitrary Heisenberg exchange interactions $J_{ij}$ between arbitrary neighbors~$j$ of a given spin~$i$ without recourse to magnetic sublattices.  The Weiss temperature $\theta_{\rm p}$ in the Curie-Weiss law is written in terms of the $J_{ij}$ values and $T_{\rm N}$ in terms of the $J_{ij}$ values and an assumed AF structure.  Other magnetic and thermal properties are then expressed in terms of quantities easily accessible from experiment as laws of corresponding states for a given spin~$S$.  For collinear ordering these properties are the reduced temperature $t = T/T_{\rm N}$, the ratio $f = \theta_{\rm p}/T_{\rm N}$ and $S$\@.  For planar noncollinear helical or cycloidal ordering, an additional parameter is the wavevector of the helix or cycloid.  The MFT is also applicable to AFs with other AF structures.  The MFT predicts that $\chi(T \leq T_{\rm N})$ of noncollinear 120$^\circ$ spin structures on triangular lattices is isotropic and independent of~$S$ and~$T$ and thus clarifies the origin of this universally observed behavior.  The high-field magnetization and heat capacity for fields applied perpendicular to the ordering axis (collinear AFs) and ordering plane (planar noncollinear AFs) are also calculated and expressed for both types of AF structures as laws of corresponding states for a given~$S$, and the reduced perpendicular field versus reduced temperature phase diagram is constructed.

\end{abstract}


\maketitle


\clearpage

\section{\label{Sec:Intro} Introduction}

The Curie law\cite{Curie1895} $\chi\equiv M/H = C/T$ for noninteracting spins and the Curie-Weiss law\cite{Weiss1907, Kittel2005} 
\be
\chi = \frac{C}{T-\theta_{\rm p}}
\label{CWlaw0}
\ee
for interacting spins describe the low-field magnetic susceptibility $\chi$ in the paramagnetic (PM) regime above any long-range ordering temperatures.  Here $M$ is the magnetization, $H$ is the applied magnetic field, $C$ is the Curie constant, $T$ is the absolute temperature and $\theta_{\rm p}$ is the Weiss temperature which is positive for ferromagnets (FMs) and generally negative for antiferromagnets (AFs).  N\'eel showed that the molecular field theory (MFT) for FMs developed by Weiss to derive Eq.~(\ref{CWlaw0}) allowed collinear AF ordering to occur.\cite{Neel1936} The simplest model of an AF structure has each ``up'' spin with only ``down'' nearest neighbor spins, and vice versa, called a ``bipartite'' AF structure.  A magnetically-ordered (single-domain) FM in zero field has a net magnetic moment whereas an AF does not.  The magnetic ordering temperature of a FM is denoted as the Curie temperature $T_{\rm C}$ and of an AF as the N\'eel temperature $T_{\rm N}$. Van~Vleck used the Weiss MFT to calculate for a bipartite two-sublattice model the anisotropic $\chi(T)$ for $T < T_{\rm N}$ in small magnetic fields applied parallel ($\chi_\parallel$) and perpendicular $(\chi_\perp)$ to the magnetic moment ordering axis (easy axis) of collinear AF structures for equal Heisenberg nearest-neighbor interactions.\cite{VanVleck1941}  He deduced $T_{\rm N}=-\theta_{\rm p}$ where $\theta_{\rm p} < 0$, but this equality is rarely observed quantitatively in real AFs.  The subject of collinear AF is usually discussed in terms of N\'eel's two-sublattice MFT model.\cite{Nagamiya1955, Kittel2005, Keffer1966} Because large deviations from  Van~Vleck's requirement $T_{\rm N}=-\theta_{\rm p}$ are observed for real materials, Van Vleck's theory has not been used much in the past to fit experimental $\chi_\parallel(T\leq T_{\rm N})$ data for collinear AFs.

The anisotropic $\chi(T\leq T_{\rm N})$ and other properties of planar noncollinear AF ordering were investigated by Yoshimori based on MFT for the ``proper screw'' helix magnetic structure,\cite{Yoshimori1959, Nagamiya1967} as shown in Fig.~1 of Ref.~\onlinecite{Johnston2012}. A proper screw helix is an AF structure in which planes of magnetic moments that are ferromagnetically aligned within each plane rotate their ordered moment directions along the helix axis ($z$-axis here) with the tips of the magnetic moment vectors tracing out the ridges on a screw (see Fig.~1 of Ref.~\onlinecite{Johnston2012}).  Thus the ordered moment directions are perpendicular to the screw $z$-axis with a fixed angle between the ordered moments in adjacent planes along this axis.  On the other hand, when the helix $z$-axis is in the $xy$~plane of the coplanar magnetic moments, the heads of the ordered moment vectors along the $z$-axis trace out points on a cycloid, and hence Yoshimori termed this a ``cycloidal'' AF structure\cite{Yoshimori1959} as shown in Fig.~1 of Ref.~\onlinecite{Goetsch2014}.  Like Van~Vleck's theory, Yoshimori's predictions were very restrictive and have been little used by experimentalists to fit their $\chi(T\leq T_{\rm N})$ data for helical or cycloidal AFs.

This paper is a followup to our 2012 Letter,\cite{Johnston2012} where we formulated a generic version of MFT for Heisenberg spin systems containing identical crystallographically-equivalent spins that improves on previous MFT treatments in ways that will be discussed.  The paper is organized as follows.  Our MFT notation and the definition of the exchange field in MFT are given in Sec.~\ref{Sec:IntroMFT}.  In Sec.~\ref{Sec:PropsBelowTN}  we introduce the Brillouin function and derive expressions for the exchange field,  reduced ordered moment and other properties versus temperature which are the same for collinear and noncollinear AFs and for FMs.  The static critical exponents and associated dimensionless reduced amplitudes are derived in Sec.~\ref{CritExps}.  A Curie-Weiss law of corresponding states is derived in Sec.~\ref{Sec:CWLaw}.  In Sec.~\ref{Sec:ChiParThy} we present our generic MFT formulation of $\chi(T\leq T_{\rm N})$ of collinear AFs and in Sec.~\ref{Sec:ChianisNoncoll} for the anisotropic $\chi(T)$ of planar noncollinear AFs.  In Sec.~\ref{Sec:ChiPerp} a generic expression for $\chi_\perp$ of collinear AFs and planar noncollinear AFs is derived.  Our MFT calculation of $\chi_{xy}(T\leq T_{\rm N})$ of planar noncollinear AFs is presented in Sec.~\ref{Sec:ChiPar}.  In Sec.~\ref{Sec:J0J1zJ2zModel} we formulate a generic minimal and powerful $J_0$-$J_{z1}$-$J_{z2}$ model for $\chi_{xy}(T)$ of helical or cycloidal AFs that is widely applicable to real materials and contains as parameters only directly measurable quantities.  The magnetization, internal energy and magnetic heat capacity below $T_{\rm N}$ of both collinear AFs and planar noncollinear AFs in high magnetic fields aligned perpendicular to the ordering axis or plane of the AF structure, respectively, are derived for both types of AF structures for which we obtain the same generic laws of corresponding states, respectively, in Sec.~\ref{Sec:HiFieldPerpM_hA1zero}.   In concluding Sec.~\ref{Sec:Discussion} the MFT predictions are discussed with respect to non-mean-field behaviors observed for real systems.

\section{\label{Sec:IntroMFT} Exchange Field}

We consider the Heisenberg model with no anisotropy terms except that due to an infinitesimal {\bf H}\@. The part ${\cal H}_i$ of the spin Hamiltonian associated with a particular central spin ${\bf S}_i$ interacting with its neighbors ${\bf S}_j$ with respective exchange constants $J_{ij}$ is
\be
{\cal H}_i = \frac{1}{2}{\bf S}_{i}\cdot\sum_j J_{ij}{\bf S}_j + g\mu_{\rm B}{\bf S}_i\cdot{\bf H},
\label{Eq:Hamili1}
\ee
where the factor of 1/2 appears in the first term because the exchange energy is evenly split between the two members of each pair of interacting spins, $g$ is the spectroscopic splitting factor ($g$-factor) of a magnetic moment $\vec{\mu}$ and $\mu_{\rm B}$ is the Bohr magneton.  

In the Weiss MFT, one only considers the thermal-average directions of ${\bf S}_i$ and ${\bf S}_j$ when calculating their interaction.  Furthermore, it is the magnetic moment $\vec{\mu}$ that interacts with a magnetic field and not the angular momentum {\bf S} \emph{per se}.  The relationship between these two quantities for an electronic spin and magnetic moment is 
\be
{\bf S} = -\frac{\vec{\mu}}{g\mu_{\rm B}},
\label{Eq:Sfrommu}
\ee
where the minus sign arises from the negative sign of the electron charge.  Throughout the remainder of this paper, the symbol $\vec{\mu}$ refers to the thermal-average value of the magnetic moment, as is appropriate in MFT\@. Then the energy $E_{{\rm mag}\,i}$ of interaction of magnetic moment $\vec{\mu}_i$ with its neighbors $\vec{\mu}_j$ is given by Eq.~(\ref{Eq:Hamili1}) as
\be
E_{{\rm mag}\,i} = \frac{1}{2g^2\mu_{\rm B}^2}\vec{\mu}_i\cdot\sum_j J_{ij}\vec{\mu}_j - \vec{\mu}_i \cdot{\bf H}.
\label{Eq:Hamili2}
\ee

In MFT, one replaces the sum of the exchange interactions acting on $\vec{\mu}_i$ in the first term by an effective magnetic field called the Weiss molecular field or ``exchange field'' ${\bf H}_{\rm exch}$ that is defined by the usual relationship for the rotational potential energy of a magnetic moment in a magnetic field, as in the second term of Eq.~(\ref{Eq:Hamili2}), as
\be
E_{{\rm exch}\,i} = -\frac{1}{2}\vec{\mu}_i\cdot{\bf H}_{{\rm exch}\,i},
\label{Eq:EexchtoHexch}
\ee
 where the factor of 1/2 again arises because in MFT all of the exchange energy between $\vec{\mu}_i$ and $\vec{\mu}_j$ is attributed to $\vec{\mu}_j$, thus canceling out the factor of 1/2 in Eq.~(\ref{Eq:Hamili2}).  From the first term in Eq.~(\ref{Eq:Hamili2}) one then obtains 
\be
{\bf H}_{{\rm exch}\,i} = -\frac{1}{g^2\mu_{\rm B}^2}\sum_j J_{ij} \vec{\mu}_{j}.\\*
\label{Eq:HexchiDef}
\ee
Using $\vec{\mu}_{j} = \mu_j\hat{\mu}_j$ where $\mu_j = |\vec{\mu}_{j}|$, the component of ${\bf H}_{{\rm exch}\,i}$ in the direction of $\vec{\mu}_{i}$ is
\be
\bs
H_{{\rm exch}\,i} &= \hat{\mu}_i\cdot {\bf H}_{{\rm exch}\,i} = -\frac{1}{g^2\mu_{\rm B}^2}\sum_j J_{ij}\mu_j\hat{\mu}_i\cdot\hat{\mu}_j\\*
&= -\frac{1}{g^2\mu_{\rm B}^2}\sum_j J_{ij}\mu_j\cos\alpha_{ji},
\end{split}
\label{Eq:HexchDef3}
\ee
where $\alpha_{ji}$ is the angle between $\vec{\mu}_{j}$ and $\vec{\mu}_{i}$ when $H\neq0$. If $H=0$ we denote this angle instead by $\phi_{ji}$. 

In the ordered magnetic state at {\bf H} = 0, the lowest energy of the spin system occurs when each magnetic moment is in the same direction as the local exchange field it sees.  Therefore the component of the local ${\bf H}_{{\rm exch}\,i0}$ in the direction of $\vec{\mu}_i$, and also its magnitude, is
\be
\bs
H_{{\rm exch}\,i0} &= \hat{\mu}_i\cdot {\bf H}_{{\rm exch}\,i0} \\*
&= -\frac{\mu_0}{g^2\mu_{\rm B}^2}\sum_j J_{ij}\cos\phi_{ji},
\end{split}
\label{Eq:Hexch0Def3}
\ee
where the subscript~0 in $H_{{\rm exch}\,i0}$ designates that $H=0$ and $\mu_0$ is the magnitude of the $T$-dependent ordered magnetic moment in {\bf H} = 0 observed, {\it e.g.}, by neutron diffraction measurements which is the same for all spins because of their crystallographic equivalence.

\section{\label{Sec:PropsBelowTN} Some Properties at Temperatures below the N\'eel Temperature}

\subsection{Brillouin Function, N\'eel Temperature, Ordered Moment, Laws of Corresponding States and Magnetic Energy}

In general, in MFT the equilibrium (thermal-average) direction of a specific ordered local moment $\vec{\mu}_i$ is always in the direction of its local magnetic induction ${\bf B}_i$.  The magnitude $\mu_i$ of $\vec{\mu}_i$ in that direction is determined by the Brillouin function $B_S(y)$ according to\cite{Johnston2012,Johnston2011}

\begin{subequations}
\label{Eq:BS(y)}
\be
\mu_i = \mu_{\rm sat}B_S(y_i)
\label{Eq:muvsBrill}
\ee
where
\be
y_i = \frac{g\mu_{\rm B}B_i}{k_{\rm B}T},
\label{Eq:yDef}
\ee
the saturation moment of each spin is
\be
\mu_{\rm sat} = gS\mu_{\rm B},
\label{Eq:musat}
\ee
\end{subequations}
and $g\approx2$ for many $3d$ transition metal ions due to quenching of the $z$-component of the orbital angular momentum, and also for spin-only Gd$^{+3}$ and Eu$^{+2}$ ions with $S = 7/2$ and orbital angular momentum $L=0$.

Our unconventional definition of the Brillouin function is
\begin{subequations} 
\label{Eqs:BS}
\be
B_S(y) = \frac{1}{2S} \left\{(2S+1)\coth\left[(2S+1)\frac{y}{2}\right]-\coth\left(\frac{y}{2}\right)\right\}
\label{Eq:BrillouinFunction},
\ee
for which the Taylor series expansion about $y=0$ is
\be
B_S(y) = \frac{(S+1)y}{3} -\frac{1}{90}(1+3S+4S^2+2S^3)y^3 + {\cal O}(y^5).
\label{Eq:BSyTaylor}
\ee
For $y\gg1$ and finite~$S$ one obtains
\be
B_S(y) \approx 1-\frac{e^{-y}}{S}.
\label{Eq:BSTto0}
\ee
\ese

The derivative of $B_S(y)$ is
\bea
&&B_S^\prime(y) \equiv \frac{dB_S(y)}{dy}\label{Eq:dBSy0}\\*
 &&=  \frac{1}{4S}\bigg\{{\rm csch}^2\left(\frac{y}{2}\right) -\ (2S+1)^2{\rm csch}^2\left[(2S+1)\frac{y}{2}\right]\bigg\}.\nonumber
\eea
From Eq.~(\ref{Eq:BSyTaylor}), the lowest-order terms of a Taylor series expansion of $B_S^\prime(y)$ about $y=0$ are
\be
B_S^\prime(y) = \frac{S+1}{3}-\frac{1+3S+4S^2+2S^3}{30}\,y^2 + {\cal O}(y^4).
\label{Eq:BSprimeExpand}
\ee

The magnetic induction in Eq.~(\ref{Eq:yDef}) is
\be
B_i = H_{{\rm exch}\,i} + H_{\parallel i},
\label{Eq:Bi0}
\ee
where $H_{{\rm exch}\,i}$ is the component of the exchange field parallel to magnetic moment~$\vec{\mu}_i$ and $H_{\parallel i} = \hat{\mu}_i\cdot{\bf H}$ is the component of the applied magnetic field in the direction of $\vec{\mu}_i$.  We define the direction of approach to a transition temperature by superscript $+$ and~$-$ symbols.  Thus on approaching the AF ordering temperature from below, denoted as $T\to T_{\rm N}^-$, an infinitesimal nonzero ordered moment develops even in the absence of an applied magnetic field.  One can Taylor expand the Brillouin function for small arguments using Eq.~(\ref{Eq:BSyTaylor}), 
and then Eq.~(\ref{Eq:muvsBrill}) becomes
\be
\mu_i = \frac{g^2\mu_{\rm B}^2S(S+1)}{3k_{\rm B}T_{\rm N}} B_i = \frac{g^2\mu_{\rm B}^2S(S+1)}{3k_{\rm B}T_{\rm N}} (H_{{\rm exch}\,i} + H).
\label{Eq:muvsBrillLowy}
\ee
For $H=0$ one obtains
\be
\mu_0 = \frac{g^2\mu_{\rm B}^2S(S+1)}{3k_{\rm B}T_{\rm N}}H_{{\rm exch}\,i0}.
\label{Eq:muvsBrillLowy2}
\ee
Substituting Eq.~(\ref{Eq:Hexch0Def3}) for $H_{{\rm exch}\,i0}$ into~(\ref{Eq:muvsBrillLowy2}) gives the most general expression for the AF ordering temperature in MFT for a system of identical crystallographically equivalent spins interacting by Heisenberg exchange as
\be
T_{\rm N} = -\frac{S(S+1)}{3k_{\rm B}}\sum_j J_{ij}\cos\phi_{ji}.
\label{Eq:TmGeneral}
\ee
This equation also predicts the magnetic ordering temperature (Curie temperature $T_{\rm C}$) of a ferromagnet where $\phi_{ji}=0$ and $\sum_j J_{ij}<0$.  By comparing Eqs.~(\ref{Eq:Hexch0Def3}) and~(\ref{Eq:TmGeneral}), one can write the zero-field exchange field ${\bf H}_{{\rm exch}\,i0}$ seen by each magnetic moment $\vec{\mu}_{i0}$ as
\begin{subequations}
\label{Eqs:Hexchi050}
\be
\bs
{\bf H}_{{\rm exch}\,i0} &= \frac{T_{\rm N}}{C_1}\vec{\mu}_{i0}\\*
H_{{\rm exch}\,0} &= \frac{T_{\rm N}}{C_1}{\mu}_{0},
\label{Eq:HexchioTm} 
\end{split}
\ee
where the magnitude $H_{{\rm exch}\,0}$ of the exchange field in \mbox{{\bf H} = 0} seen by each spin is the same for all spins because of their crystallographic equivalence, hence the subscript~$i$ is dropped, and the single-spin Curie constant $C_1$ is defined as\cite{Kittel2005}
\be
C_1 = \frac{g^2\mu_{\rm B}^2S(S+1)}{3k_{\rm B}}.
\label{Eq:CurieConst2}
\ee
\end{subequations}

Equations~(\ref{Eq:HexchDef3}), (\ref{Eq:Hexch0Def3}) and~(\ref{Eq:HexchioTm}) for the exchange field do not make any reference to magnetic moments other than the central magnetic moment $\vec{\mu}_i$ and its generic neighbors.  In particular, the exchange field and relative magnetic moment direction in {\bf H} = 0 are determined solely by the local interactions of each magnetic moment with its neighbors.  Thus we do not define or identify distinct magnetic sublattices in our formulation of MFT, in contrast to traditional approaches.  

We define the reduced zero-field ordered moment and reduced temperature respectively as
\bse
\label{Eqs:Redmu0t}
\be
\bar{\mu}_0 = \frac{\mu_0}{\mu_{\rm sat}}=\frac{\mu_0}{gS\mu_{\rm B}} ,
\label{Eq:barmu0Def}
\ee
\be
t = \frac{T}{T_{\rm N}} ,
\label{Eq:tDef}
\ee
\ese 
where the saturation moment $\mu_{\rm sat}$ of spin~$S$ is given by Eq.~(\ref{Eq:musat}).  The zero-field exchange field in the direction of $\vec{\mu}_i$ in Eq.~(\ref{Eq:HexchioTm}) becomes
\be
H_{{\rm exch}\,0} = \frac{3k_{\rm B}T_{\rm N}\bar{\mu}_0}{(S+1)g\mu_{\rm B}}. 
\label{Eq:HexchioTm2}
\ee
Then Eq.~(\ref{Eq:muvsBrill}) for calculating the ordered moment $\mu_0$ versus $T$ in $H = 0$ can be compactly written as\cite{Johnston2011}
\be
\bar{\mu}_0 = B_S(y_0), \quad{\rm with}\quad  y_0 =\frac{3\bar{\mu}_0}{(S+1)t},
\label{Eq:mubar0}
\ee
and the Brillouin function $B_S(y_0)$ is given in Eq.~(\ref{Eq:BrillouinFunction}).  This zero-field expression is valid within MFT for ferromagnets and both collinear and noncollinear AFs.  Plots of the zero-field reduced ordered moment $\bar{\mu}_0$ versus reduced temperature $t$ for several spin $S$ values according to Eq.~(\ref{Eq:mubar0}) are shown in Fig.~10 of Ref.~\onlinecite{Johnston2011}.  The order parameter for an AF transition is the single-spin ordered moment.  From Fig.~10 of Ref.~\onlinecite{Johnston2011}, one sees that the ordered moment increases continuously from zero on entering the AF state from above.  Thus the transition is a continuous (second-order) transition with no latent heat. 

The total temperature derivative $d\bar{\mu}_0/dt$ is calculated from Eq.~(\ref{Eq:mubar0}) as
\be
\frac{d\bar{\mu}_0}{dt} = -\frac{\bar{\mu}_0(t)}{t\Big[\frac{(S+1)t}{3B_S^\prime(y_0)} - 1\Big]},
\label{Eq:dbarmu0dt}
\ee
where $B_S^\prime(y_0)\equiv dB_S(y)/dy|_{y=y_0}$ and the function $B_S^\prime(y)$ is given in Eq.~(\ref{Eq:dBSy0}).

The expression for $\bar{\mu}_0$ versus $t$ in Eq.~(\ref{Eq:mubar0}) is an example of a so-called ``law of corresponding states'' for a given spin~$S$\@.  Spin systems are said to be in corresponding states when their reduced state variables such as $t$ and $\bar{\mu}_0$ have the same values, respectively.  Thus when an equation in reduced variables such as Eq.~(\ref{Eq:mubar0}) is a law of corresponding states, the equation applies equally well to different spin systems with the same~$S$ but with, e.g., different exchange constants and N\'eel temperatures, which are implicitly contained in the reduced variables $t$ and~$\bar{\mu}_0$.  Many other laws of corresponding states for spin systems with the same~$S$ are obtained in later sections because we usually write MFT predictions in terms of universal reduced variables.

Using the Taylor series expansion in Eq.~(\ref{Eq:BSyTaylor}) of the Brillouin function for small arguments to order $y_0^3$ appropriate for $t\to1^-$ and solving for $\bar{\mu}_0(t)$ yields the  behaviors on approaching the N\'eel temperature $t=1$ to the lowest two orders as
\bse
\bea
{\bar{\mu}_0}^2 &=& \frac{10(1+S)^2}{3(1+2S+2S^2)}\,(1-t)\qquad\quad (t\to1^-) \label{Eq:mu02Tto0}\\*
&&\hspace{-0.4in} +\  \frac{25(1+S)^2(3 + 12S + 28 S^2 + 32 S^3 + 16 S^4)}{21(1 + 2S + 2S^2)^3} \,(1-t)^2 \nonumber\\*
\bar{\mu}_0 &=& \frac{\sqrt{\frac{10}{3}}(1+S)}{\sqrt{1+2S+2S^2}}(1-t)^{1/2}\qquad\quad (t\to1^-)\label{Eq:mu0Tto0}\\*
&&\hspace{-0.4in} +\ \frac{5\sqrt{\frac{5}{6}}(1+S)(1+2S+4S^2)(3+6S+4S^2)}{14(1+2S+2S^2)^{5/2}} (1-t)^{3/2}.\nonumber
\eea
\ese
The leading $\sqrt{1-t}$ temperature dependence of the order parameter [the ordered moment in Eq.~(\ref{Eq:mu0Tto0}) in this case] is characteristic of the critical behavior predicted by mean-field theories of second-order phase transitions on approach to the ordering temperature from below.

In $H=0$, the magnetic energy per spin $E_{\rm mag}/N$ is defined within MFT by Eq.~(\ref{Eq:EexchtoHexch}) as
\be
\frac{E_{\rm mag}}{N} = -\frac{1}{2}\vec{\mu}_i\cdot {\bf H}_{{\rm exch}\,i},
\ee
where $N$ is the number of spins.  Then using the relations $\vec{\mu}_i \parallel {\bf H}_{\rm exch\,i}$ for $H=0$ and therefore $\vec{\mu}_i\cdot {\bf H}_{{\rm exch}\,i} = \mu_0 H_{\rm exch\,0}$ and also using Eqs.~(\ref{Eqs:Redmu0t}) and~(\ref{Eq:HexchioTm2}) one obtains the magnetic energy per spin as
\be
\frac{E_{\rm mag}}{Nk_{\rm B}} = -\frac{3ST_{\rm N}}{2(S+1)}\,\bar{\mu}_0^2.
\label{Eq:Emag}
\ee

\subsection{\label{Sec:BSmu0TNCmag} Magnetic Heat Capacity}

Using Eqs.~(\ref{Eq:tDef}) and~(\ref{Eq:Emag}), the molar magnetic contribution $C_{\rm mag}(t)$ to the heat capacity in zero applied magnetic field is given in MFT by
\be
\frac{C_{\rm mag}(t)}{R} = -\frac{3S}{2(S+1)}\frac{d{\bar{\mu}_0}^2(t)}{dt}= -\frac{3S}{S+1}\bar{\mu}_0(t)\frac{d{\bar{\mu}_0}(t)}{dt},
\label{Eq:Cmag0}
\ee
where we have set $N= N_{\rm A}$, $N_{\rm A}$ is Avogadro's number and $R = N_{\rm A}k_{\rm B}$ is the molar gas constant.  Substituting $d\bar{\mu}_0/dt$ from Eq.~(\ref{Eq:dbarmu0dt}) into the second equality in Eq.~(\ref{Eq:Cmag0}) yields
\be
\frac{C_{\rm mag}(t)}{R} = \frac{3S\bar{\mu}_0^2(t)}{(S+1)t\Big[\frac{(S+1)t}{3B_S^\prime(y_0)} - 1\Big]}. 
\label{Eq:Cmag}
\ee
Thus $C_{\rm mag}(t)$ for $H=0$ is determined solely by the spin~$S$ and by the temperature dependence of the reduced ordered moment and hence is a law of corresponding states for a given $S$\@.  Plots of $C_{\rm mag}(t)/R$ for various values of $S$ from the minimum quantum value $S=1/2$ to the classical limit ($S \to \infty$) are shown in Fig.~11 of Ref.~\onlinecite{Johnston2011}.   The magnetic entropy $S_{\rm mag}$ at $T_{\rm N}$ calculated from $C_{\rm mag}(T)$ for each of the finite $S$ values satisfies the quantum statistical mechanics prediction $S_{\rm mag}(T\to\infty) = R\ln(2S+1)$.

One can obtain the behavior of $C_{\rm mag}$ for $t\to1^-$ by taking the temperature derivative of $\bar{\mu}_0^2(t)$ in Eq.~(\ref{Eq:mu02Tto0}) and inserting the result into the first equality in Eq.~(\ref{Eq:Cmag0}), yielding
\bse
\label{Eqs:Cmag}
\bea
\frac{C_{\rm mag}}{R} &=& \frac{5S(1+S)}{1+2S+2S^2}\label{Eq:CmagtTo1} \qquad\qquad (t\to 1^-)\label{Eq:CmagTto0-}\\*
&&\hspace{-0.5in} +\ \frac{25S(1+S)(3 + 12S + 28 S^2 + 32 S^3 + 16 S^4)}{7(1+2S+2S^2)^3}\, (1-t).\nonumber
\eea
Since $C_{\rm mag}=0$ for $t>1$ as seen in Fig.~11 of Ref.~\onlinecite{Johnston2011}, the heat capacity jump on cooling below $T_{\rm N}$ in Fig.~11 of Ref.~\onlinecite{Johnston2011} is given by Eq.~(\ref{Eq:CmagtTo1}) as
\be
\frac{\Delta C_{\rm mag}}{R} = \frac{5S(1+S)}{1+2S+2S^2},
\label{Eq:CmagJump}
\ee
\ese
which has the narrow range $\Delta C_{\rm mag}/R = 3/2$ for $S=1/2$ to $\Delta C_{\rm mag}/R = 5/2$ for $S\to \infty$.

Equations~(\ref{Eq:mubar0}), (\ref{Eq:Cmag}) and~(\ref{Eqs:Cmag}) are generally applicable to Heisenberg magnets containing identical crystallographically equivalent spins in $H=0$ within MFT including FMs and both collinear and noncollinear AFs.

\subsection{Staggered Magnetization versus Staggered Magnetic Field Isotherms}

\begin{figure}
\includegraphics[width=3.3in]{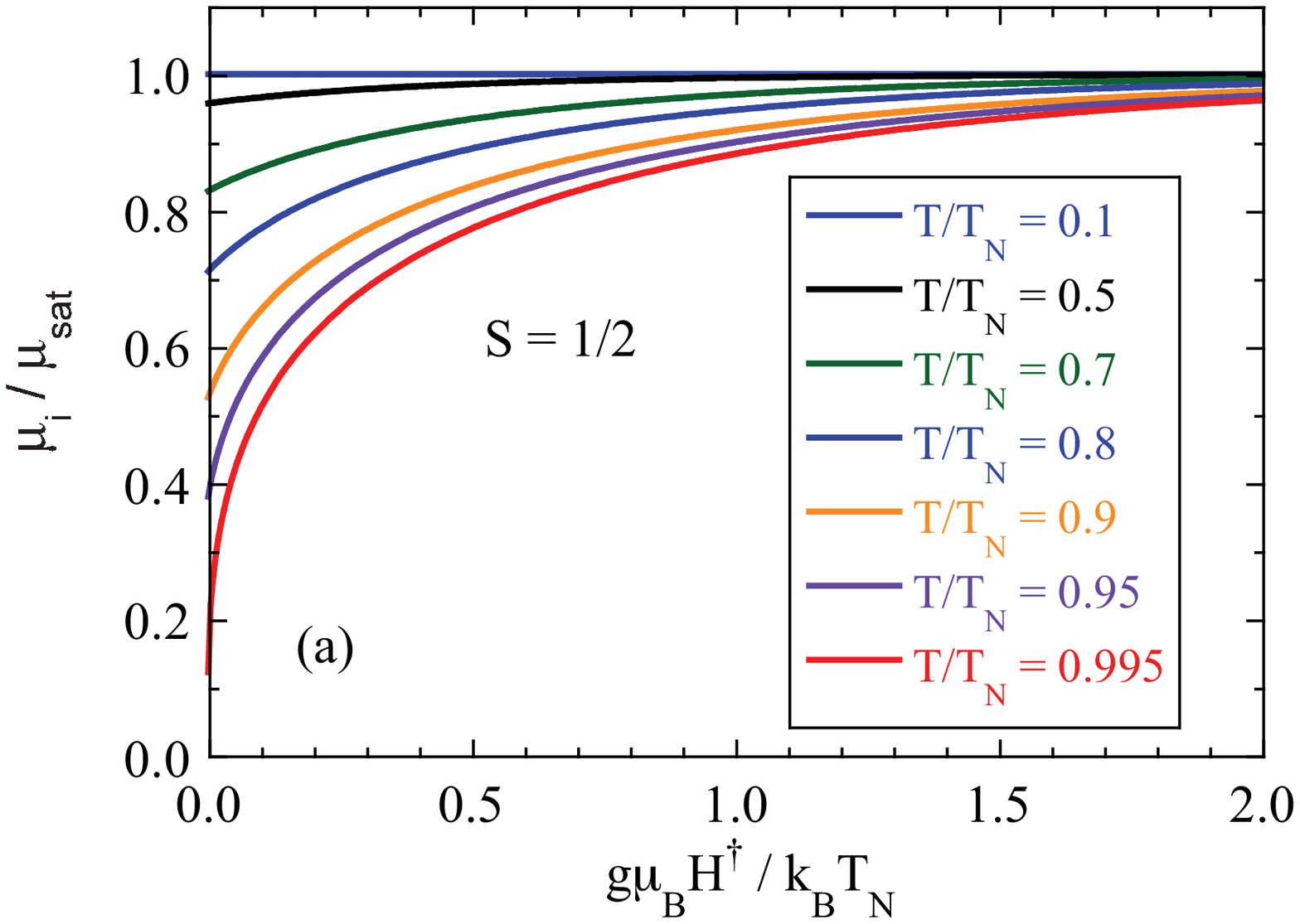}
\includegraphics[width=3.3in]{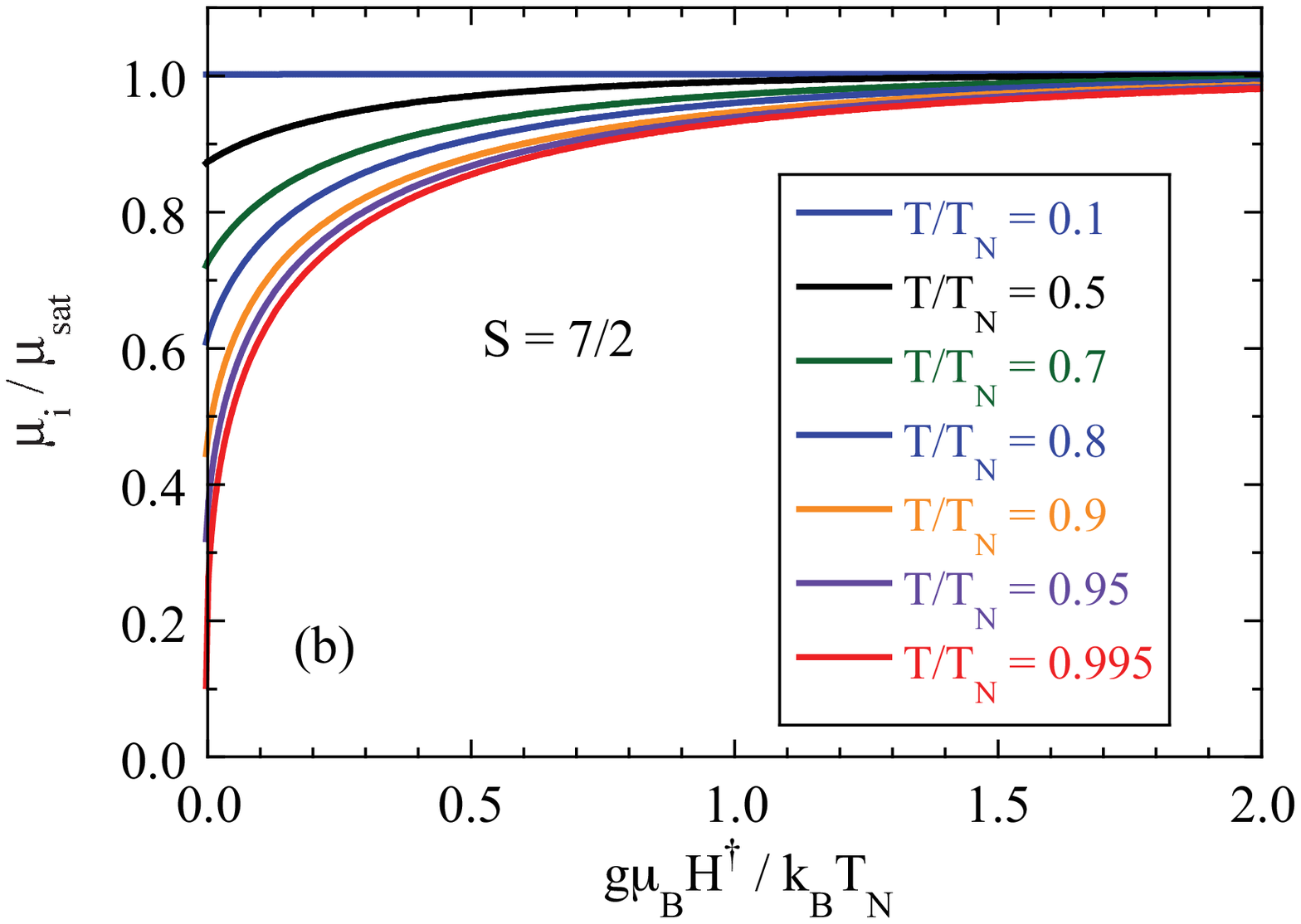}
\caption{(Color online) Reduced ordered plus induced moment $\bar{\mu}_i\equiv \mu_i/\mu_{\rm sat}$ versus reduced staggered magnetic field $h^\dagger \equiv g\mu_{\rm B}H^\dagger/k_{\rm B}T_{\rm N}$ at the indicated reduced  temperatures $t\equiv T/T_{\rm N}$ for (a) spin $S = 1/2$ and (b) spin $S = 7/2$.  The curves were calculated using Eqs.~(\ref{Eq:muiFM}) and~(\ref{Eq:yFM1}).  The temperatures of the curves from top to bottom are the same as in the figure legends.  The behaviors for a single-domain ferromagnet are identical to those shown, with the reduced uniform field $h\equiv g\mu_{\rm B}H/(k_{\rm B}T_{\rm C})$ replacing the reduced staggered field~$h^\dagger$, where $T_{\rm C}$ is the ferromagnetic ordering (Curie) temperature. }
\label{Fig:ChiParallel_FM_S1_2} 
\end{figure}

If one applies a parallel field to a single-domain FM below the Curie temperature, at zero field the ordered moment is $\mu_i$.  On increasing the field $\mu_i$ increases because it accrues a field-induced moment that increases with increasing field.  Similarly, in an AF, one can imagine a staggered magnetic field ${\bf H}_i^\dagger$ for each ordered moment $\vec{\mu}_i$ that is applied in the direction of each moment in the sample and is therefore also in the direction of the exchange field ${\bf H}_{{\rm exch}\,i}$ for each moment.  Thus ${\bf H}_i^\dagger$ does not change the angles of the spins with respect to each other, irrespective of the magnitude of ${\bf H}_i^\dagger$.  Due to the assumed crystallographic equivalence of each spin, the magnitude $H_i^\dagger$ is independent of~$i$ and hence we write it as $H^\dagger$.  The expression for the exchange field ${\bf H}_{{\rm exch}\,i}$ for that moment is therefore the same as that for ${\bf H}_{{\rm exch}0\,i}$ in Eq.~(\ref{Eq:HexchioTm2}) but with a field-dependent $\bar{\mu}_i$ replacing $\bar{\mu}_0$.  The magnitude $H_{{\rm exch}0\,i}$ is the same for each moment and hence we drop the index~$i$.  Within MFT, the dependence of $\mu_i$ on $H$ in a FM is identical to the dependence of $\mu_i$ on $H^\dagger$ in an AF.  This equivalence applies to both collinear and noncollinear AFs.  The calculations in the present and following section are not usually presented when the predictions of MFT are discussed.

Here we calculate the ordered plus induced moment $\mu_i$ of each spin~$i$ versus $H^\dagger$ for an AF below its $T_{\rm N}$.  From Eqs.~(\ref{Eq:BS(y)}) and~(\ref{Eq:Bi0}), one has 
\bse
\label{Eqs:mu0Calc}
\be
\bar{\mu}_i = B_S(y),
\label{Eq:muiFM}
\ee
where
\be
\qquad y = \frac{g\mu_{\rm B}}{k_{\rm B}T}(H_{{\rm exch}0} + H^\dagger).
\label{Eq:yFM0}
\ee
Using Eq.~(\ref{Eq:HexchioTm2}) and defining the reduced staggered field
\be
h^\dagger \equiv \frac{g\mu_{\rm B}H^\dagger}{k_{\rm B}T_{\rm N}},
\ee
the variable~$y$ in Eq.~(\ref{Eq:yFM0}) becomes
\be
y = \frac{3\bar{\mu}_i}{(S+1)t}+ \frac{h^\dagger}{t},
\label{Eq:yFM1}
\ee
\ese
where we have used the definition of the reduced temperature~$t$ in Eq.~(\ref{Eq:tDef}).

Numerical solutions of Eqs.~(\ref{Eq:muiFM}) and~(\ref{Eq:yFM1}) for $\bar{\mu}_i$ versus $h^\dagger$ were obtained for spins $S = 1/2$ and $S = 7/2$ and the results are plotted in Fig.~\ref{Fig:ChiParallel_FM_S1_2} for seven values of~$t< 1$.  The values of $\bar{\mu}_i$ at $h^\dagger=0$ are the ordered moments for these spin values at the respective temperatures as plotted in Fig.~\ref{Fig:OrderedMomentMFT3} below.  The initial slope of $\bar{\mu}_i$ versus $h^\dagger$ increases with increasing $t$ and diverges to~$\infty$ for $t\to 1^-$. This means that the reduced staggered susceptibility $\chi^\dagger \equiv (\bar{\mu}_i/h^\dagger)|_{h^\dagger\to0}$ increases with increasing~$t$ and diverges for $t\to1^-$, which also means that $\chi \equiv (\bar{\mu}_i/h)|_{h\to0}$ diverges for a single-domain FM on approaching its Curie temperature from below.  Indeed, the $t$-dependent values of $\chi(t)$ for a single-domain FM and for $\chi^\dagger(t)$ of collinear and noncollinear AFs for $t<1$ are identical within MFT\@.  For a bulk FM $\chi(t)$ is difficult to measure in the FM-ordered state due to formation of multiple FM domains and their relative size and number dependence on field, which introduces a contribution to the uniform $M(H)$ behavior beyond that predicted by MFT\@.  For AFs, it is usually not possible to apply a real staggered magnetic field.  However, the staggered susceptibility of an AF can be determined indirectly from inelastic neutron scattering measurements.

At $t=1$, the system is in the PM state since the ordered moment is zero at that temperature.  We define the reduced uniform magnetic field $h$ for a paramagnet as
\be
h \equiv \frac{g\mu_{\rm B}H}{k_{\rm B}T_{\rm N}}.
\label{Eq:barhDef}
\ee
By expanding Eq.~(\ref{Eq:muiFM}) at $t=1$ to third order in $\bar{\mu}_i$ and first order in $h$ with $y$ given by Eq.~(\ref{Eq:yFM1}) with $h$ replacing $h^\dagger$ and solving for $\bar{\mu}_i$ gives the asymptotic isothermal critical magnetization versus field at the ordering temperature as
\be
\bar{\mu}_i = (S+1)\left[\frac{10}{9(1+2S+2S^2)}\right]^{1/3}{h}^{1/3}\ (t=1,\ h\to0) .
\label{Eq:muVShDagger}
\ee
This shows that the initial dependence of $\bar{\mu}_i$ versus $h$ at $t=1$ as in Fig.~\ref{Fig:M_vs_H_crit_isotherms} below has an infinite slope for $h\to0$.

\subsection{Staggered Magnetic Susceptibility}

As seen from Fig.~\ref{Fig:ChiParallel_FM_S1_2}, for $t<1$ the initial behavior of $\bar{\mu}_i$ of an AF versus $h^\dagger$ is
\be
\bar{\mu}_i(t,h^\dagger) = \bar{\mu}_0(t) + \chi^\dagger(t) h^\dagger,
\ee
where the reduced staggered susceptibility $\chi^\dagger$ is the initial slope of $[\bar{\mu}_i(t,h^\dagger) - \bar{\mu}_0(t)]$ versus $h^\dagger$.  Since $\bar{\mu}_i(t,h^\dagger)$ versus $h^\dagger$ is nonanalytic at $t<1$ and $h^\dagger = 0$ according to Fig.~\ref{Fig:ChiParallel_FM_S1_2} , one cannot utilize a Taylor series expansion of Eq.~(\ref{Eq:muiFM}) about $h^\dagger=0$ to calculate $\chi^\dagger(t)$. Instead one must obtain  numerical values from the expression
\be
\chi^\dagger(t) = \frac{\bar{\mu}_i(t,h^\dagger) - \bar{\mu}_0(t)}{h^\dagger},
\ee
where $h^\dagger$ has a value sufficiently small to obtain the required accuracy for $\chi^\dagger(t)$ at a given~$t$.  The zero-field ordered moment $\bar{\mu}_0$ is calculated from Eq.~(\ref{Eq:mubar0}) and $\bar{\mu}_i(h^\dagger)$ from Eqs.~(\ref{Eq:muiFM}) and~(\ref{Eq:yFM1}).  Here, we used the fixed value $h^\dagger=10^{-9}$, which gave an accuracy for the calculated $\chi^\dagger(t)$ of better than 0.01\% for $t \leq 0.995$.

\begin{figure}[t]
\includegraphics[width=3.3in]{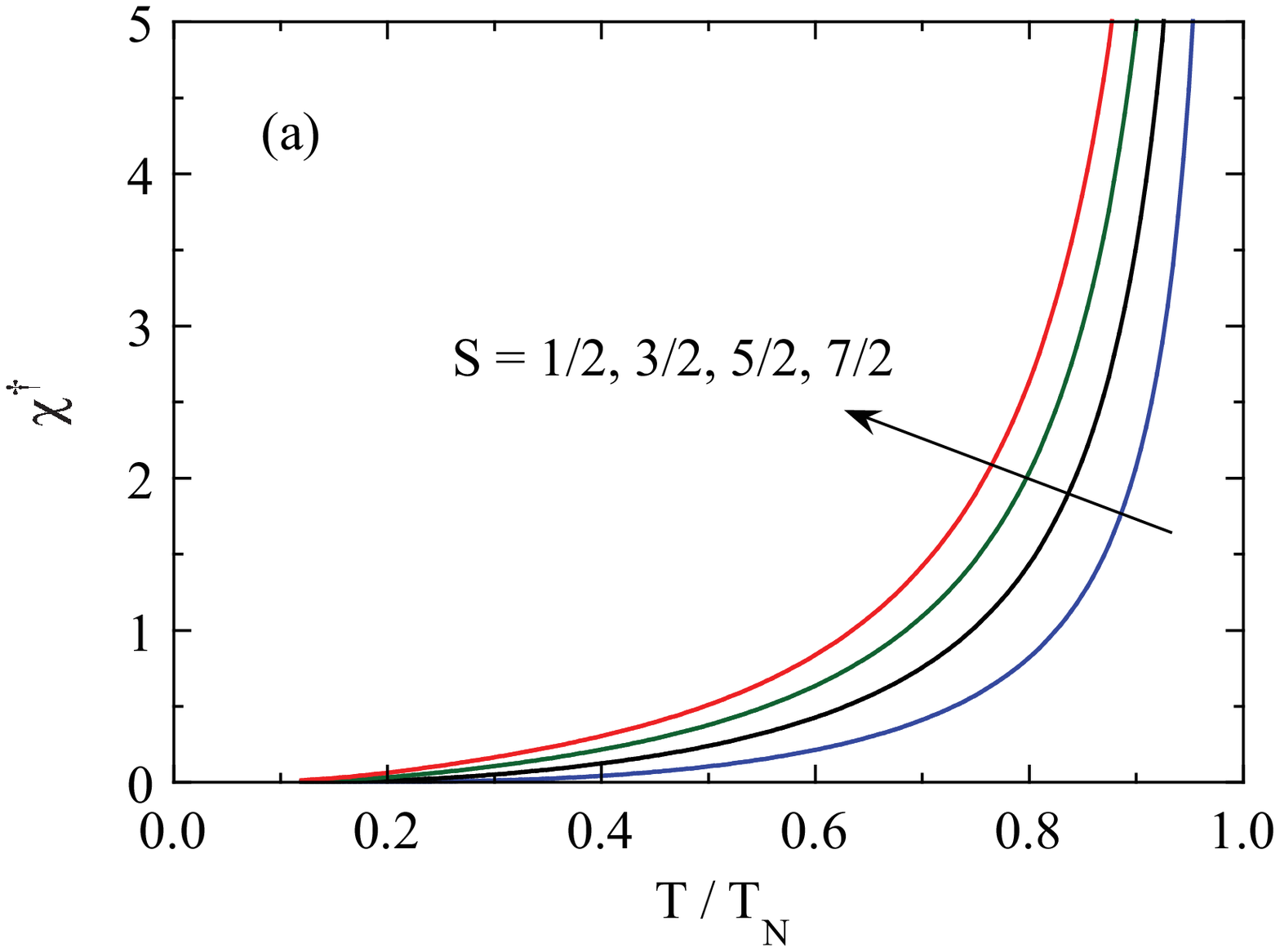}
\includegraphics[width=3.3in]{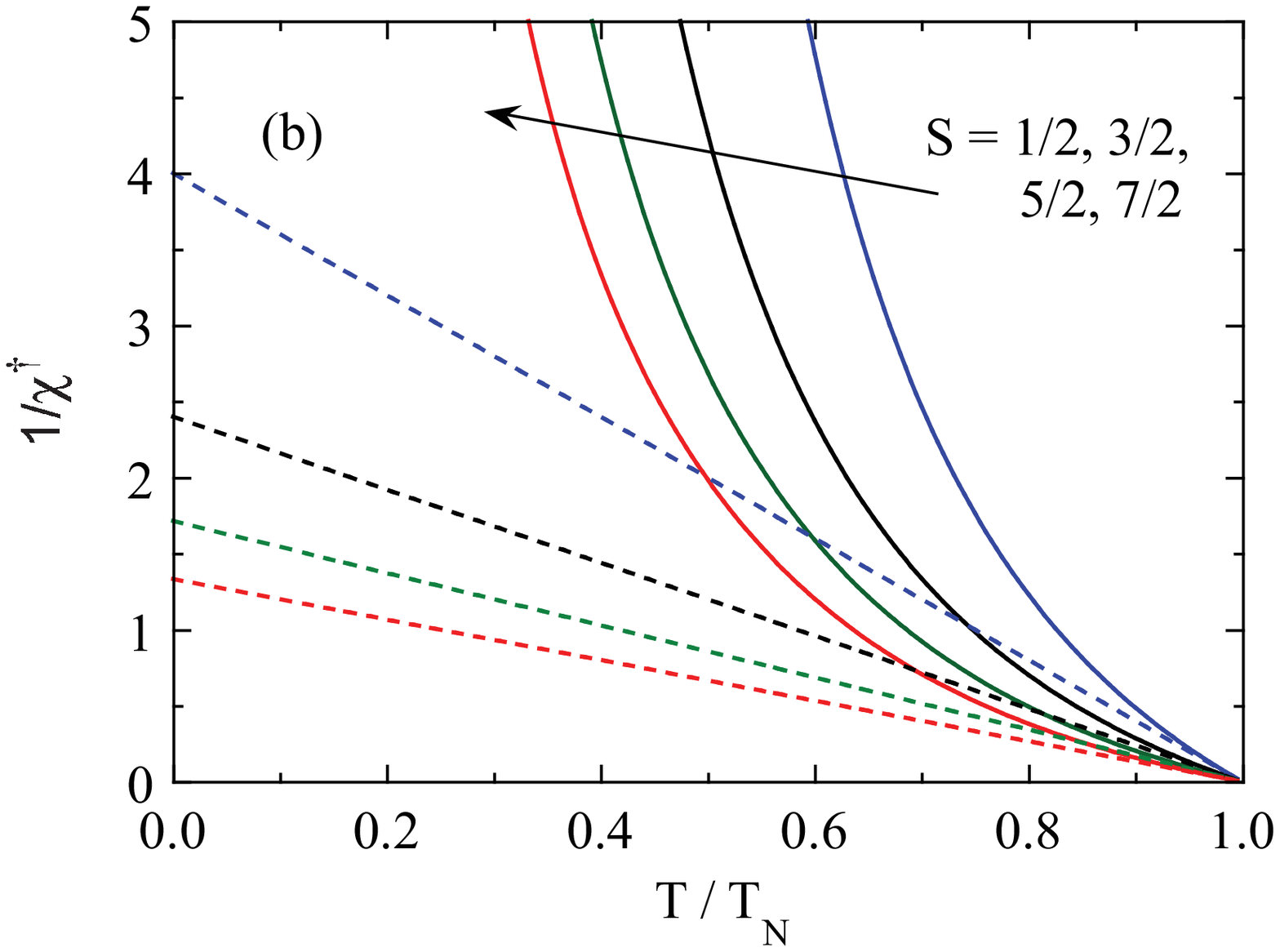}
\caption{(Color online) (a) Reduced staggered magnetic susceptibility $\chi^\dagger$ and (b) its inverse for Heisenberg antiferromagnets with spins $S = 1/2$, 3/2, 5/2 and 7/2 versus reduced temperature $t = T/T_{\rm N}$\@.  The straight dashed lines in (b) are extrapolations of the respective asymptotic Curie-Weiss-like critical behaviors for $1/\chi^\dagger$ at $t\to1^-$ in Eq.~(\ref{Eq:chiDagCW}).  These asymptotic critical behaviors are seen to be followed only at temperatures very close to $T_{\rm N}$.}
\label{Fig:chi_for_T_less_TN} 
\end{figure}

The results for $\chi^\dagger(t<1)$ are shown in Fig.~\ref{Fig:chi_for_T_less_TN}(a) for spins $S=1/2$, 3/2, 5/2 and~7/2.  The inverse staggered susceptibilities are shown in Fig.~\ref{Fig:chi_for_T_less_TN}(b), where the respective dashed straight lines are the inverses of the asymptotic Curie-Weiss-like critical behaviors given in Eq.~(\ref{Eq:chiDagCW}) below.  The uniform $\chi(t<1)$ for a single-domain Heisenberg FM containing  spins~$S$ is identical to the above result for $\chi^\dagger(t<1)$ for the Heisenberg AF with the same~$S$, with the changes in notation given in the caption of Fig.~\ref{Fig:ChiParallel_FM_S1_2}.

\section{\label{CritExps} Static Critical Exponents and Amplitudes}

The static critical exponents $\alpha$, $\alpha^\prime$, $\beta$, $\gamma$, $\gamma^\prime$ and~$\delta$ and the corresponding dimensionless reduced amplitudes $a,\ a^\prime,\ b,\ g,\ g^\prime$ and $d$ for magnetic systems are defined by Stanley, where the values obtained in mean-field theory for the critical exponents are given, together with the critical amplitudes for spin $S=1/2$.\cite{Stanley1971}  Here we calculate the critical exponents and amplitudes and give the general dependences of the critical amplitudes on the spin~$S$, which upon setting $S = 1/2$ are found to agree with the corresponding values calculated in Ref.~\onlinecite{Stanley1971} for $S = 1/2$. The described critical behaviors are the same for collinear and noncollinear AFs.  Furthermore, because the thermodynamic properties at $H=0$ in MFT of a single-domain ferromagnet and an antiferromagnet are the same, with the reduced uniform field~$h$ and magnetic susceptibility~$\chi$ and the Curie temperature~$T_{\rm C}$ of a ferromagnet replacing the reduced staggered field $h^\dagger$ and staggered magnetic susceptibility $\chi^\dagger$ and the N\'eel temperture $T_{\rm N}$ of an antiferromagnet, respectively, the static critical exponents and amplitudes are the same within MFT for FM and AF ordering.

\subsubsection{Magnetic Heat Capacity}

The critical behaviors of the molar magnetic heat capacity are defined by
\bse
\bea
\frac{C_{\rm mag}}{R} &=& a(t-1)^\alpha \qquad (t\to1^+),\\*
\frac{C_{\rm mag}}{R} &=& a^\prime(1-t)^{\alpha^\prime} \qquad (t\to1^-).
\eea
\ese
From Sec.~\ref{Sec:BSmu0TNCmag}, $C_{\rm mag}$ has the constant value of zero for $t\to1^+$.  Therefore one obtains
\be
\alpha = 0, \qquad a = 0 \qquad (t\to1^+).
\ee
Equation~(\ref{Eq:CmagTto0-}) shows that $C_{\rm mag}$ approaches the given finite value on approaching $T_{\rm N}$ from below, yielding
\be
\alpha^\prime = 0, \qquad a^\prime = \frac{5S(1+S)}{1+2S+2S^2}  \qquad (t\to1^-).
\ee

\subsubsection{Order Parameter}

The order parameter for a FM is the uniform magnetization and that for an AF is the staggered magnetization (the ordered moment per spin).  In a finite uniform field there is no FM phase transition because the order parameter for that transition (the uniform magnetization) is greater than zero at all finite temperatures.  In either case, for $H=0$ and $H^\dagger=0$, respectively, and $t\to1^-$ one has the same equation defining the critical exponent and amplitude given by
\be
\bar{\mu}_0 = b\,(1-t)^\beta.
\label{Eq:MCritExp}
\ee
From Eq.~(\ref{Eq:mu0Tto0}), the asymptotic critical behavior for $t\to1^-$ is
\be
\bar{\mu}_0 = \frac{\sqrt{\frac{10}{3}}(1+S)}{\sqrt{1+2S+2S^2}}(1-t)^{1/2}\qquad\quad (t\to1^-).
\label{Eq:mu0Asympt}
\ee
Comparing Eq.~(\ref{Eq:mu0Asympt}) with~(\ref{Eq:MCritExp}) gives the critical exponent and amplitude as
\be
\beta = \frac{1}{2},\qquad b = \frac{\sqrt{\frac{10}{3}}(1+S)}{\sqrt{1+2S+2S^2}}.
\ee

\begin{figure}[t]
\includegraphics [width=3.3in]{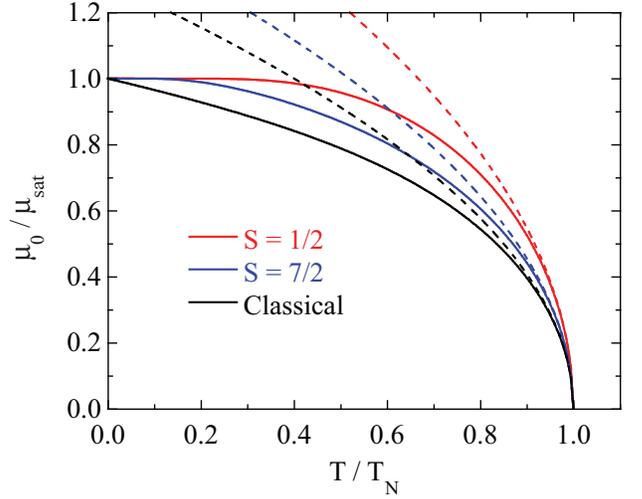}
\caption{(Color online) Comparisons of the normalized ordered moment $\bar{\mu}_0\equiv \mu_0/\mu_{\rm sat}$ versus reduced temperature $t = T/T_{\rm N}$ for spins $S = 1/2$, 3/2 and~$\infty$ (classical) (solid curves) from Fig.~10 of Ref.~\onlinecite{Johnston2011} with the respective asymptotic critical behaviors predicted by Eq.~(\ref{Eq:mu0Asympt}) (dashed curves).  The asymptotic critical behaviors describe the calculations rather well for $t\gtrsim0.9$.  The order of the curves from top to bottom is the same as in the figure legend. }
\label{Fig:OrderedMomentMFT3}
\end{figure}

Comparisons of $\bar{\mu}_0$ versus $t$ for spins $S = 1/2$, 3/2 and~$\infty$ (classical) from Fig.~10 of Ref.~\onlinecite{Johnston2011} with the asymptotic critical behaviors predicted by Eq.~(\ref{Eq:mu0Asympt}) are shown in Fig.~\ref{Fig:OrderedMomentMFT3}.  One sees that the calculations follow the critical behavior for $t\gtrsim 0.9$.  Quantitatively, the critical behavior values are larger than the calculations by 1\% at $t \approx 0.97\ (S = 1/2,\ \infty)$ and by 5\% at $t \approx 0.89$ for $S = 1/2$ and at $t\approx 0.84$ for $S = \infty$.

\subsubsection{Critical Magnetization versus Staggered Field Isotherm}

\begin{figure}[t]
\includegraphics [width=3.3in]{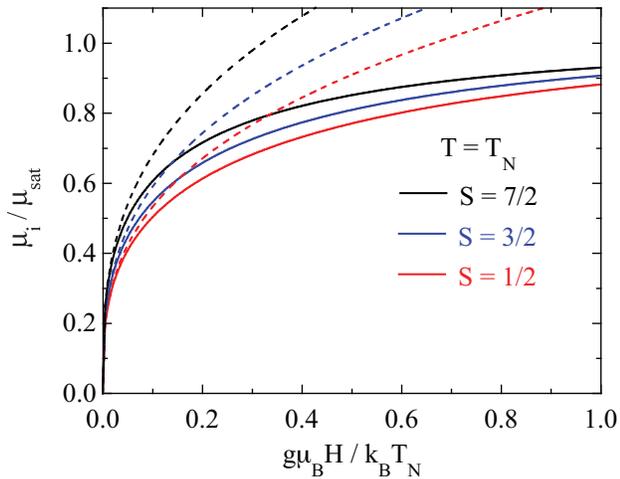}
\caption{(Color online) Comparisons of the critical ($t=1$) normalized induced moment $\bar{\mu}_i\equiv \mu_i/\mu_{\rm sat}$ versus reduced magnetic field $h \equiv g\mu_{\rm B}H/k_{\rm B}T_{\rm N}$ for spins $S = 1/2$, 3/2 and~7/2 (solid curves) calculated from Eqs.~(\ref{Eq:muiFM}) and~(\ref{Eq:yFM1}) together with the respective asymptotic critical behaviors predicted by Eq.~(\ref{Eq:muVShDagger}) (dashed curves).  The order of each set of dashed and solid curves from top to bottom is the same as in the figure legend. }
\label{Fig:M_vs_H_crit_isotherms}
\end{figure}

At the critical temperature $t=1$, there is no spontaneous (ordered) moment in zero field but a nonzero moment can be induced in the direction of an applied field $H$\@.  The critical exponent~$\delta$ and amplitude~$d$ for the critical ($t=1$) magnetization versus field isotherm are defined in terms of our dimensionless reduced units by
\be
h = d\,|\bar{\mu}_i|^\delta {\rm sgn}(\bar{\mu}_i) \qquad(\bar{\mu}_i\to0,\ t=1),
\ee
where the reduced field~$h$ is defined in Eq.~(\ref{Eq:barhDef}).  If $\delta$ is an integer, this relation becomes
\be
h = d\,\bar{\mu}_i^\delta\qquad(\bar{\mu}_i\to0,\ t=1).
\label{Eq:a_delta}
\ee
Within MFT, Eq.~(\ref{Eq:muVShDagger}) yields
\be
h = \left[\frac{9(1+2S+2S^2)}{10(1+S)^3}\right]{\bar{\mu}_i}^3.
\label{Eq:hdagVSbarmui}
\ee

Comparing Eq.~(\ref{Eq:hdagVSbarmui}) with~(\ref{Eq:a_delta}) gives the critical exponent and amplitude as
\be
\delta = 3,\qquad d = \frac{9(1+2S+2S^2)}{10(1+S)^3}.
\label{Eq:delta_d_values}
\ee

Critical magnetization versus field isotherms at~$t=1$ for spins $S=1/2$, 3/2 and~7/2 calculated with Eqs.~(\ref{Eq:muiFM}) and~(\ref{Eq:yFM1}) with $h$ replacing $h^\dagger$ are compared with the corresponding asymptotic critical behaviors predicted from Eq.~(\ref{Eq:muVShDagger}) in Fig.~\ref{Fig:M_vs_H_crit_isotherms}.  There is no ordered moment at $t=1$ and hence the induced moment is in the PM regime with all induced moments lined up with the field~{\bf H}\@.  One sees that the asymptotic critical behaviors are followed by the corresponding $\bar{\mu}_i$ versus $h$ calculations only very close to $h=0$.

\subsubsection{Magnetic Susceptibility}

To obtain the asymptotic magnetic susceptibilities for $t\to1^\pm$ we follow Stanley's exposition  for a single-domain ferromagnet\cite{Stanley1971} and first expand Eq.~(\ref{Eq:muiFM}) using~(\ref{Eq:yFM1}) in a Taylor series to first order in $h^\dagger$ and $\Delta t\equiv t-1$ and to third order in $\bar{\mu}_i$ and obtain
\bea
0 &=& \left(\frac{1+S}{3}\right)(1-\Delta t)h^\dagger  -\Delta t\,\bar{\mu}_i \label{Eq:BshExpand}\\*
&& -\ \left[\frac{3(1+2S+2S^2)}{10(1+S)^2}\right](1 - 3\Delta t)\bar{\mu}_i^3.\nonumber
\eea

For $t<1$ one has $\bar{\mu}_i>0$ for $h=0$.  Taking the partial derivative of both sides with respect to $h^\dagger$ and recognizing that $\partial \bar{\mu}_i/\partial h^\dagger\equiv \chi^\dagger$, where $\chi^\dagger$ is the dimensionless reduced staggered susceptibility, gives
\bea
0 &=& \left(\frac{1+S}{3}\right)(1-\Delta t) - \chi^\dagger \Delta t \label{Eq:hDeriv}\\*
&& -\ \chi^\dagger\left[\frac{9(1+2S+2S^2)}{10(1+S)^2}\right](1 - 3\Delta t)\bar{\mu}_i^2.\nonumber
\eea
Inserting the asymptotic critical behavior for $\bar{\mu}_i$ in Eq.~(\ref{Eq:mu0Asympt}) into~(\ref{Eq:hDeriv}) one obtains
\bse
\label{Eqs:ChiCritExpsCalc}
\be
\chi^\dagger = \frac{(1+S)/6}{1-t} \qquad (t\to1^-),
\label{Eq:chiDagCW}
\ee
where we have used the definition $\Delta t = -(1-t)$.  This has the form of a Curie-Weiss-like law even though it applies to the ordered state. 

In the PM temperature regime $t\geq 1$, the third term in Eq.~(\ref{Eq:BshExpand}) is negligible compared to the second, and from Eq.~(\ref{Eq:BshExpand}) one obtains 
\be
\chi^\dagger\equiv \frac{\bar{\mu}_i}{h} = \frac{(1+S)/3}{t-1}\qquad (t\to1^+),
\label{Eq:chiCritt>1}
\ee
\ese
which is a Curie-Weiss law where the Curie constant is a factor of two larger than in Eq.~(\ref{Eq:chiDagCW}) for the temperature regime $t<1$ as noted by Stanley.\cite{Stanley1971} 

The critical exponents and amplitudes for the isothermal staggered susceptibility of an AF are defined by
\bse
\label{Eqs:ChiCritExpsDef}
\bea
\chi^\dagger &=& g^\prime\,(1-t)^{-\gamma^\prime}\quad (t\to1^-),\\*
\chi^\dagger &=& g\,(t-1)^{-\gamma}\quad (t\to1^+),
\eea
\ese
where the direction of the staggered field for each spin for $t>1$ is the same as for $t<1$.  Comparing Eqs.~(\ref{Eqs:ChiCritExpsCalc}) with~(\ref{Eqs:ChiCritExpsDef}) gives the respective exponents and amplitudes as
\bse
\bea
\gamma &=& 1,\hspace{0.57in} \gamma^\prime = 1,\\*
g &=& \frac{S+1}{3},\qquad g^\prime = \frac{S+1}{6}.
\eea
\ese

The straight dashed lines in Fig.~\ref{Fig:chi_for_T_less_TN}(b) are plots of the asymptotic critical behaviors at $t < 1$ of the inverse staggered susceptibility of an AF obtained from Eq.~(\ref{Eq:chiDagCW}) for spins $S=1/2$, 3/2, 5/2 and~7/2.  As seen from the figure, the asymptotic critical behavior for each spin value is only realized at temperatures very near~$T_{\rm N}$.

Thus the staggered susceptibility of an AF diverges on approaching $T_{\rm N}$ both from below and above.  For $T \geq T_{\rm C}$, the uniform susceptibility of a FM diverges for $T\to T_{\rm C}^+$, whereas as discussed in the following section, the uniform susceptibility of an AF does not diverge at $T_{\rm N}$.  Another way of saying this is that a uniform applied magnetic field does not directly couple to the AF order parameter, which is the staggered magnetization instead of the uniform magnetization as for a FM\@.

\section{\label{Sec:CWLaw} The Curie-Weiss Law for Temperatures in the Paramagnetic Regime}

In the PM state at temperatures above $T_{\rm N}$, the thermal average of each magnetic moment is in the direction of the applied field.  Hence $\alpha_{ji}=0$ in Eq.~(\ref{Eq:HexchDef3}) and one obtains 
\be
H_{{\rm exch}\,i} = -\frac{\mu_i}{g^2\mu_{\rm B}^2}\sum_j J_{ij},
\label{Eq:Hexchipara}
\ee
where $\mu_i$ is the thermal-average magnetic moment in the direction of {\bf H}, which is the same for all spins and can therefore be taken out of the sum.  Then Eqs.~(\ref{Eq:BSyTaylor}), (\ref{Eq:muiFM}), (\ref{Eq:yFM1}) with $h$ replacing $h^\dagger$, and~(\ref{Eq:Hexchipara}) yield the Curie-Weiss law
\begin{subequations}
\label{CWlaw}
\be
\chi(T) = \frac{\mu_i}{H} = \frac{C_1}{T-\theta_{\rm p}}
\label{Eq:CWLaw}
\ee
with
\be
\theta_{\rm p} = -\frac{S(S+1)}{3k_{\rm B}}\sum_j J_{ij},
\label{Eq:WeissTemp}
\ee
\end{subequations}
where the single-spin Curie constant $C_1$ is given above in Eq.~(\ref{Eq:CurieConst2}) and $\theta_{\rm p}$ is the Weiss temperature.  It is possible for a system of interacting spins to have a Curie-law susceptibility ($\theta_{\rm p}=0$).  From Eq.~(\ref{Eq:WeissTemp}), this can happen if the sum of the exchange constants accidentally satisfies $\sum_j J_{ij} = 0$.

One can write calculations of $\chi(T)$ for local moment Heisenberg AFs within MFT in terms of the physically measurable ratio 
\be
f\equiv \frac{\theta_{\rm p}}{T_{\rm N}} = \frac{\sum_j J_{ij} }{\sum_j J_{ij}\cos\phi_{ji}},
\label{Eq:fRatioDef}
\ee
where for the second equality Eqs.~(\ref{Eq:TmGeneral}) and~(\ref{Eq:WeissTemp}) were used.  For a FM, $\phi_{ji}=0$ for all~$j$, and hence $f=1$.  For AFs, at least one of the $J_{ij}$ has to be positive (AF interaction) and at least one of the $\phi_{ji}\neq0$, leading to $f<1$.  Thus within MFT, if AF ordering is caused solely by exchange interactions, one requires
\be
-\infty<f<1.
\label{Eq:fRange}
\ee
By definition $T_{\rm N}>0$, whereas $\theta_{\rm p}$ for an AF can be either negative (the usual case) or positive, leading via the first equality in Eq.~(\ref{Eq:fRatioDef}) to a corresponding negative or positive value of~$f$.  The latter result occurs when the dominant $J_{ij}$ interactions are FM (negative), but where AF (positive) interactions cause the overall magnetic structure to be AF.  For AFs, $|f|$ is called the ``frustration parameter'' for AF ordering.\cite{Ramirez1994, Ramirez2001, Moessner2006}  A value $|f| \gg 1$ means that $T_{\rm N}$ is suppressed far below the value $|\theta_{\rm p}|$ expected from MFT for bipartite AFs with equal nearest-neighbor interactions, which is suggestive of strong frustration effects for AF ordering that arise from geometric and/or bond frustration.

The Curie-Weiss law in Eq.~(\ref{Eq:CWLaw}) can be written as a law of corresponding states
\bse
\label{Eqs:ChiPM}
\be
\frac{\chi(t)T_{\rm N}}{C_1} = \frac{1}{t-f}\qquad (T \geq T_{\rm N}),
\label{Eq:CWLawRed}
\ee
where the reduced temperature~$t$ was previously defined in Eq.~(\ref{Eq:tDef}).  The right side of Eq.~(\ref{Eq:CWLawRed}) has no explicit dependence on $S$, on the detailed type of spin lattice, or on the exchange constants in the system.  These quantities are implicitly contained in $t$  and $f$.  At the ordering temperature $T = T_{\rm N}$ ($t=1$), Eq.~(\ref{Eq:CWLawRed}) gives
\be
\frac{\chi(T_{\rm N})T_{\rm N}}{C_1} = \frac{1}{1-f}.\qquad (T = T_{\rm N})
\label{Eq:Chi(TN)}
\ee
\ese
The ratio of the isotropic $\chi(T>T_{\rm N})$ to $\chi(T=T_{\rm N})$ is given by Eqs.~(\ref{Eqs:ChiPM}) as
\be
\frac{\chi(t)}{\chi(T_{\rm N})} = \frac{1-f}{t-f}\qquad (T \geq T_{\rm N}).
\label{Eq:ChiPMNorm}
\ee

Since the left-hand side of Eq.~(\ref{Eq:Chi(TN)}) must necessarily be positive, MFT and the Heisenberg model require the right-hand side also to be positive.  This constrains $f$ to be in the range already given in Eq.~(\ref{Eq:fRange}).  This equality can be violated in practice if the Heisenberg model and MFT are inadequate to describe the spin system in the PM state above $T_{\rm N}$. 

From Eqs.~(\ref{Eq:TmGeneral}), (\ref{Eq:CurieConst2}) and~(\ref{Eq:WeissTemp}), one obtains
\be
T_{\rm N} - \theta_{\rm p} = \frac{S(S+1)}{3k_{\rm B}}\sum_j J_{ij}(1-\cos\phi_{ji}),
\label{Eq:TmTheta}
\ee
where $\phi_{ji}$ is the angle between ordered moments $j$ and~$i$ in the ordered AF state with $H=0$.   Using Eqs.~(\ref{Eq:CurieConst2}) and~(\ref{Eq:TmTheta}), the (isotropic) PM susceptibility at the N\'eel temperature is given by the Curie-Weiss law~(\ref{Eq:CWLaw}) as
\begin{subequations}
\label{Eqs:CWTNs}
\bea
\chi(T_{\rm N}) &=& \frac{C_1}{T_{\rm N} - \theta_{\rm p}}\label{Eq:CWTN}\\*
&=& \frac{g^2\mu_{\rm B}^2}{\sum_j J_{ij}(1-\cos\phi_{ji})},\label{Eq:TmJs}
\eea
\end{subequations}
which, perhaps surprisingly, is independent of $S$.

\subsection{Van Vleck's Solution for $T_{\rm N}$ and $\theta_{\rm p}$}

Van Vleck's solution for $T_{\rm N}$ and $\theta_{\rm p}$ for the bipartite AF with identical nearest-neighbor AF interactions $J>0$ between identical spins using the two-sublattice formulation of MFT theory is\cite{VanVleck1941}
\begin{subequations}
\label{Eqs:CWVV}
\bea
T_{\rm N} &=& \frac{S(S+1)zJ}{3k_{\rm B}},\label{Eq:TNVV}\\*
\theta_{\rm p} &=& -\frac{S(S+1)zJ}{3k_{\rm B}},\label{Eq:thetaVV}\\*
f &=& \frac{\theta_{\rm p}}{T_{\rm N}} = -1\label{Eq:fVV},
\eea
\end{subequations}
where $z$ is the nearest-neighbor coordination number of a magnetic moment by magnetic moments in the opposite sublattice and $f$ is defined in the first equality in Eq.~(\ref{Eq:fRatioDef}).  Thus $T_{\rm N} = -\theta_{\rm p}$, where $\theta_{\rm p}<0$ .  By comparing Eqs.~(\ref{Eqs:CWVV}) with~(\ref{Eq:TmGeneral}) and~(\ref{Eq:WeissTemp}), one sees that in going from Van~Vleck's theory to the general formulation of the MFT, one replaces $zJ$ in Eq.~(\ref{Eq:TNVV}) by $-\sum_jJ_{ij}\cos{\phi_{ji}}$ and $zJ$ in Eq.~(\ref{Eq:thetaVV}) by $\sum_jJ_{ij}$.  Van~Vleck's value $f=-1$ is very restrictive compared with the range of values in Eq.~(\ref{Eq:fRange}) allowed by the general expression~(\ref{Eq:fRatioDef}).

\subsection{Van Vleck's Solution for Anisotropic Magnetic Susceptibility in the Antiferromagnetic State}

In this paper we consider Heisenberg spin systems containing identical crystallographically equivalent spins in which the magnetic structure contains ordered magnetic moments that are all aligned within the same plane.  Collinear spin systems and also planar helical and cycloidal noncollinear magnetic structures all fall into this category, and therefore have either one axis (for planar noncollinear magnetic structures) or two axes (for collinear magnetic structures) that are perpendicular to the plane or axis of the ordered moments, respectively.  The MFT prediction for the perpendicular susceptibility per spin $\chi_\perp$ of such systems for $T\leq T_{\rm N}$ all have the same behavior, and as shown in Sec.~\ref{Sec:ChiPerp} below is given by
\be
\chi_\perp(T\leq T_{\rm N}) = \chi(T_{\rm N}) = \frac{C_1}{T_{\rm N}-\theta_{\rm p}},
\label{Eq:ChiPerp(T)}
\ee
which is the same as has been previously derived for several special cases,\cite{Johnston2011, VanVleck1941, Yoshimori1959} where the second equality is obtained from the Curie-Weiss law in Eq.~(\ref{Eq:CWLaw}).

Van~Vleck's MFT solution\cite{VanVleck1941} for $\chi_\parallel(T\leq T_{\rm N})$ per spin of a collinear bipartite AF with only nearest-neighbor interactions in his Eq.~(15) is given in our dimensionless notation as
\bse
\be
\frac{\chi_\parallel(t) T_{\rm N}}{C_1} = \frac{1}{\tau^\ast(t) + 1}, 
\label{Eq:ChiParf-1}
\ee
where we define the dimensionless variable $\tau^\ast$ containing the reduced temperature $t$ as
\be
\tau^\ast(t) = \frac{(S+1)t}{3B_S^\prime(y_0)}
\label{Eq:tauastDef}
\ee
and $B_S^\prime(y_0)$, $t$ and $y_0$ are defined above in Eqs.~(\ref{Eq:dBSy0}), (\ref{Eq:tDef}) and~(\ref{Eq:mubar0}), respectively. Using the first term in the Taylor series expansion of $B_S^\prime(y_0)$ in Eq.~(\ref{Eq:BSprimeExpand}), when $T = T_{\rm N}$ ($t = 1$), Eq.~(\ref{Eq:tauastDef}) becomes
\be
\tau^\ast(t = 1) = 1
\label{Eq:taustaratTN}
\ee
\ese
and Eq.~(\ref{Eq:ChiParf-1}) yields a value of $\chi_\parallel(t=1)$ that is the same as  predicted at $T_{\rm N}$ by the Curie-Weiss law~(\ref{Eq:Chi(TN)}) using $f=-1$ in Eq.~(\ref{Eq:fVV}), as required.  From Eqs.~(\ref{Eq:ChiParf-1}) and~(\ref{Eq:taustaratTN}) one obtains
\be
\frac{\chi_\parallel(T_{\rm N}) T_{\rm N}}{C_1} = \frac{1}{2}\qquad (T=T_{\rm N}),
\ee
which together with Eq.~(\ref{Eq:ChiParf-1}) gives
\be
\frac{\chi_\parallel(T)}{\chi_\parallel(T_{\rm N})} = \frac{2}{\tau^\ast(t) + 1}.
\label{Eq:ChiParNormVV}
\ee

When $t\to0$, $B_S^\prime(y,t\to0) \to 0$ exponentially, $\tau^\ast(t\to0)=\infty$, and from Eq.~(\ref{Eq:ChiParf-1}) one gets
\be
\chi_\parallel(t\to0) = 0.
\ee
Plots of $\chi_\parallel(t) T_{\rm N}/C_1$ versus $t$ for $f=-1$ and spins $S=1/2$ and $S=7/2$ for $t<1$ and $t>1$ derived from Eqs.~(\ref{Eq:ChiParf-1}) and~(\ref{Eq:CWLawRed}), respectively, are given below in Fig.~\ref{Fig:ChiPar} and corresponding plots of the ratio $\chi_\parallel(T)/\chi_\parallel(T_{\rm N})$ versus $t$ obtained from Eqs.~(\ref{Eq:ChiParNormVV}) and~(\ref{Eq:ChiPMNorm}) are given in Fig.~\ref{Fig:ChiParNorm} below.

\subsection{Magnetization versus Field in the Paramagnetic State}

The PM state is a state in which there is no long-range magnetic order induced by interactions between the moments.  Let the applied field be in the $+z$ direction according to convention.  In the PM state, each thermal-average magnetic moment points in the direction of {\bf H}, and the exchange field~(\ref{Eq:HexchiDef}) thus also points in the direction of {\bf H} with $z$-component
\be
H_{{\rm exch}\,z} = -\frac{\mu_z}{g^2\mu_{\rm B}^2}\sum_j J_{ij},
\label{Eq:HexchPara}
\ee
which is the same for all spins~$i$ and hence the subscript~$i$ has been dropped.  Defining the reduced magnetic moment
\be
\bar{\mu}_z \equiv \frac{\mu_z}{\mu_{\rm sat}},
\label{Eq:muBarDef}
\ee
the exchange field can be written
\be
H_{{\rm exch}\,z} = -\frac{\bar{\mu}_zS}{g\mu_{\rm B}}\sum_j J_{ij}.
\label{Eq:HexchPara2}
\ee
Using Eq.~(\ref{Eq:WeissTemp}), Eq.~(\ref{Eq:HexchPara2}) becomes
\bse
\be
H_{{\rm exch}\,z} = \frac{3\bar{\mu}_zk_{\rm B}\theta_{\rm p}}{g\mu_{\rm B}(S+1)}
\label{Eq:Hexch:z0}
\ee
and we therefore have
\be
\frac{g\mu_{\rm B}H_{{\rm exch}\,z}}{k_{\rm B}T} = \frac{3\bar{\mu}_z\theta_{\rm p}}{(S+1)T}.
\label{Eq:Hexch:z}
\ee
\ese
Now Eqs.~(\ref{Eq:BS(y)}) using Eq.~(\ref{Eq:Bi0}) give
\be
\bar{\mu}_z = B_S\left[\frac{3\bar{\mu}_z\theta_{\rm p}}{(S+1)T} + \frac{g\mu_{\rm B}H}{k_{\rm B}T}\right].
\label{Eq:muvsBrill2}
\ee
For $H\to0$, using only the first term in the Taylor series expansion~(\ref{Eq:BSyTaylor}), Eq.~(\ref{Eq:muvsBrill2}) becomes the Curie-Weiss law $\mu_z = C_1H/(T-\theta_{\rm p})$ in Eq.~(\ref{Eq:CWLaw}).

In terms of the reduced temperature in Eq.~(\ref{Eq:tDef}) and the reduced magnetic field in Eq.~(\ref{Eq:barhDef}), Eq.~(\ref{Eq:muvsBrill2}) becomes
\be
\bar{\mu}_z = B_S\left[\frac{3\bar{\mu}_zf}{(S+1)t} + \frac{h}{t}\right],\qquad (t \geq 1)
\label{Eq:muvsBrill3}
\ee
where the measurable ratio $f = \theta_{\rm p}/T_{\rm N}$ is given in terms of the exchange constants and the AF structure in Eq.~(\ref{Eq:fRatioDef}).  This is the equation of state in MFT, in the form of a law of corresponding states for a given value of~$S$, for the PM phase that relates the measurable reduced state variables $\bar{\mu}_z,\ t$ and~$h$ to each other.  Equation~(\ref{Eq:muvsBrill3}) must be solved numerically for $\bar{\mu}_z$ for given values of $S$, $f$, $h$ and $t$.

\section{\label{Sec:ChiParThy} Uniform Parallel Susceptibility of Collinear Antiferromagnets below Their N\'eel Temperatures}

Here we generalize Van~Vleck's MFT calculation of $\chi_\parallel(T)$ at $T \leq T_{\rm N}$ for collinear AF structures\cite{VanVleck1941} to include cases where the spin lattice can have a discrete distribution of exchange interactions with its neighbors including possibly frustrating interactions.  As in Van~Vleck's theory, we consider the spins to be identical and crystallographically equivalent.  Most physical realizations of Heisenberg spin lattices showing collinear spin ordering are in this general category.  In particular, very few, if any, real collinear AFs exactly satisfy the Van Vleck theory requirement that $f = -1$. Indeed, Eq.~(\ref{Eq:fRange}) shows that a large range of $f$ values is possible.

In order to develop a formulation of MFT that does not use the concept of magnetic sublattices, one must self-consistently calculate the exchange field ${\bf H}_{{\rm exch}\,i}$ seen by a representative ordered moment $\vec{\mu}_i$, where both $\vec{\mu}_i$ and ${\bf H}_{{\rm exch}\,i}$ are changed by the applied magnetic field {\bf H}\@.  When {\bf H} is applied along the axis of a collinear magnetic structure at temperatures $0 < T < T_{\rm N}$, the magnetic field increases the magnitudes of the ordered moments parallel to {\bf H} and decreases those antiparallel to {\bf H}.  In the limit of small $H$, one can express this qualitative expectation for the magnitude $\mu_j$ of an arbitrary magnetic moment $\vec{\mu}_j$ as
\be
\mu_j = \mu_0 + \delta_{\rm max}\hat{\mu}_j\cdot {\bf H} = \mu_0 + \delta_{\rm max}H\cos\phi_j,
\label{Eq:muj}
\ee
where $\mu_0$ is the temperature-dependent magnitude of the ordered moment in $H=0$, $\delta_{\rm max}$ is a constant to be determined and $\phi_j$ is the angle between $\vec{\mu}_j$ and {\bf H} for $H = 0$.  For a collinear AF structure one has the two possibilities $\phi_j = 0^\circ$ or 180$^\circ$.  In this section, without loss of generality our central magnetic moment $\vec{\mu}_i$ in the collinear AF structure is chosen to be in the direction of {\bf H}, {\it i.e.}, $\phi_i = 0$.  Thus, for the central magnetic moment $\vec{\mu}_i$ one has 
\be
\mu_i-\mu_0 = \delta_{\rm max}H.
\label{Eq:mui0}
\ee
Furthermore, the angle $\phi_{ji}$ between magnetic moments $\vec{\mu}_j$ and $\vec{\mu}_i$ is the same as $\phi_j$ and Eq.~(\ref{Eq:muj}) becomes 
\be
\mu_j = \mu_0 + \delta_{\rm max}H\cos\phi_{ji}.
\label{Eq:muj2}
\ee
Since $\phi_{ji} = 0^\circ$ or $180^\circ$, the component of the exchange field~$H_{{\rm exch}\,i}$ in the direction of $\vec{\mu}_i$ is given by Eq.~(\ref{Eq:HexchDef3}) with $\alpha_{ji}=\phi_{ji}$ as
\bea
H_{{\rm exch}\,i} &=& -\frac{\mu_0}{g^2\mu_{\rm B}^2}\sum_j J_{ij}\cos\phi_{ji} - \frac{\delta_{\rm max}H}{g^2\mu_{\rm B}^2}\sum_jJ_{ij}\cos^2\phi_{ji}\nonumber\\*
&=& H_{{\rm exch}\,i0} - \frac{\delta_{\rm max}H}{g^2\mu_{\rm B}^2}\sum_jJ_{ij},
\label{Eq:HexchiCol}
\eea
where we have used Eq.~(\ref{Eq:Hexch0Def3}) for the first term and $\cos^2\phi_{ ji}=1$ in the second.

Using the definition of $\bar{\mu}$ in Eq.~(\ref{Eq:barmu0Def}), Eq.~(\ref{Eq:mui0}) becomes 
\be
 \delta_{\rm max}H = gS\mu_{\rm B}(\bar{\mu}_i - \bar{\mu}_0).
\ee
Substituting this into Eq.~(\ref{Eq:HexchiCol}) gives
\be
H_{{\rm exch}\,i}  = H_{{\rm exch}\,i0} - \frac{S(\bar{\mu}_i - \bar{\mu}_0)}{g\mu_{\rm B}}\sum_j J_{ij}.
\label{Eq:Eexchibarmu}
\ee
From Eqs.~(\ref{CWlaw}) one has
\be
\sum_j J_{ij} = -\frac{3k_{\rm B}\theta_{\rm p}}{S(S+1)}.
\ee
Substituting this into Eq.~(\ref{Eq:Eexchibarmu}) gives
\be
H_{{\rm exch}\,i} - H_{{\rm exch}\,i0} =  \frac{3k_{\rm B}\theta_{\rm p}(\bar{\mu}_i - \bar{\mu}_0)}{(S+1)g\mu_{\rm B}}.
\label{Eq:Eexchibarmu2}
\ee

Using Eqs.~(\ref{Eq:BS(y)}) one obtains
\be
\bar{\mu}_i = B_S\left[\frac{g\mu_{\rm B}}{k_{\rm B}T}\left(H_{{\rm exch}\,i} + H\right)\right].
\label{Eq:barmufromBS}
\ee
Taylor expanding the Brillouin function about $H=0$ to first order in~$H$ gives
\bea
\bar{\mu}_i &=& B_S\left(\frac{g\mu_{\rm B}}{k_{\rm B}T}H_{{\rm exch}\,i0}\right)\label{barmui3}\\*
&& +\ \left[\frac{g\mu_{\rm B}}{k_{\rm B}T}(H_{{\rm exch}\,i} - H_{{\rm exch}\,i0} + H)\right]B_S^\prime(y_0),\nonumber
\eea
where $y_0$ is defined in Eq.~(\ref{Eq:mubar0}) and the expression for $B_S^\prime(y_0)$ is given in Eq.~(\ref{Eq:dBSy0}).  From Eq.~(\ref{Eq:mubar0}), the first term is just $\bar{\mu}_0$ and we substitute Eq.~(\ref{Eq:Eexchibarmu2}) into the second term to obtain
\be
\bar{\mu}_i - \bar{\mu}_0 = \left[ \frac{3(\bar{\mu}_i - \bar{\mu}_0)\theta_{\rm p}}{(S+1)T}+ \frac{g\mu_{\rm B}H}{k_{\rm B}T}\right]B_S^\prime(y_0).
\ee
Solving for $\bar{\mu}_i - \bar{\mu}_0$ gives
\be
\bar{\mu}_i - \bar{\mu}_0 = \frac{\frac{(S+1)g\mu_{\rm B}H}{3k_{\rm B}}}{\frac{(S+1)T}{3B_S^\prime(y_0)}-\theta_{\rm p}}.
\ee
Utilizing the definition $\bar{\mu}_i - \bar{\mu}_0 = (\mu_i - \mu_0)/(g\mu_{\rm B}S)$ as in Eq.~(\ref{Eq:barmu0Def}), one obtains
\be
\mu_i - \mu_0 = \frac{C_1H}{\frac{(S+1)T}{3B_S^\prime(y_0)}-\theta_{\rm p}},
\label{Eq:muimu0}
\ee
where the single-spin Curie constant $C_1$ is defined in Eq.~(\ref{Eq:CurieConst2}).

The parallel susceptibility per spin is obtained from Eq.~(\ref{Eq:muimu0}) as
\be
\chi_\parallel(T) = \frac{\mu_i - \mu_0}{H} = \frac{C_1}{\frac{(S+1)T}{3B_S^\prime(y_0)}-\theta_{\rm p}}.
\label{Eq:ChiPar101}
\ee
Multiplying both sides of Eq.~(\ref{Eq:ChiPar101}) by $T_{\rm N}$ and dividing both sides by $C_1$ gives the dimensionless law of corresponding states for the parallel susceptibility for a given~$S$ as 
\bse
\label{Eqs:ChiParColl}
\be
\frac{\chi_\parallel(t)T_{\rm N}}{C_1} = \frac{1}{\tau^\ast(t)-f},
\label{Eq:ChiParTCol}
\ee
where the definition of $\tau^\ast(t)$ is given in Eq.~(\ref{Eq:tauastDef}) and is a function of $S$ in addition to~$t$.  Equation~(\ref{Eq:ChiParTCol}) becomes identical to Van~Vleck's prediction  in Eq.~(\ref{Eq:ChiParf-1}) by setting $f$ to his value $f=-1$.  Another previous special case described by Eq.~(\ref{Eq:ChiParTCol}) is the two-sublattice collinear AF with equal couplings between spins in the same and opposite sublattices, respectively [see Eq.~(4.18) in Ref.~\onlinecite{Nagamiya1955}].

As noted previously in Eq.~(\ref{Eq:taustaratTN}), $\tau^\ast(t=1) = 1$, so the isotropic susceptibility at $T_{\rm N}$ is predicted by Eq.~(\ref{Eq:ChiParTCol}) to be
\be
\frac{\chi(t=1)T_{\rm N}}{C_1} = \frac{1}{1-f}\qquad (T = T_{\rm N}).
\label{Eq:ChiParTNCol}
\ee
Equation~(\ref{Eq:ChiParTNCol}) for $\chi(T_{\rm N})$ is identical with the prediction of the Curie-Weiss law at $T_{\rm N}$ in Eq.~(\ref{Eq:Chi(TN)}), as required.  This is an important consistency check.

The parallel susceptibility normalized by the isotropic value at $T_{\rm N}$ is obtained by dividing Eq.~(\ref{Eq:ChiParTCol}) by~(\ref{Eq:ChiParTNCol}), yielding
\be
\frac{\chi_\parallel(T)}{\chi(T_{\rm N})}  =\frac{1-f}{\tau^\ast(t)-f},
\label{Eq:chiparonchiTN}
\ee
\ese
which only depends on the experimentally accessible parameters $t$, $f$ and $\chi(T_{\rm N})$, and the spin $S$ that one can often estimate from chemical or other considerations.  The temperature dependence of $\chi_\parallel$ comes only from $\tau^\ast(t)$, which also depends on $S$.  The exchange constants and spin lattice geometry do not appear explicitly in Eqs.~(\ref{Eq:ChiParTCol}) or~(\ref{Eq:chiparonchiTN}) but are implicit in the values of $f$ and $t$, so these are laws of corresponding states for a given~$S$.  By expanding Eq.~(\ref{Eq:chiparonchiTN}) in a Taylor series about $t=1$ to first order in $1-t$, one obtains
\be
\frac{\chi_\parallel(t)}{\chi(T_{\rm N})} = 1-\frac{2(1-t)}{1-f} \qquad (t\to1^-),
\label{Eq:ChiParNearTN}
\ee
where again the spin does not appear explicitly in this expression.  The initial slope $d[\chi_\parallel(t)/\chi(T_{\rm N})]/dt = 2/(1-f)$ near $T_{\rm N}$ increases as $f$ increases, where the allowable range is $-\infty<f<1$ as given in Eq.~(\ref{Eq:fRange}).  We also obtain
\be
\chi_\perp - \chi_\parallel(t) = \frac{2(1-t)}{1-f}\,\chi(T_{\rm N}) \qquad (t\to1^-),
\ee
where $\chi(T_{\rm N}) = \chi_\perp(T\leq T_{\rm N})$ in MFT\@.

\begin{figure}
\includegraphics [width=3.1in]{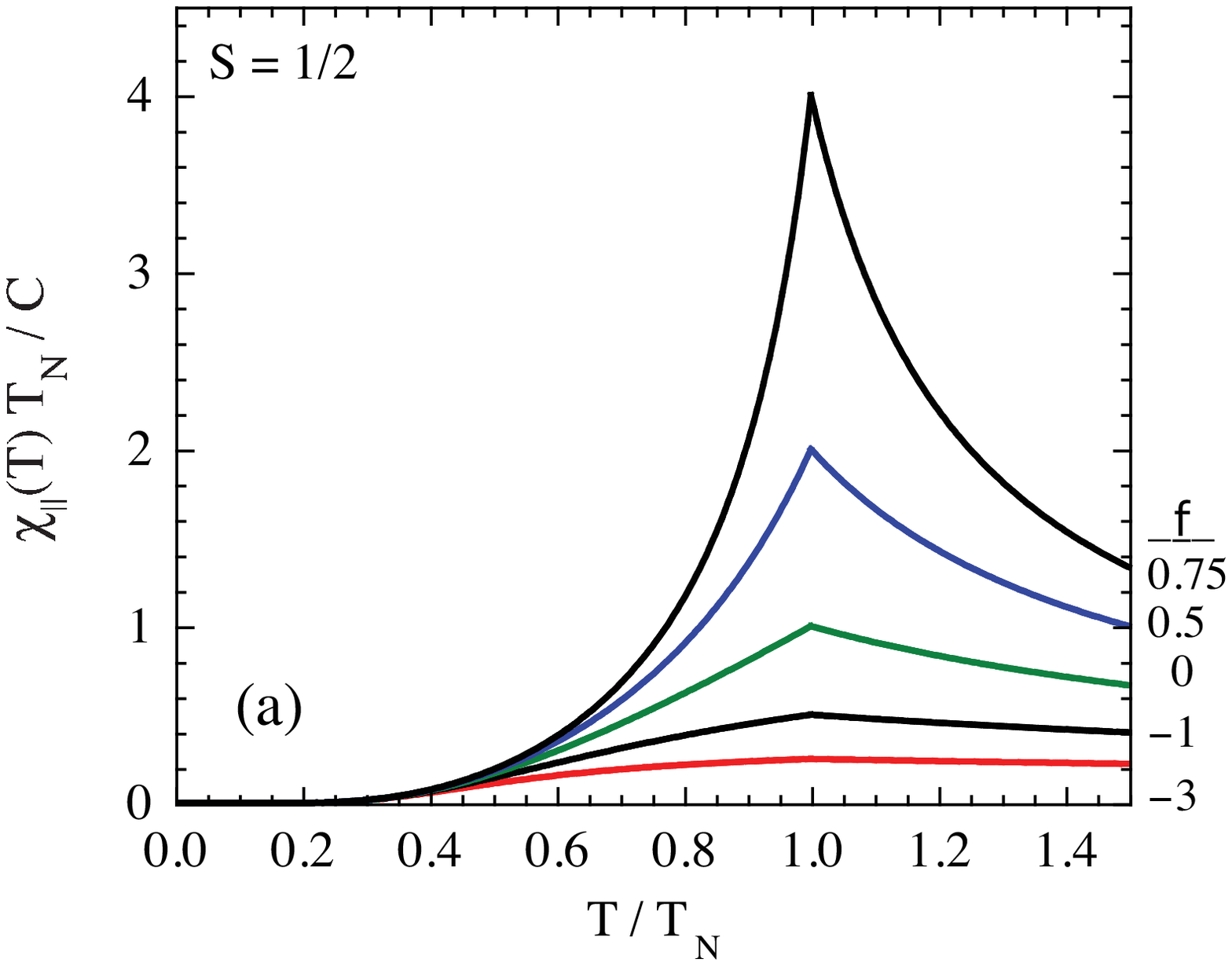}
\includegraphics [width=3.1in]{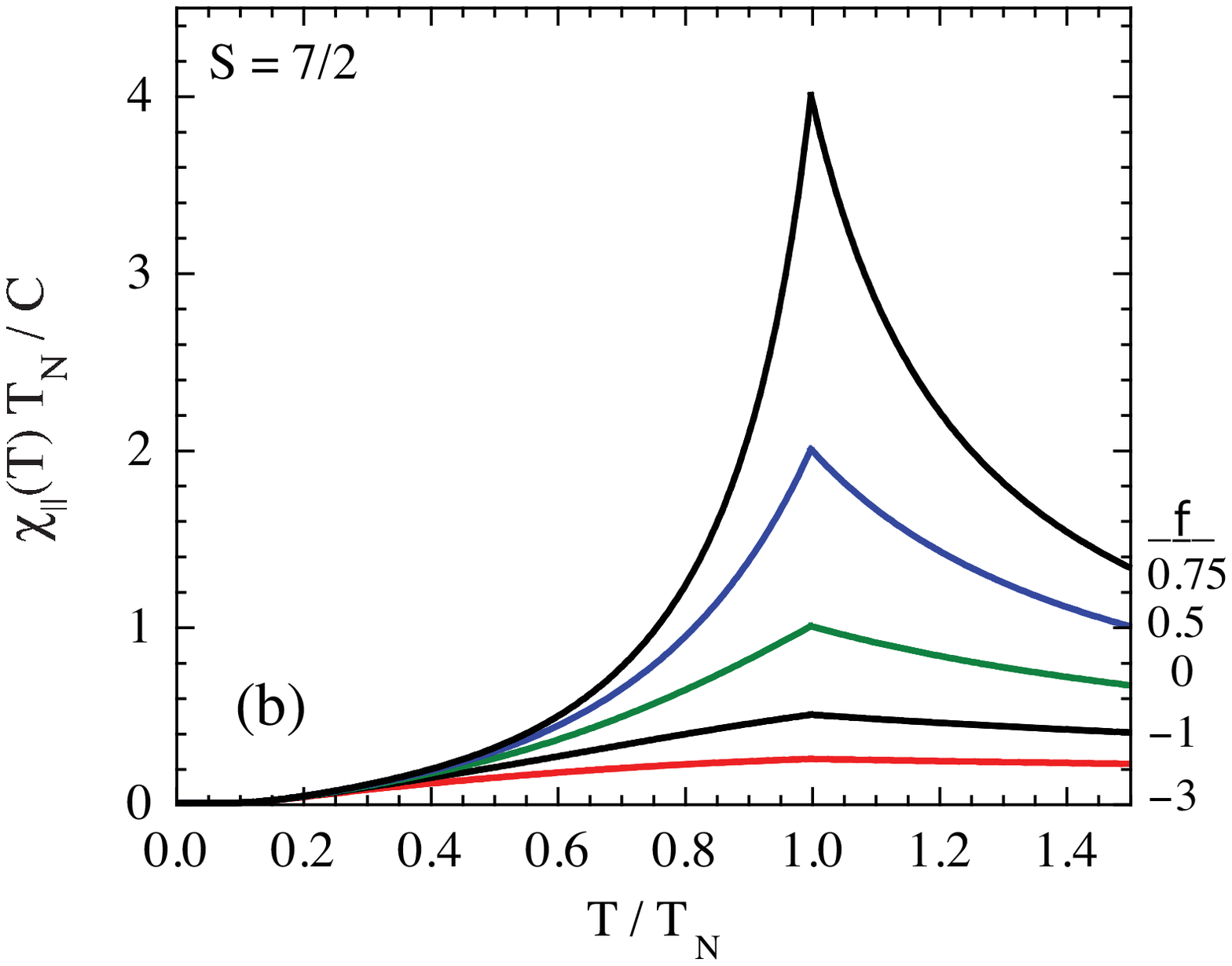}
\caption{(Color online) Normalized magnetic susceptibility parallel to the easy axis, $\chi_\parallel(T)T_{\rm N}/C$, versus temperature $T$ divided by the N\'eel temperature $T_{\rm N}$ for the listed values of $f \equiv \theta_{\rm p}/T_{\rm N}$ and for spins (a) $S = 1/2$ and (b) $S = 7/2$.  A negative value of $f$ reflects the dominance of antiferromagnetic interactions, and a positive value ferromagnetic interactions.  All curves shown correspond to collinear antiferromagnetic ordering at $T < T_{\rm N}$.  The data shown for $T/T_{\rm N} \geq 1$ and $T/T_{\rm N} \leq 1$ were obtained from Eqs.~(\ref{Eq:CWLawRed}) and~(\ref{Eq:ChiParTCol}), respectively.  The data in the former range do not depend on the spin~$S$, but in the latter range they do.  The maximum range of $f$ for Heisenberg antiferromagnets in MFT is given by Eq.~(\ref{Eq:fRange}) as $-\infty < f < 1$.}
\label{Fig:ChiPar}
\end{figure}

\begin{figure}
\includegraphics [width=3.3in]{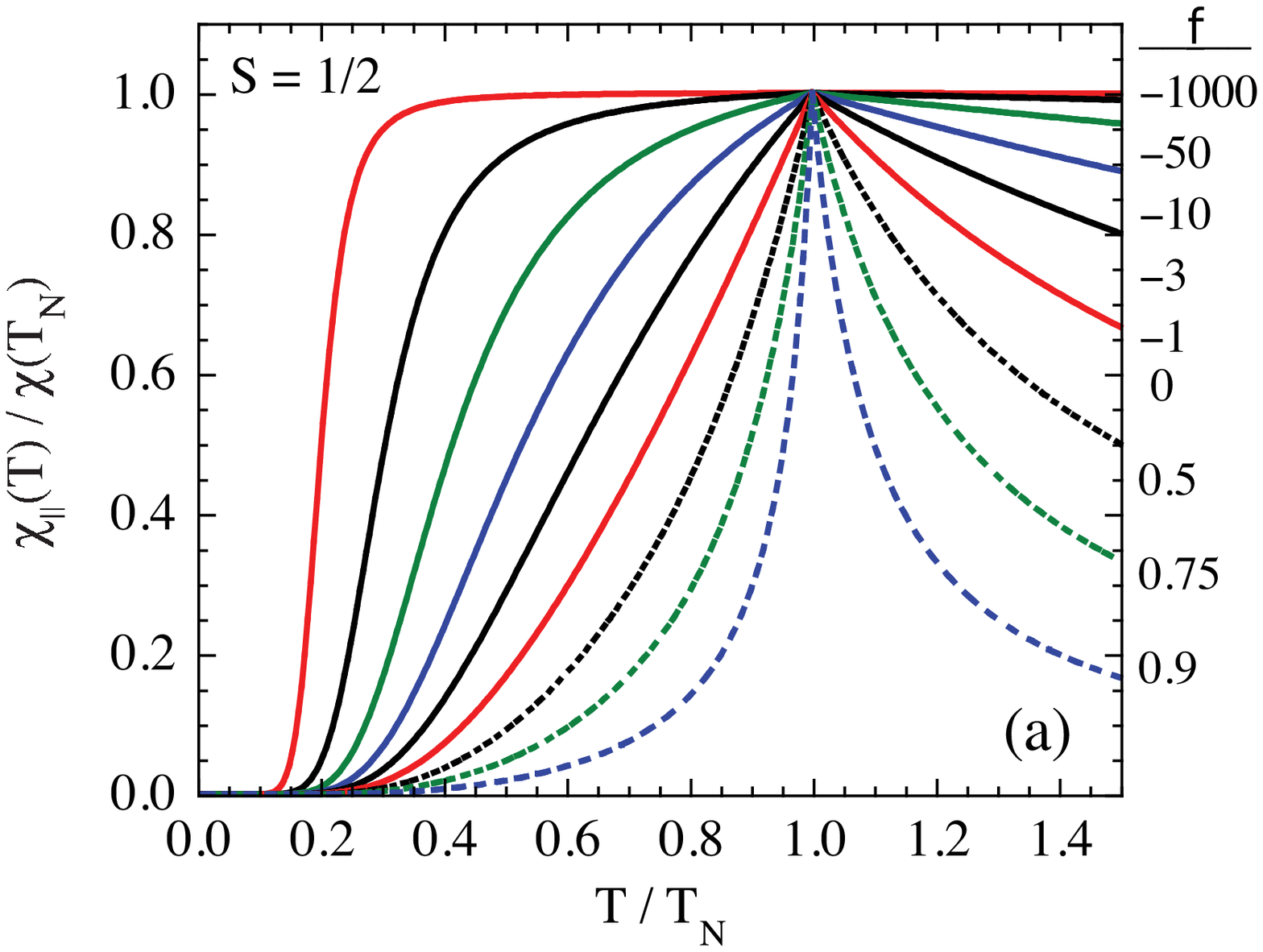}
\includegraphics [width=3.3in]{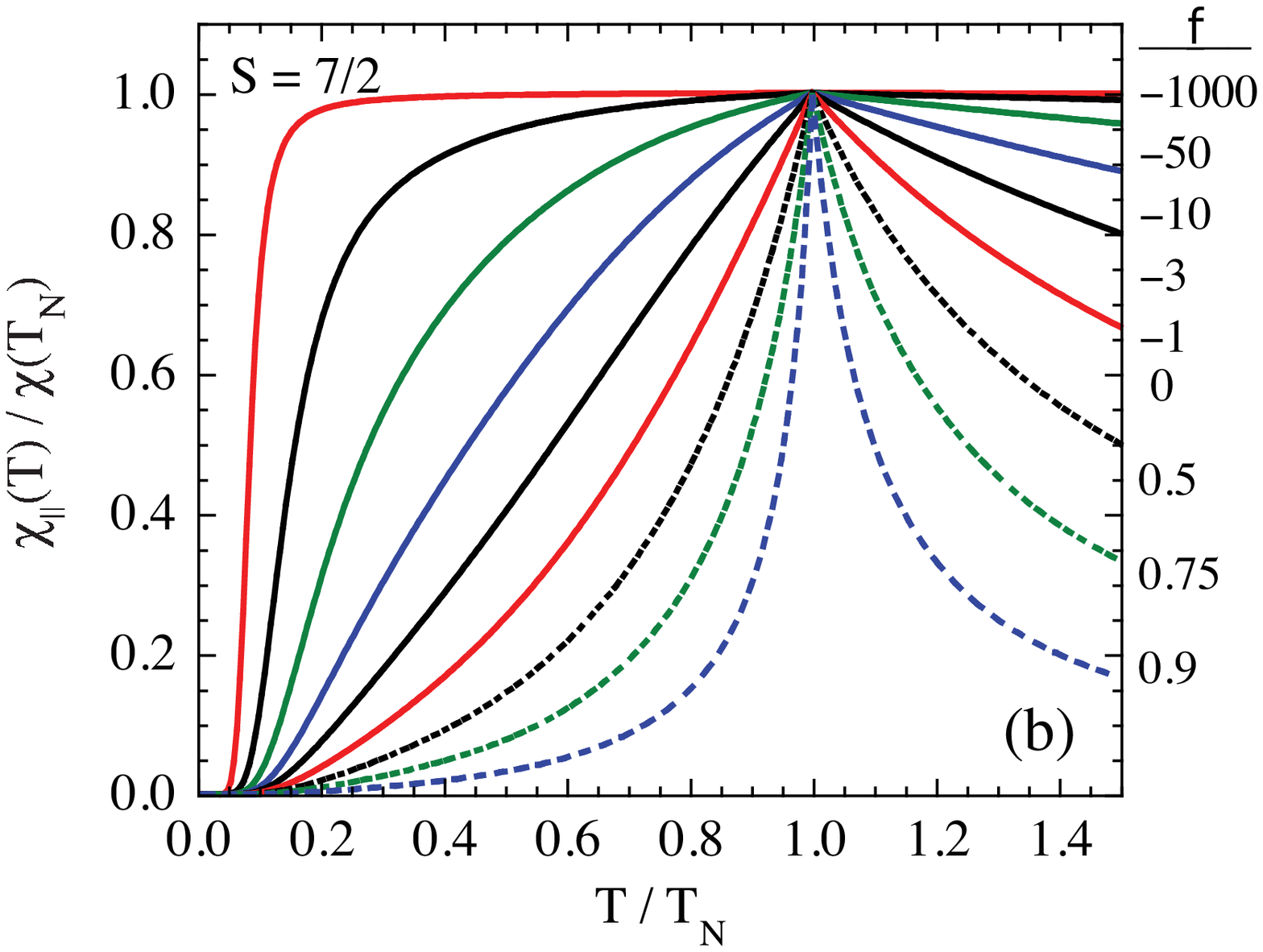}
\caption{(Color online) Normalized magnetic susceptibility parallel to the easy axis, $\chi_\parallel(T)/\chi(T_{\rm N})$, versus temperature $T$ divided by the N\'eel temperature $T_{\rm N}$ for the listed values of $f \equiv \theta_{\rm p}/T_{\rm N}$ and for spins (a) $S = 1/2$ and (b) $S = 7/2$.  The plots shown for $T/T_{\rm N} \geq 1$ and $T/T_{\rm N} \leq 1$ were obtained from Eqs.~(\ref{Eq:ChiPMNorm}) and~(\ref{Eq:chiparonchiTN}), respectively.}
\label{Fig:ChiParNorm}
\end{figure}

Plots of $\chi_\parallel(T)T_{\rm N}/C_1$ versus $T/T_{\rm N}$ for collinear antiferromagnets at $T/T_{\rm N}\geq 1$ and $T/T_{\rm N}\leq 1$ using Eqs.~(\ref{Eq:CWLawRed}) and~(\ref{Eq:ChiParTCol}), respectively, for several allowed values of $f$ are shown in Figs.~\ref{Fig:ChiPar}(a) and \ref{Fig:ChiPar}(b) for spins $S = 1/2$ and $S = 7/2$, respectively.  The plots for a given $f$ are the same above $T_{\rm N}$ for the two spin values, but not below.  Plots of normalized $\chi_\parallel(T)/\chi(T_{\rm N})$ versus $T/T_{\rm N}$ for $T/T_{\rm N} \geq 1$ and $T/T_{\rm N} \leq 1$ using Eqs.~(\ref{Eq:ChiPMNorm}) and~(\ref{Eq:chiparonchiTN}), respectively, for a large range of $f$ values are shown in Figs.~\ref{Fig:ChiParNorm}(a) and \ref{Fig:ChiParNorm}(b) for spins $S = 1/2$ and $S = 7/2$, respectively. One sees that the plots in Figs.~\ref{Fig:ChiPar} and~\ref{Fig:ChiParNorm} are not particularly sensitive to the value of $S$, but are very sensitive to the value of $f$.


\section{\label{Sec:ChianisNoncoll} Magnetic Susceptibility of Planar Noncollinear Antiferromagnets}

In the above-considered single-domain collinear Heisenberg AFs, the orientations of the ordered moments all lie along a single axis.  In the present section we generalize the MFT treatment to include noncollinear Heisenberg AFs where the ordered moments lie in a specified plane that we denote as the $xy$-plane.  The $z$~axis is defined in different ways depending on the type of magnetic structure and is not necessarily perpendicular to the $xy$~plane.  For example, from Fig.~1 of Ref.~\onlinecite{Johnston2012} and Fig.~1 of Ref.~\onlinecite{Goetsch2014}, the proper helix $z$~axis is perpendicular to the $xy$~plane and the cycloidal $z$~axis is parallel to the $xy$~plane.  Because of the different definitions of the $z$~axis, the out-of-plane direction is defined here as the ``perpendicular'' ($\perp$) direction, where $\hat{\bf i}\times\hat{\bf j} = \hat{\perp}$.  When the theory is applied to specific compounds, the $x$, $y$, $z$ and $\perp$~axes are assigned to the appropriate crystallographic directions. 

We follow Yoshimori\cite{Yoshimori1959} and calculate in MFT both the out-of-plane ($\chi_\perp$) and in-plane ($\chi_{xy}$) susceptibilities by solving for the conditions under which the equilibrium torque on a magnetic moment is zero in the presence of the net sum of the exchange and applied magnetic fields.  Yoshimori calculated these susceptibilities specifically for a proper helix magnetic structure for the body-centered tetragonal spin sublattice and for a specific configuration of exchange interactions.  In the following Secs.~\ref{Sec:ChiPerp} and~\ref{Sec:ChiPar} we generalize his treatment for calculating $\chi_{\perp}$ and $\chi_{xy}$, respectively.  

\subsection{\label{Sec:ChiPerp} Magnetic Susceptibility Perpendicular to the Ordering Plane}

\begin{figure}[t]
\includegraphics [width=1.75in]{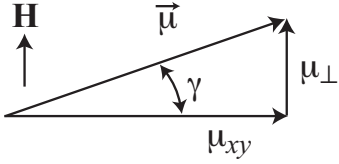}
\caption{Diagram showing the influence of an infinitesimal magnetic field {\bf H} along the $\perp$~axis on each spin originally aligned in the $xy$~plane.  The {\bf H} induces a tilting of each ordered magnetic moment $\vec{\mu}$ towards the magnetic field direction by an angle $\gamma$ (radian measure), which results in an induced $\perp$-axis component $\mu_\perp$ of $\vec{\mu}$.  The angle $\gamma$ is greatly exaggerated for clarity.  To first order in $\gamma\propto H$ the magnitude of the ordered moment $\mu_0$ in $H=0$ is unaffected by {\bf H}.}
\label{Fig:chiPerp}
\end{figure}

Since a collinear AF is a special case of a planar noncollinear AF, the generic predictions for the perpendicular susceptibility $\chi_\perp$ of the two types of ordering are identical.  The only assumptions made in this section for planar AF ordering, in which the ordered moments for $H=0$ lie in the same $xy$~plane, are that the spins are identical and crystallographically equivalent.  The spins themselves do not have to occupy the same plane.  The crystallographic equivalence assumption means that the spin coordination and exchange bond environment of every spin are the same.  To calculate the equilibrium conditions on the parameters, we calculate the conditions under which the net torque $\vec{\tau}_i$ on a representative magnetic moment $\vec{\mu}_i$ is zero
\be
\vec{\tau}_i = \vec{\mu}_i\times {\bf B}_i = 0
\ee
or
\be
\vec{\tau}_i = \vec{\mu}_i\times ({\bf H}_{{\rm exch}\,i} + {\bf H}) = 0,
\label{Eq:torqueperp0}
\ee
where the magnetic induction seen by $\vec{\mu}_i$ is
\be
{\bf B}_i = {\bf H}_{{\rm exch}\,i} + {\bf H}.
\label{Eq:Bi}
\ee
The $\chi_\perp$ is calculated with the magnetic field applied along the $\hat{\perp} = \hat{\bf i}\times \hat{\bf j}$ direction, i.e.,
\[
{\bf H} = H\hat{\bf \perp}.
\]

When calculating magnetic susceptibilities, we consider a representative central ordered moment $\vec{\mu}_i$ that interacts with its neighboring ordered moments $\vec{\mu}_j$.  To calculate $\chi_\perp$ we use cylindrical coordinates for the moment directions where the $\perp$ axis is the cylindrical axis and the moments in $H=0$ are aligned within the $xy$~plane.  To first order in the deviation angle $\gamma$ in Fig.~\ref{Fig:chiPerp}, one has
\be
\bs
\vec{\mu}_i &= \mu_0(\cos\phi_i\,\hat{\bf i} + \sin\phi_i\,\hat{\bf j} + \gamma\,\hat{\perp})\\*
\vec{\mu}_j &= \mu_0(\cos\phi_j\,\hat{\bf i} + \sin\phi_j\,\hat{\bf j} + \gamma\,\hat{\perp}),
\end{split}
\label{Eq:mui,muj}
\ee
where $\mu_0$ is the magnitude of the ordered moment of each spin in zero field at the particular temperature $T < T_{\rm N}$ of interest and $\phi_i$ and $\phi_j$ are the respective azimuthal angles of $\vec{\mu}_i$ and $\vec{\mu}_j$ with respect to the positive $x$~axis.  From Eqs.~(\ref{Eq:mui,muj}) the ordered moment is independent of $\gamma$ to first order in $\gamma$ (or to first order in $H$, since we will find that $H\propto \gamma$).  In this and the following section we express the azimuthal angle $\phi_j$ of a neighboring magnetic moment $\vec{\mu}_j$ in terms of the azimuthal angle $\phi_i$ of the central magnetic moment $\vec{\mu}_i$ and the azimuthal angle $\phi_{ji} = \phi_j-\phi_i$ between them.  Thus we write
\be
\bs
\phi_j&=\phi_i+\phi_{ji},\\*
\sin\phi_j &= \sin\phi_i\cos\phi_{ji} + \cos\phi_i\sin\phi_{ji},\\*
\cos\phi_j &= \cos\phi_i\cos\phi_{ji} - \sin\phi_i\sin\phi_{ji}.\\*
\end{split}
\label{Eq:sincosthetaj}
\ee
Inserting Eqs.~(\ref{Eq:mui,muj}) and~(\ref{Eq:sincosthetaj}) into Eq.~(\ref{Eq:HexchiDef}) for the exchange field ${\bf H}_{{\rm exch}\,i}$ and keeping only terms to order $\gamma$ gives the torque on $\vec{\mu}_i$ due to ${\bf H}_{{\rm exch}\,i}$ as
\begin{widetext}
\bea
\vec{\mu}_i\times {\bf H}_{{\rm exch}\,i} &=& -\frac{\gamma\mu_0^2}{g^2\mu_{\rm B}^2}\bigg\{(\sin\phi_i\,\hat{\bf i}-\cos\phi_i\,\hat{\bf j})\Big[\sum_j J_{ij}(1-\cos\phi_{ji})\Big] - (\cos\phi_i\,\hat{\bf i} + \sin\phi_i\,\hat{\bf j})\sum_j J_{ij}\sin\phi_{ji} \bigg\}
\label{Eq:mucrossHexchi}\\*
&& -\frac{\mu_0^2}{g^2\mu_{\rm B}^2} \hat{\perp}\sum_j J_{ij}\sin\phi_{ji}. \nonumber
\eea
\end{widetext}
This equation gives a torque on $\vec{\mu}_i$ even in zero field ($\gamma=0$)  unless the last term vanishes:
\be
\sum_jJ_{ij}\sin\phi_{ji} = 0.
\label{Eq:SumSinZero}
\ee
This condition must be satisfied by any planar AF structure in $H=0$ so that the structure is stable.  Condition~(\ref{Eq:SumSinZero}) is satisfied identically by collinear AFs, since for them $\phi_{ji} = 0^\circ$ or~180$^\circ$.

Using Eq.~(\ref{Eq:SumSinZero}), the torque contribution due to the exchange field in Eq.~(\ref{Eq:mucrossHexchi}) simplifies to
\begin{subequations}
\be
\vec{\mu}_i\times {\bf H}_{{\rm exch}\,i} = -\frac{\gamma\mu_0^2}{g^2\mu_{\rm B}^2}(\sin\phi_i\,\hat{\bf i}-\cos\phi_i\,\hat{\bf j})\sum_j J_{ij}(1-\cos\phi_{ji}).
\label{Eq:muiCrossHexchi}
\ee
Using Eqs.~(\ref{Eq:TmGeneral}) and~(\ref{Eq:WeissTemp}) one has 
\[
\sum_j J_{ij}(1-\cos\phi_{ji}) = \frac{3k_{\rm B}}{S(S+1)}(T_{\rm N} - \theta_{\rm p}).
\]
Inserting this expression into Eq.~(\ref{Eq:muiCrossHexchi}) and then using Eqs.~(\ref{Eq:CurieConst2}) and~(\ref{Eq:CWTN}), Eq.~(\ref{Eq:muiCrossHexchi}) becomes
\bea
\vec{\mu}_i\times {\bf H}_{{\rm exch}\,i} &=& -\gamma\mu_0^2\frac{T_{\rm N} - \theta_{\rm p}}{C_1}(\sin\phi_i\,\hat{\bf i}-\cos\phi_i\,\hat{\bf j})\nonumber\\*
&=& -\frac{\gamma\mu_0^2}{\chi(T_{\rm N})}(\sin\phi_i\,\hat{\bf i}-\cos\phi_i\,\hat{\bf j}).
\label{Eq:mucrossHexchi2}
\eea
\end{subequations}
The contribution of the applied magnetic field to the torque in Eq.~(\ref{Eq:torqueperp0}) to first order in $H$ is
\be
\vec{\mu}_i\times {\bf H} = \mu_0 H (\sin\phi_i\,\hat{\bf i}-\cos\phi_i\,\hat{\bf j}).
\label{Eq:HTorquei}
\ee

Setting the sum of the torque terms~(\ref{Eq:mucrossHexchi2}) and~(\ref{Eq:HTorquei}) equal to zero according to Eq.~(\ref{Eq:torqueperp0}) yields
\be
\frac{\mu_0\gamma}{H} = \chi(T_{\rm N}).
\label{Eq:gammaSoln}
\ee
The $\chi_\perp$ per spin is obtained to first order in $\gamma$ from Eq.~(\ref{Eq:gammaSoln}) and Fig.~\ref{Fig:chiPerp} as
\be
\chi_\perp(T\leq T_{\rm N}) = \frac{\mu_\perp}{H} = \frac{\mu_0\gamma}{H} = \chi(T_{\rm N}).
\label{Eq:ChiPerpSoln3}
\ee
The $T$-dependent ordered moment $\mu_0(T)$ canceled out of the calculation, so $\chi_\perp$ is independent of $T$ below $T_{\rm N}$.  The standard result~(\ref{Eq:ChiPerp(T)}) for collinear Heisenberg AFs obtained from MFT\cite{VanVleck1941} is of course identical to this general result~(\ref{Eq:ChiPerpSoln3}) for planar noncollinear AF structures with Heisenberg interactions since the former is a special case of the latter.

\subsection{\label{Sec:ChiPar} Magnetic Susceptibility Parallel to the Plane of the Ordered Magnetic Moments}

\subsubsection{Introduction}

\begin{figure}[t]
\includegraphics [width=1.5in]{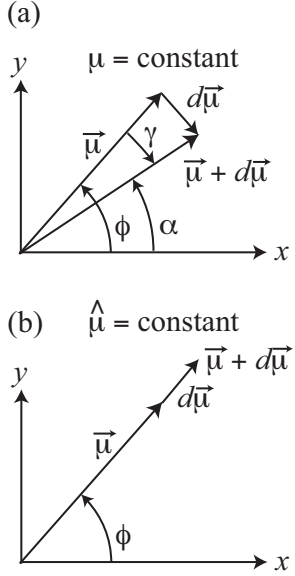}
\caption{A change $d\vec{\mu}$ in an ordered magnetic moment $\vec{\mu}$ due to an infinitesimal magnetic field ${\bf H}=H\,\hat{\bf i}$ can come about through (a) a change in direction $\hat{\mu}$ of the moment at constant magnitude $\mu$ and/or by (b) a change in the magnitude $\mu$ of the moment at constant direction $\hat{\mu}$. Positive azimuthal angles are measured in the counterclockwise direction.  Thus in panel~(a) $\gamma$ is negative and $\phi$ and $\alpha$ are positive.}
\label{Fig:Magnetic_susceptibility2}
\end{figure}

In this section we consider noncollinear ordered magnetic moments lying in the $xy$-plane as in Fig.~\ref{Fig:Magnetic_susceptibility2}, with the magnetic field applied in the azimuthal positive \mbox{$x$-axis} ($\hat{\bf i}$) direction
\be
{\bf H} = H\hat{\bf i}.
\label{Eq:vecH}
\ee
As in the previous section, the direction of the third Cartesian $\perp$ axis is defined by the right-hand rule as $\hat{\bf i}\times\hat{\bf j}=\hat{\perp}$, which in Fig.~\ref{Fig:Magnetic_susceptibility2} is pointed out of the page.  The direction of the moment $\vec{\mu}_i$ in $H=0$ is defined by the azimuthal angle $\phi_i$ with respect to the $\hat{\bf i}$ direction.  In the absence of any anisotropy, upon application of the infinitesimal {\bf H} the plane of the ordered moments would flop to a perpendicular orientation to lower the magnetic free energy of the system.  Therefore we assume that there is an infinitesimal XY anisotropy present that prevents this from happening; this anisotropy has no observable affect on any magnetic behaviors of the spin system predicted from the Heisenberg model.

On applying the field {\bf H}, the contribution to the torque on $\vec{\mu}_i$ due to {\bf H} is $\vec{\tau}_i=\vec{\mu}_i\times {\bf H}$ with magnitude $\tau_i = \mu_i H |\sin\phi_i|$.  This torque rotates $\vec{\mu}_i$ towards the direction of {\bf H} by an infinitesimal angle $\gamma_i$, as shown in exaggeration in  Fig.~\ref{Fig:Magnetic_susceptibility2}(a).  The maximum magnitude of the torque occurs for $|\phi_i| = \pi/2$~rad, at which $|\gamma_i|\equiv\gamma_{\rm max}$ where $\gamma_{\rm max}>0$.  The exchange field provides the restoring torque.  Since all spins are crystallographically equivalent by definition and the local exchange field in $H=0$ at each spin position is therefore the same, the restoring torque for each spin with the same value of $\phi$ is also the same.  Therefore the tilt angle for an arbitrary spin~$i$ due to the applied infinitesimal field is given by
\be
\gamma_i = -\gamma_{\rm max}\sin\phi_i,
\label{Eq:Gammai_planar}
\ee
which takes into account the negative sign of $\gamma_i$ in Fig.~\ref{Fig:Magnetic_susceptibility2}(a) and the angle $\phi_i$ of particular magnetic moment $\vec{\mu}_i$ with respect to {\bf H}\@.  On the other hand, if $\sin\phi_i$ is negative, then $\gamma_i$ is positive.  It will turn out that $\gamma_{\rm max} \propto H$, as expected, so we write Eq.~(\ref{Eq:Gammai_planar}) as
\be
\gamma_i = -\left(\frac{\gamma_{\rm max}}{H}\right)H\sin\phi_i,
\label{Eq:gammai}
\ee
where the quantity in parentheses is independent of $H$.  This gives rise to an in-plane susceptibility component $\chi_{xy\,i}\big|_{\mu}$ arising from the rotation of $\vec{\mu}_i$ due to the field at constant moment magnitude $\mu$ given by Fig.~\ref{Fig:Magnetic_susceptibility2}(a) and Eq.~(\ref{Eq:gammai}) as
\bea
\chi_{xy\,i}\big|_{\mu} &=& \frac{d\mu_{i}\big|_{\mu}\sin\phi_i}{H} = -\frac{\mu_0\gamma_i\sin\phi_i}{H}\nonumber\\*
 &=& \mu_0\left(\frac{\gamma_{\rm max}}{H}\right)\sin^2\phi_i,
\eea
where $\mu_0$ is the $T$-independent ordered moment for \mbox{$H = 0$}, from Fig.~\ref{Fig:Magnetic_susceptibility2} the change $d\mu_{i}\big|_{\mu}\sin\phi_i$ is the component of $d\vec{\mu}_{i}\big|_{\mu}$ in the direction of {\bf H}, and $d\mu_{i}\big|_{\mu} = -\mu_0\gamma_i$ where the negative sign comes from the sign convention for~$\gamma$ in Fig.~\ref{Fig:Magnetic_susceptibility2}.  Then the average susceptibility per spin for the entire spin system is obtained by averaging over~$i$, yielding
\be
\chi_{xy}\big|_{\mu} = \mu_0\left(\frac{\gamma_{\rm max}}{H}\right)\langle\sin^2\phi_i\rangle,
\label{Eq:ChiXYFixmu}
\ee
where $\langle\cdots\rangle$ denotes the average of the enclosed quantity over all magnetic moments $\vec{\mu}_i$.

As shown in Fig.~\ref{Fig:Magnetic_susceptibility2}(b), the applied field can also change the magnitude $\mu_i$ of an ordered moment.  For infinitesimal applied fields, we expect that the previous Eq.~(\ref{Eq:muj}) applies to planar noncollinear as well as collinear AFs, i.e., 
\be
\mu_i = \mu_0 + \delta_{\rm max}H\cos\phi_i
\nonumber
\ee
or
\be
d\mu_i\big|_{\hat{\mu}} = \mu_i - \mu_0  = \delta_{\rm max}H\cos\phi_i,
\label{Eq:dmu}
\ee
where $\delta_{\rm max}H$ is the maximum change in the magnitude of $\vec{\mu}$ due to {\bf H} that occurs when $\vec{\mu}\parallel {\bf H}$. One can now define a susceptibility contribution at fixed magnetic moment direction for spin~$i$ as
\bea
\chi_{xy\,i}\big|_{\hat\mu} &=& \frac{d\mu_{xy\,i}\big|_{\hat{\mu}}\cos\phi_i}{H} = \frac{\delta_{\rm max}H\cos^2\phi_i}{H}\nonumber\\*
 &=& \delta_{\rm max}\cos^2\phi_i.
\label{Eq:ChiXYConstMu}
\eea
Here the component of $d\vec{\mu}_i\big|_{\hat{\mu}}$ parallel to {\bf H} at fixed~$\phi_i$ is $d\mu_{xy\,i}\big|_{\hat{\mu}}\cos\phi_i$ according to Fig.~\ref{Fig:Magnetic_susceptibility2}(b). Then the average of the contribution~(\ref{Eq:ChiXYConstMu}) over the whole spin system per spin is
\be
\chi_{xy}\big|_{\hat\mu} = \delta_{\rm max}\langle\cos^2\phi_i\rangle.
\label{Eq:ChiXYFixmuHat}
\ee
The total average susceptibility of the system per spin is then obtained from Eqs.~(\ref{Eq:ChiXYFixmu}) and~(\ref{Eq:ChiXYFixmuHat}) as
\bse
\label{Eqs:ChiCompsDefined}
\bea
&&\hspace{-0.55in}\chi_{xy}  = \chi_{xy}\big|_{\mu} + \chi_{xy}\big|_{\hat\mu}\label{Eq:chiTot}\\*
&&\hspace{-0.3in} = \mu_0\left(\frac{\gamma_{\rm max}}{H}\right)\langle\sin^2\phi_i\rangle + \delta_{\rm max}\langle\cos^2\phi_i\rangle.\label{Eq:ChiParts0}
\eea
\ese

Equations~(\ref{Eq:Gammai_planar}) and~(\ref{Eq:dmu}) are the keys to calculating the in-plane susceptibility of large classes of planar noncollinear Heisenberg AFs without needing to define magnetic sublattices.  As in the calculation of $\chi_\parallel(T\leq T_{\rm N})$ for collinear AFs in Sec.~\ref{Sec:ChiParThy}, one must self-consistently calculate the exchange field ${\bf H}_{{\rm exch}\,i}$ seen by $\vec{\mu}_i$, where ${\bf H}_{{\rm exch}\,i}$ is itself changed by~{\bf H}.  We solve for the two unknowns $\gamma_{\rm max}$ and $\delta_{\rm max}$ in Eq.~(\ref{Eq:ChiParts0}) in the following two sections, respectively.  The resulting two simultaneous equations each contain both $\gamma_{\rm max}$ and $\delta_{\rm max}$, which allows us to solve for these two unknowns.

\subsubsection{Calculation of $\gamma_{\rm max}$}

In this section we solve for an expression relating $\gamma_{\rm max}$ and $\delta_{\rm max}$ derived from the condition that in equilibrium, the net torque on $\vec{\mu}_i$ in the presence of both the exchange field and the applied field at fixed moment magnitude is zero.  The magnetic induction ${\bf B}_i$ seen by $\vec{\mu}_i$ is
\be
{\bf B}_i = {\bf H}_{{\rm exch}\,i} + {\bf H}.
\ee
The net torque on $\vec{\mu}_i$ is therefore
\be
\bs
\vec{\tau}_i &= \vec{\mu}_i \times {\bf B}_i\\*
&= \vec{\mu}_i\times {\bf H}_{{\rm exch}\,i} + \vec{\mu}_i\times{\bf H} = 0.
\end{split}
\label{Eq:mucrossB}
\ee
The ordered moments are oriented within the $xy$~plane (the spatial spin lattice is not specified), and due to the assumed infinitesimal XY anisotropy, the moments remain in the $xy$~plane when the infinitesimal {\bf H} along the $x$~axis in Eq.~(\ref{Eq:vecH}) is applied.

The cross product $\vec{\mu}_i\times \vec{\mu}_j$ for $H > 0$ is given from its definition as
\be
\vec{\mu}_i\times \vec{\mu}_j = \mu_i\mu_j\sin\alpha_{ji}\hat{\perp},
\label{Eq:muiXmuj}
\ee
where the angle between $\vec{\mu}_j$ and $\vec{\mu}_i$ in $H>0$ is denoted by $\alpha_{ji} = \alpha_{j} - \alpha_{i}$ and $\alpha_{j},\ \alpha_{i}$ are the respective azimuthal angles of $\vec{\mu}_j$ and $\vec{\mu}_i$ with respect to the positive $x$~axis in $H>0$ [see Fig.~\ref{Fig:Magnetic_susceptibility2}(a)].  The direction of the cross product is in the direction of $\hat{\perp} = {\bf i}\times{\bf j}$ for $\alpha_{ji} > 0$ and is in the direction of $-\hat{\perp}$ for $\alpha_{ji} < 0$.  Using Eq.~(\ref{Eq:muiXmuj}) and the expression in Eq.~(\ref{Eq:HexchiDef}) for ${\bf H}_{{\rm exch}\,i}$, one obtains
\bea
\vec{\mu}_i\times {\bf H}_{{\rm exch}\,i} = -\frac{\hat{\perp}}{g^2\mu_{\rm B}^2}\sum_j J_{ij}\mu_i\mu_j\sin\alpha_{ji} .
\label{mucrossHi}
\eea

We now express all angles in terms of the azimuthal angle $\phi_i$ of the central magnetic moment $\vec{\mu}_i$ in $H = 0$ and the zero-field angle $\phi_{ji}$ between magnetic moments $\vec{\mu}_j$ and $\vec{\mu}_i$.  Referring to Fig.~\ref{Fig:Magnetic_susceptibility2}(a), for $H = 0$ one has
\be
\phi_{ji} \equiv \phi_j-\phi_i.
\label{Eq:phiji}
\ee
Similarly, when $H > 0$ we define
\be
\bs
\alpha_j &= \phi_j+\gamma_j, \qquad \alpha_{ji} \equiv\alpha_j-\alpha_i,\\*
\gamma_{ji}&\equiv\gamma_j-\gamma_i, \qquad\alpha_{ji} = \phi_{ji}+\gamma_{ji},
\end{split}
\label{Eq:alphajgammaji}
\ee
where $\gamma_j$ and $\gamma_i$ are defined in Fig.~\ref{Fig:Magnetic_susceptibility2}(a) and expressions for them are given in Eq.~(\ref{Eq:Gammai_planar}), yielding
\begin{subequations}
\be
\alpha_{ji} = \phi_{ji} - \gamma_{\rm max}\left[\sin\phi_i\cos\phi_{ji} + \cos\phi_i\sin\phi_{ji} - \sin\phi_i \right].
\ee
Using trig identities, to first order in $\gamma_{\rm max}$ one then obtains
\bea
\sin\alpha_{ji} &=& \sin\phi_{ji} -\gamma_{\rm max}(\sin\phi_i\cos\phi_{ji}\label{Eq:SinAlphaji}\\*
&&\hspace{0.2in} +\ \cos\phi_i\sin\phi_{ji} - \sin\phi_i)\cos\phi_{ji},\nonumber\\*
\cos\alpha_{ji}&=& \cos\phi_{ji} +\gamma_{\rm max}(\sin\phi_i\cos\phi_{ji}\label{Eq:CosAlphaji}\\*
&&\hspace{0.2in} +\ \cos\phi_i\sin\phi_{ji} - \sin\phi_i)\sin\phi_{ji}.\nonumber
\eea
\end{subequations}
Using Eqs.~(\ref{Eq:muj}) and~(\ref{Eq:phiji}) and a trig identity one obtains to first order in $H$
\be
\bs
\mu_i\mu_j 
&= \mu_0^2 +\delta_{\rm max}\mu_0H\cos\phi_i\\*
&\hspace{0.33in} +\ \delta_{\rm max}\mu_0H(\cos\phi_i\cos\phi_{ji} - \sin\phi_i\sin\phi_{ji}),
\end{split}
\label{Eq:muimuj}
\ee
where $\mu_0(T)$ is the ordered moment for $H=0$ and $\delta_{\rm max}\geq 0$ is a variable to be determined that depends on $T$ but not on $H$ (see below).

Substituting Eq.~(\ref{Eq:SinAlphaji}) for $\sin\alpha_{ji}$ and~(\ref{Eq:muimuj}) for $\mu_i \mu_j$ into Eq.~(\ref{mucrossHi}), to first order in $H$ one obtains
\begin{widetext}
\bea
\vec{\mu}_i\times {\bf H}_{{\rm exch}\,i} &=&-\frac{\hat{\perp}}{g^2\mu_{\rm B}^2}\Bigg\{\mu_0^2\bigg[\sum_jJ_{ij}\sin\phi_{ji} - \gamma_{\rm max}\sin\phi_i\sum_jJ_{ij}(\cos^2\phi_{ji} - \cos\phi_{ji})-\gamma_{\rm max}\cos\phi_i\sum_j J_{ij}\sin\phi_{ji}\cos\phi_{ji}\bigg]\nonumber\\*
&& \hspace{0.6in}+\ \mu_0\delta_{\rm max}H\bigg[ \cos\phi_i\Big(\sum_j J_{ij}\sin\phi_{ji} + \sum_j J_{ij}\sin\phi_{ji}\cos\phi_{ji}\Big) - \sin\phi_i\sum_jJ_{ij}\sin^2\phi_{ji}\bigg]\Bigg\}.\label{Eq:HexchTorque0}
\eea
\end{widetext}
The second term in Eq.~(\ref{Eq:mucrossB}) for the torque on $\vec{\mu}_i$ due to {\bf H} is, to first order in $H$,
\be
\vec{\mu}_i\times{\bf H} = -\mu_0H\sin\phi_i \hat{\perp}.
\label{Eq:muCrossH}
\ee
The net torque on $\vec{\mu}_i$ according to Eq.~(\ref{Eq:mucrossB}) is the sum of the two torques in Eqs.~(\ref{Eq:HexchTorque0}) and~(\ref{Eq:muCrossH}).

In order for the equilibrium net torque on $\vec{\mu}_i$ to be zero for $H=0$ and $\gamma_{\rm max} = 0$ (since $\gamma_{\rm max} \propto H$, see below) requires that the first term in Eq.~(\ref{Eq:HexchTorque0}) be zero, which gives
\be
\sum_jJ_{ij}\sin\phi_{ji} = 0.
\label{Eq:SumJijSinji}
\ee
This condition for the stability of the AF structure is the same as already given in Eq.~(\ref{Eq:SumSinZero}).  Furthermore, we only consider AF structures for which 
\be
\sum_j J_{ij}\sin\phi_{ji}\cos\phi_{ji} = \sum_j J_{ij}\sin(2\phi_{ji}) =0.
\label{Eq:sums=0}
\ee
This is a weak constraint.  Terms in the sum are zero if $\phi_{ji} = 0$ (FM alignment of moments $\vec{\mu}_i$ and~$\vec{\mu}_j$) or $180^\circ$ (AF alignment of $\vec{\mu}_i$ and~$\vec{\mu}_j$).  Therefore Eq.~(\ref{Eq:sums=0}) is satisfied identically for collinear AFs.  More generally, the sum is also zero for AF structures with inversion symmetry for which the AF structure consists of pairs of ordered moments $\vec{\mu}_i$ and~$\vec{\mu}_j$ and $\vec{\mu}_i$ and~$\vec{\mu}_k$ with couplings $J_{ij} = J_{ik}$ and with orientations with respect to the central moment $\vec{\mu}_i$ given by $\phi_{ji} = -\phi_{ki}$.  The latter situation occurs between moments in neighboring FM-aligned layers along the axes of helical and cycloidal AF structures within the $J_0$-$J_{z1}$-$J_{z2}$ model shown in Fig.~1 of Ref.~\onlinecite{Johnston2012} and Fig.~1 of Ref.~\onlinecite{Goetsch2014}, respectively.  Equation~(\ref{Eq:sums=0}) is also satisfied by some  AF structures and exchange models where the magnetic and structural unit cells are the same.\cite{Johnston2012}

Setting $\sum_jJ_{ij}\sin\phi_{ji} = \sum_j J_{ij}\sin\phi_{ji}\cos\phi_{ji} = 0$ in Eq.~(\ref{Eq:HexchTorque0}) according to Eqs.~(\ref{Eq:SumJijSinji}) and~(\ref{Eq:sums=0}) yields a simple expression for $\vec{\mu}_i\times {\bf H}_{{\rm exch}\,i}$ given by
\bea
\vec{\mu}_i\times {\bf H}_{{\rm exch}\,i} &=&\frac{\mu_0^2\gamma_{\rm max}\sin\phi_i\hat{\perp}}{g^2\mu_{\rm B}^2}\sum_j J_{ij}(\cos^2\phi_{ji}- \cos\phi_{ji}) \nonumber\\*
&&\hspace{-0.05in} +\ \frac{\mu_0\delta_{\rm max}H\sin\phi_i\hat{\perp}}{g^2\mu_{\rm B}^2}\sum_j J_{ij}\sin^2\phi_{ji}.
\label{Eq:muicrossHexchicalc}
\eea
Inserting Eqs.~(\ref{Eq:muCrossH}) and~(\ref{Eq:muicrossHexchicalc}) into~(\ref{Eq:mucrossB}) and solving for $\gamma_{\rm max}$ gives
\be
\frac{\gamma_{\rm max}\mu_0}{g^2\mu_{\rm B}^2H} = \frac{1-\frac{\delta_{\rm max}}{g^2\mu_{\rm  B}^2}\sum_j J_{ij}\sin^2\phi_{ji}}{\sum_j J_{ij}(\cos^2\phi_{ji}-\cos\phi_{ji})},
\label{Eq:gammaMaxSoln}
\ee
which is valid for AF structures and applied magnetic field directions such that the angles between the ordered magnetic moments and the applied field satisfy $\langle\sin^2\phi_i\rangle \neq 0$ in Eq.~(\ref{Eq:ChiXYFixmu}).  As expected, we find that $\gamma_{\rm max}\propto H$ to lowest order in $H$.  The maximum tilt angle $\gamma_{\rm max}$ due to $H$ depends in part on the maximum change $\delta_{\rm max}H$ of the moment magnitude, and hence includes a component arising from the changes in magnitudes of the magnetic moments as they rotate in response to the field.  Since $\delta_{\rm max}$ depends on temperature (see below), so does $\gamma_{\rm max}$.

In addition to planar noncollinear AF structures, Eq.~(\ref{Eq:gammaMaxSoln}) also applies to the special case of collinear AFs for an applied field direction perpendicular to the ordering axis, because in that case the moment magnitudes do not change as a result of a small applied field and hence $\delta_{\rm max} = 0$.  Then substituting $\cos^2\phi_{ij}=1$ into Eq.~(\ref{Eq:gammaMaxSoln}) for a collinear AF and then substituting Eq.~(\ref{Eq:gammaMaxSoln}) with $\langle\sin^2\phi_i\rangle=1$ into~(\ref{Eq:ChiXYFixmu}) and using Eq.~(\ref{Eq:TmJs}) gives the expression $\chi_\perp(T\leq T_{\rm N}) = \chi(T_{\rm N})$ already derived in Eq.~(\ref{Eq:ChiPerpSoln3}) for the collinear AF and thus provides an important consistency check.

Equation~(\ref{Eq:gammaMaxSoln}) contains two unknowns $\gamma_{\rm max}$ and~$\delta_{\rm max}$.  In the following section an independent equation is obtained in these two unknowns, which allows us to solve for both separately in Sec.~\ref{Sec:SolveChiXY} and thereby obtain the in-plane susceptibility of a planar noncollinear AF structure utilizing Eqs.~(\ref{Eq:ChiXYFixmu}), (\ref{Eq:ChiXYFixmuHat}) and~(\ref{Eqs:ChiCompsDefined}).

\subsubsection{Calculation of $\delta_{\rm max}$}

Here we obtain an expression for $\delta_{\rm max}$ by using the Brillouin function $B_S(y)$ in Eqs.~(\ref{Eqs:BS}) to determine the response of the magnitudes of the ordered magnetic moments to the temperature and infinitesimal applied field.

The equilibrium magnetic induction in Eq.~(\ref{Eq:Bi}) seen by magnetic moment $\vec{\mu}_i$ in the presence of {\bf H} must be parallel to the equilibrium ordered magnetic moment $\vec{\mu}_i$.  Thus one can obtain the component $H_{{\rm exch}\,i}$ of the local exchange field in the direction of $\hat{\mu}_i$ by taking the dot product of the two, yielding Eq.~(\ref{Eq:HexchDef3}) which we reproduce here for clarity
\be
H_{{\rm exch}\,i} = -\frac{1}{g^2\mu_{\rm B}^2}\sum_j J_{ij}\mu_j\cos\alpha_{ji}.
\label{Eq:HexchDef333}
\ee
We need an expression for $\mu_j\cos\alpha_{ji}$ at infinitesimal $H>0$ to insert into Eq.~(\ref{Eq:HexchDef333}).  From Eqs.~(\ref{Eq:muj}) and~(\ref{Eq:sincosthetaj}), one has
\be
\mu_j = \mu_0 + \delta_{\rm max}H(\cos\phi_i\cos\phi_{ji} - \sin\phi_i\sin\phi_{ji}).
\ee
The expression for $\cos\alpha_{ji}$ is given in Eq.~(\ref{Eq:CosAlphaji}). Keeping only terms to first order in $\gamma_{\rm max}$ and $H$ in the product $\mu_j\cos\alpha_{ji}$ that survive the sum in Eq.~(\ref{Eq:HexchDef333}) according to Eqs.~(\ref{Eq:SumSinZero}) and~(\ref{Eq:sums=0}), one obtains
\be
\bs
\mu_j\cos\alpha_{ji} &= \mu_0(\cos\phi_{ji} + \gamma_{\rm max}\cos\phi_i\sin^2\phi_{ji})\\*
&\hspace{0.2in} + \delta_{\rm max}H\cos\phi_i\cos^2\phi_{ji}
\end{split}
\ee
and therefore Eq.~(\ref{Eq:HexchDef333}) becomes
\be
\bs
H_{{\rm exch}\,i} &= -\frac{1}{g^2\mu_{\rm B}^2}\Big(\mu_0\sum_jJ_{ij}\cos\phi_{ji}\\*
&\hspace{0.2in}+ \mu_0\gamma_{\rm max}\cos\phi_i\sum_jJ_{ij}\sin^2\phi_{ji}\\*
&\hspace{0.2in}+ \delta_{\rm max}H\cos\phi_i\sum_jJ_{ij}\cos^2\phi_{ji}\Big).
\end{split}
\label{Eq:HexchFindDmax}
\ee
Then using $H_{{\rm exch}\,i0}$ from Eq.~(\ref{Eq:Hexch0Def3}) to replace the first term in Eq.~(\ref{Eq:HexchFindDmax}) one obtains
\be
\bs
H_{{\rm exch}\,i} &= H_{{\rm exch}\,i0}-\frac{\mu_0\gamma_{\rm max}}{g^2\mu_{\rm B}^2} \cos\phi_i\sum_jJ_{ij}\sin^2\phi_{ji}\\*
&\hspace{0.64in}-\frac{\delta_{\rm max}H}{g^2\mu_{\rm B}^2} \cos\phi_i\sum_jJ_{ij}\cos^2\phi_{ji}.
\end{split}
\label{Eq:Hexchi52}
\ee
Using Eq.~(\ref{Eq:muj}) and the definition $\bar{\mu}\equiv \mu/gS\mu_{\rm B}$ as in Eq.~(\ref{Eq:barmu0Def}) one obtains
\be
\bs
\mu_i - \mu_0 = gS\mu_{\rm B}(\bar{\mu}_i - \bar{\mu}_0) = \delta_{\rm max}H\cos\phi_i.
\end{split}
\label{Eq:mim0}
\ee
Using the second equality, Eq.~(\ref{Eq:Hexchi52}) becomes
\be
\bs
H_{{\rm exch}\,i} &= H_{{\rm exch}\,i0}-\frac{{\mu}_0\gamma_{\rm max}}{g^2\mu_{\rm B}^2} \cos\phi_i\sum_jJ_{ij}\sin^2\phi_{ji}\\*
&\hspace{0.64in}-\frac{(\bar{\mu}_i - \bar{\mu}_0)S}{g\mu_{\rm B}} \sum_jJ_{ij}\cos^2\phi_{ji}.
\end{split}
\label{Eq:Hexchi57}
\ee

The magnitude of the reduced ordered moment $\bar{\mu}_i \equiv \mu_i/\mu_{\rm sat} = \mu_i/(gS\mu_{\rm B})$ is given by the Brillouin function $B_S(y)$ of the magnetic induction ${\bf B}_i$ in Eqs.~(\ref{Eq:BS(y)}) as
\be
\bar{\mu}_i = B_S\left[\frac{g\mu_{\rm B}}{k_{\rm B}T}(H_{{\rm exch}\,i} + H_{\parallel i})\right],
\label{Eq:muimagnitude0}
\ee
where only the components of ${\bf H}_{{\rm exch}\,i}$ and {\bf H} that are parallel to $\vec{\mu}_i$ are relevant here.  To first order in $H$ one has
\be
H_{\parallel i}=H\cos\phi_i.
\label{Eq:Hparalleli}
\ee
Inserting Eqs.~(\ref{Eq:Hexchi57}) and~(\ref{Eq:Hparalleli}) into~(\ref{Eq:muimagnitude0}) and expanding $B_S(y)$ in a Taylor series about $y_0\equiv g\mu_{\rm B}H_{{\rm exchi}\,i0}/k_{\rm B}T$ to first order in $H$ (and $\gamma_{\rm max}$, which is proportional to $H$) gives
\bea
\bar{\mu}_i - \bar{\mu}_0  &=& B_S^\prime(y_0)\bigg[-\frac{{\mu}_0\gamma_{\rm max}}{g\mu_{\rm B}k_{\rm B}T} \cos\phi_i\sum_jJ_{ij}\sin^2\phi_{ji}\nonumber\\*
&&\hspace{-0.2in} -\ \frac{S(\bar{\mu}_i-\bar{\mu}_0)}{k_{\rm B}T}\sum_j J_{ij}\cos^2\phi_{ji} + \frac{g\mu_{\rm B}}{k_{\rm B}T}H\cos\phi_i\bigg],\nonumber
\eea
where we used $B_S(y_0) = \bar{\mu}_0$ from Eq.~(\ref{Eq:mubar0}).  Solving for $\bar{\mu}_i - \bar{\mu}_0$ gives
\be
\bar{\mu}_i - \bar{\mu}_0 = \frac{\frac{g\mu_{\rm B}}{S}H\cos\phi_i- \frac{{\mu}_0\gamma_{\rm max}}{g\mu_{\rm B}S}\cos\phi_i\sum_jJ_{ij}\sin^2\phi_{ji}}{\frac{k_{\rm B}T}{B_S^\prime(y_0)S}+ \sum_j J_{ij}\cos^2\phi_{ji}}.
\label{Eq:dMui}
\ee
Using Eq.~(\ref{Eq:mim0}) this can be written
\bea
\mu_i - \mu_0 &=& \delta_{\rm max}H\cos\phi_i \\*
&&\hspace{-0.25in} =\frac{g^2\mu_{\rm B}^2H\cos\phi_i- {\mu}_0\gamma_{\rm max}\cos\phi_i\sum_jJ_{ij}\sin^2\phi_{ji}}{\frac{k_{\rm B}T}{B_S^\prime(y_0)S} + \sum_j J_{ij}\cos^2\phi_{ji}}.\nonumber
\eea
Thus, under the condition that $\langle\cos^2\phi_i\rangle \neq 0$ in Eq.~(\ref{Eq:ChiXYFixmuHat}), solving for $\delta_{\rm max}$ yields
\be
\frac{\delta_{\rm max}}{g^2\mu_{\rm B}^2} = \frac{1 - \frac{\mu_0\gamma_{\rm max}}{g^2\mu_{\rm B}^2H}\sum_jJ_{ij}\sin^2\phi_{ji}}{\frac{k_{\rm B}T}{B_S^\prime(y_0)S}  + \sum_j J_{ij}\cos^2\phi_{ji}}.
\label{Eq:deltamaxSoln}
\ee
Note that $\gamma_{\rm max}$ in the numerator is proportional to $H$ according to Eq.~(\ref{Eq:gammaMaxSoln}), and hence $H$ cancels out, leaving $\delta_{\rm max}$ a function of~$T$ but not of~$H$\@.  Thus the change in magnitude of a magnetic moment due to the presence of the magnetic field depends both on the temperature and on the change in the angle that the spin makes with the applied magnetic field direction due to the magnetic field.

One expects Eq.~(\ref{Eq:deltamaxSoln}) to be applicable for the magnetic field applied along the easy axis of collinear AFs if one sets $\gamma_{\rm max}=0$, $\sin^2\phi_{ji}=0$ and $\cos^2\phi_{ji}=1$, and Eq.~(\ref{Eq:deltamaxSoln}) becomes
\be
\frac{\delta_{\rm max}}{g^2\mu_{\rm B}^2} = \frac{1}{\frac{k_{\rm B}T}{B_S^\prime(y_0)S}  + \sum_j J_{ij}}.
\ee
Then Eq.~(\ref{Eq:ChiXYFixmuHat}) with $\langle\cos^2\phi_i\rangle = 1$ gives the parallel susceptibility per spin of a collinear AF as 
\be
\chi_\parallel = \delta_{\rm max} = \frac{g^2\mu_{\rm B}^2}{\frac{k_{\rm B}T}{B_S^\prime(y_0)S}  + \sum_j J_{ij}}.
\ee
Using the expression for $\theta_{\rm p}$ in Eq.~(\ref{Eq:WeissTemp}), one sees that this expression for $\chi_\parallel$ is identical to that for the collinear AF already derived in Eq.~(\ref{Eq:ChiParTCol}). This is an important consistency check.

\subsubsection{\label{Sec:SolveChiXY} Solving for the In-Plane Susceptibility}

The two simultaneous equations~(\ref{Eq:gammaMaxSoln}) and~(\ref{Eq:deltamaxSoln}) in the two unknowns $\gamma_{\rm max}$ and $\delta_{\rm max}$, respectively, allow one to solve for these two unknowns, yielding
\bse
\label{Eqs:ChiComponentsSolved}
\bea
\frac{\gamma_{\rm max}\mu_0}{g^2\mu_{\rm B}^2H} &=& \frac{\tau+B-A}{(\tau+B)(B-E)-A^2}\\*
\frac{\delta_{\rm max}}{g^2\mu_{\rm B}^2} &=& \frac{B-A-E}{(\tau+B)(B-E)-A^2},
\eea
\ese
where
\bse
\be
\tau = \frac{k_{\rm B}T}{SB^\prime_S(y_0)} 
\label{Eq:tauDef1}
\ee
and
\be
\bs
y_0 &= \frac{3\bar{\mu}_0}{(S+1)t}, \qquad A = \sum_j J_{ij}\sin^2\phi_{ji}, \\* 
B &= \sum_j J_{ij}\cos^2\phi_{ji}, \qquad E = \sum_j J_{ij}\cos\phi_{ji}.
\end{split}
\label{Eq:ABEDefs}
\ee
\ese

The in-plane magnetic susceptibility components are obtained by substituting Eqs.~(\ref{Eqs:ChiComponentsSolved})  into~(\ref{Eqs:ChiCompsDefined}), yielding 
\be
\bs
\frac{\chi_{xy}\big|_{\mu}}{g^2\mu_{\rm B}^2} &= \frac{\mu_0\gamma_{\rm max}}{g^2\mu_{\rm B}^2H}\langle\sin^2\phi_i\rangle\\*
&= \frac{\tau+B-A}{(\tau+B)(B-E)-A^2}\langle\sin^2\phi_i\rangle,\\*
\frac{\chi_{xy}\big|_{\hat{\mu}}}{g^2\mu_{\rm B}^2} &= \frac{\delta_{\rm max}}{g^2\mu_{\rm B}^2}\langle\cos^2\phi_i\rangle\\*
&= \frac{B-E-A}{(\tau+B)(B-E)-A^2}\langle\cos^2\phi_i\rangle.
\end{split}
\label{Eq:ChiParts}
\ee
These expressions are only valid if $\langle\sin^2\phi_i\rangle\neq0$ and $\langle\cos^2\phi_i\rangle\neq0$, i.e., for planar noncollinear AF structures.  For commensurate planar noncollinear AF structures in which a hodograph of the ordered moments within a magnetic unit cell forms a regular polygon, the averages in Eqs.~(\ref{Eq:ChiParts}) over one magnetic unit cell are
\be
\langle\sin^2\phi_i\rangle = \langle\cos^2\phi_i\rangle = \frac{1}{2}.
\label{Eq:sin2cos2ave}
\ee
A magnetic unit cell that is commensurate with the underlying spin lattice is required in order for the averages in Eq.~(\ref{Eq:sin2cos2ave}) to be exact.  In practice, one can always consider the magnetic unit cell to be commensurate for a sufficiently large magnetic unit cell because the experimental resolution in measuring the incommensurability is finite. 

For $0\leq T \leq T_{\rm N}$ or equivalently $0\leq t \leq 1$, substituting Eqs.~(\ref{Eq:ChiParts}) and~(\ref{Eq:sin2cos2ave}) into (\ref{Eq:chiTot}) gives
\be
\frac{\chi_{xy}}{g^2\mu_{\rm B}^2} = \frac{\tau-E +2B-2A}{2[(\tau+B)(B-E)-A^2]}.
\label{Eq:ChiFinal}
\ee
Using Eqs.~(\ref{Eq:TmGeneral}), (\ref{Eq:WeissTemp}) and~(\ref{Eq:ABEDefs}), one can rewrite $A$ and $E$ as
\be
\bs
A &= -\frac{3k_{\rm B}\theta_{\rm p}}{S(S+1)} - B\\*
E &=-\frac{3k_{\rm B}T_{\rm N}}{S(S+1)}.
\end{split}
\ee
By multiplying both sides of Eq.~(\ref{Eq:ChiFinal}) by $\frac{3k_{\rm B}T_{\rm N}}{S(S+1)}$ and then multiplying the numerator and denominator of the right-hand side of Eq.~(\ref{Eq:ChiFinal}) by $\left[\frac{S(S+1)}{3k_{\rm B}T_{\rm N}}\right]^2$, Eq.~(\ref{Eq:ChiFinal}) can be written as a law of corresponding states for a given spin~$S$ in terms of easily measured quantities, which are $f=\theta_{\rm p}/T_{\rm N}$, $t = T/T_{\rm N}$ and additional reduced variables $\tau^\ast$ and $B^\ast$, as
\be
\frac{\chi_{xy}(T) T_{\rm N}}{C_1} = \frac{1+\tau^\ast + 2f + 4B^\ast}{2\left[(\tau^\ast + B^\ast)(1+B^\ast) - (f+B^\ast)^2\right]},
\label{Eq:ChiFinal2}
\ee
where
\begin{subequations}
\label{Eq:tauastBast}
\bea
\tau^\ast &=& \frac{S(S+1)\tau}{3k_{\rm B}T_{\rm N}} = \frac{(S+1)t}{3B_S^\prime(y_0)} \label{Eq:tauast}\\
B^\ast &=& \frac{S(S+1)B}{3k_{\rm B}T_{\rm N}} = -\frac{\sum_j {J}_{ij}\cos^2\phi_{ji}}{\sum_j {J}_{ij}\cos\phi_{ji}},\label{Eq:BstartDef2}
\eea
\end{subequations}
$y_0 = 3\bar{\mu}_0/[(S+1)t]$ from Eq.~(\ref{Eq:mubar0}) and we used Eq.~(\ref{Eq:TmGeneral}) to obtain the last equality.  At $T = T_{\rm N}$, according to Eq.~(\ref{Eq:taustaratTN}) one has $\tau^\ast = 1$ and Eq.~(\ref{Eq:ChiFinal2}) becomes
\be
\frac{\chi(T_{\rm N}) T_{\rm N}}{C_1} = \frac{1}{1-f}.
\label{Eq:ChiAtTN}
\ee
This agrees with the Curie-Weiss law prediction for $\chi(T_{\rm N})$ in Eq.~(\ref{Eq:ChiParTNCol}), an important consistency check.

Using Eqs.~(\ref{Eq:ChiFinal2}) and~(\ref{Eq:ChiAtTN}), for $T \leq T_{\rm N}$ one obtains the ratio
\be
\frac{\chi_{xy}(T)}{\chi(T_{\rm N})} = \frac{(1+\tau^\ast + 2f + 4B^\ast)(1-f)}{2\left[(\tau^\ast + B^\ast)(1+B^\ast) - (f+B^\ast)^2\right]}.
\label{Eq:ChiFinalRatio}
\ee
Using $\tau^\ast(t=0)=\infty$ obtained from Eqs.~(\ref{Eq:BSTto0}) and~(\ref{Eq:tauast}), Eq.~(\ref{Eq:ChiFinalRatio}) yields
\be
\frac{\chi_{xy}(T=0)}{\chi(T_{\rm N})} = \frac{1-f}{2(1+B^\ast)}.
\label{Eq:ChiT0Ratio}
\ee
Substituting $\tau^\ast(t=1) = 1$ at $T_{\rm N}$ from Eq.~(\ref{Eq:taustaratTN}) into Eq.~(\ref{Eq:ChiFinalRatio}) gives the identity
\be
\frac{\chi_{xy}(T=T_{\rm N})}{\chi(T_{\rm N})} = 1
\ee
 as required, irrespective of the values of $f$ and~$B^\ast$.

\section{\label{Sec:J0J1zJ2zModel} Generic $J_0$-$J_{z1}$-$J_{z2}$ Model for Planar Helical and Cycloidal Antiferromagnets}

In this section we recast our results for $\chi_{xy}(T\leq T_{\rm N})$ derived in the previous section in terms of a minimal generic $J_0$-$J_{z1}$-$J_{z2}$ model\cite{Nagamiya1967} that allows the  proper helix or cycloidal helix AF structures in Fig.~1 of Ref.~\onlinecite{Johnston2012} and Fig.~1 of Ref.~\onlinecite{Goetsch2014}, respectively, to be the AF ground states.  In this model, one sums all the exchange interactions of a given magnetic moment with other moments in the same ferromagnetically-aligned layer perpendicular to the helical or cycloidal wave vector $k_z$  and calls that sum $J_0$.  One also sums all the exchange interactions of a moment in a layer with all moments in one of the two nearest-neighbor layers and calls it $J_{z1}$ and similarly for the exchange interactions of the magnetic moment with all the magnetic moments in one of the two next-nearest-neighbor layers and calls it $J_{z2}$.  Third-nearest-neighbor or even further interlayer interactions are certainly possible but are not included in this model.  These net exchange interactions are indicated in Fig.~1 of Ref.~\onlinecite{Johnston2012} and Fig.~1 of Ref.~\onlinecite{Goetsch2014}.  One main purpose of synthesizing this model is to express the parameter $B^\ast$ in Eqs.~(\ref{Eq:tauastBast})  in terms of physically measurable quantities.  This is the only parameter in Eq.~(\ref{Eq:ChiFinalRatio}) for $\chi_{xy}(T\leq T_{\rm N})/\chi(T_{\rm N})$ that we have not yet expressed this way.  The second purpose is to synthesize a model for which the generic $J_0$, $J_{z1}$ and $J_{z2}$ exchange interactions can be expressed for specific compounds in terms of specific exchange interactions between the magnetic moments.  This is a powerful generic formulation that applies to large classes of planar noncollinear AFs.    

The competing phases in this model are a FM phase, a helical or cycloidal AF phase, and a collinear AF phase with propagation vector $(0,0,\frac{1}{2})$~r.l.u.  The latter phase is an A-type AF in which each FM-aligned layer is aligned AF with respect to its nearest-neighbor layers.  The helical and cycloidal phases are equivalent from the point of view of the theory.  For each phase, as in the previous section, the ordered magnetic moments are confined to a plane, which we designate as the $xy$~plane.  This $xy$~plane can be assigned to a particular crystal plane in a particular compound, as appropriate.

Within the $J_0$-$J_{z1}$-$J_{z2}$ model, the classical energy of interaction $E_i$ of spin ${\bf S}_i$ with its neighboring spins ${\bf S}_j$, where all spins have the same value of $S$, is given by Eq.~(\ref{Eq:Hamili1}) with $H=0$ as
\be
E_i = \frac{S^2}{2} \big[J_0 + 2J_{z1}\cos(kd) + 2J_{z2} \cos(2kd)\big],
\label{Eq:EiJ0J1J2Model}
\ee
where $d$ is the interlayer distance in Fig.~1 of Ref.~\onlinecite{Johnston2012} and Fig.~1 of Ref.~\onlinecite{Goetsch2014},  $k$~is the magnitude of the wave vector of the helix or cycloid and $\phi_{ji} = kd$ is the magnetic moment turn angle between adjacent FM-aligned layers upon moving along the positive helix or cycloid $z$~axis.  By minimizing $E_i$ with respect to $kd$ one obtains
\be
J_{z1}\sin (kd) +4J_{z2}\sin (kd) \cos (kd) = 0.
\ee
Two solutions for $kd$ are obtained by setting $kd=0$ or~$\pi$~rad, which correspond to FM and A-type AF states, respectively.  The third solution is a helical or cycloidal AF state with the turn angle~$kd$ determined by the exchange constants as
\be
\cos (kd) = -\frac{J_{z1}}{4J_{z2}}.
\label{Eq:coskzd}
\ee
Thus in general the helical or cycloidal wave vector is incommensurate with the underlying crystallographic spin lattice.  However, as discussed in the preceding section, one can always consider the wave vector to be commensurate to within experimental resolution with a sufficiently large magnetic unit cell.

Using Eq.~(\ref{Eq:EiJ0J1J2Model}) and the above three solutions for $kd$, the corresponding classical energies of the three phases are
\be
\bs
E_{\rm FM} &=  \frac{S^2}{2}(J_0 + 2J_{z1} + 2J_{z2})\\*
E_{\rm A\,type\,AF} &=  \frac{S^2}{2}(J_0 - 2J_{z1} + 2J_{z2})\\*
E_{\rm helix} &=  \frac{S^2}{2}\left(J_0 -\frac{J_{z1}^2}{4J_{z2}}-2J_{z2}\right),
\end{split}
\label{Eq:ESolnsJ0J1zJ2z}
\ee
where we used Eq.~(\ref{Eq:coskzd}) to obtain the last equality.  Note that the net intralayer exchange coupling $J_0$ has no effect on the \emph{relative} energies of the three phases, and hence is not relevant to the magnetic phase diagram.  For the helical or cycloidal phase, the condition $-1\leq \cos(kd)\leq~1$ in Eq.~(\ref{Eq:coskzd}) constrains $J_{z1}$ and $J_{z2}$ to satisfy\cite{Yoshimori1959, Nagamiya1967}
\be
J_{z2}>0,\qquad 0 \leq \frac{|J_{z1}|}{4J_{z2}} \leq 1.
\label{Eq:Jz2Restrict}
\ee

\begin{figure}
\includegraphics [width=3.in]{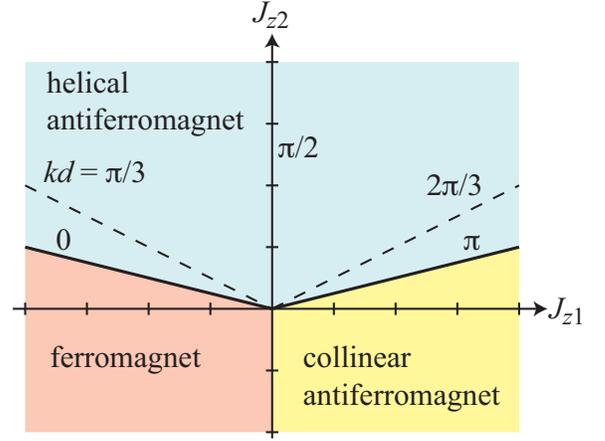}
\caption{(Color online) Classical phase diagram in the $J_{z1}$-$J_{z2}$ plane for the minimal $J_0$-$J_{z1}$-$J_{z2}$ model. The three regions are the ferromagnetic region, the collinear AF region with type-A ordering, and the helical or cycloidal AF region.  In the latter region, the wave vector of the helix or cycloid times the interlayer distance $d$ is $kd$, which is the turn angle between magnetic moments in adjacent layers along the helix or cycloid axis.  In order to obtain a helical or cycloidal magnetic structure, the net next-nearest layer interaction must be antiferromagnetic ($J_{z2} >0$).}
\label{Fig:J0_Jz1_Jz2_Phase_Diagram}
\end{figure}

The classical $T=0$ phase diagram for the $J_0$-$J_{z1}$-$J_{z2}$ model determined by finding the minimum energy solutions versus $J_{z1}$ and $J_{z2}$ in Eqs.~(\ref{Eq:ESolnsJ0J1zJ2z}) using Eqs.~(\ref{Eq:Jz2Restrict}) is shown in Fig.~\ref{Fig:J0_Jz1_Jz2_Phase_Diagram}.  For the helical or cycloidal phase, the nearest-layer interaction $J_{z1}$ can be either positive (AF) or negative (FM), but the next-nearest-neighbor interaction $J_{z2}$ must be positive (AF) as explicitly noted in Eqs.~(\ref{Eq:Jz2Restrict}).

A singular solution for the helical or cycloidal phase occurs when $J_{z1}=0$, for which the turn angle between planes would nominally be $kd = \pi/2$~rad from Eq.~(\ref{Eq:coskzd}) and Fig.~\ref{Fig:J0_Jz1_Jz2_Phase_Diagram}.  However, this solution physically corresponds to the presence of two noninteracting sublattices, each of which consists of next-nearest-neighbor magnetic moment layers that are mutually associated with exchange interaction $J_{z2}$.  Hence the turn angle between ordered moments in adjacent layers along the helix or cycloid axis is undefined for $J_{z1} = 0$.

\subsection{Alternative Expressions for the Variables in the $J_0$-$J_{z1}$-$J_{z2}$ Model}

The $\chi_{xy}(T)/\chi(T_{\rm N})$ of the planar noncollinear phase in Eq.~(\ref{Eq:ChiFinalRatio}) is expressed in terms of the quantities $S$, $\bar{\mu}_0 = \mu_0/\mu_{\rm sat}$, $t=T/T_{\rm N}$, $f\equiv \theta_{\rm p}/T_{\rm N}$ and $B^\ast$.  Usually one has experimental values of the first four quantities, whereas $B^\ast$ as defined in Eqs.~(\ref{Eq:tauastBast}) is not known without knowledge of the exchange constants, which are not directly measurable, and of the AF structure.   In the following we derive an expression for $B^\ast$ within the $J_0$-$J_{z1}$-$J_{z2}$ model in terms of the physically measurable quantities $f$ and $kd$.  To do that, we need explicit expressions for other variables in terms of the $J_0$-$J_{z1}$-$J_{z2}$ model that we now derive.

In this model, Eqs.~(\ref{Eq:WeissTemp}) and~(\ref{Eq:TmGeneral}) respectively become
\begin{subequations}
\bea
k_{\rm B}\theta_{\rm p} &=& -\frac{S(S+1)}{3}(J_0+2J_{z1}+2J_{z2}), \label{thetap}\\*
k_{\rm B}T_{\rm N} &=& -\frac{S(S+1)}{3}\\*
&&\times\ [J_0 +2J_{z1}\cos(kd)+2J_{z2}\cos(2kd)].\nonumber
\eea
\end{subequations}
From these expressions, the definitions
\bse
\be j_0 = \frac{J_0}{J_{z2}},\qquad j_1 = \frac{J_{z1}}{J_{z2}},
\ee
 and the relation 
\be
j_1=-4\cos(kd)
\ee
\ese
obtained from Eq.~(\ref{Eq:coskzd}), one obtains
\be
f \equiv \frac{\theta_{\rm p}}{T_{\rm N}} = \frac{j_0 -8\cos(kd)+2}{j_0 - 4\cos^2(kd) -2}.\label{Eq:ffromj0kz}
\ee

The parameter $B^\ast$ in Eq.~(\ref{Eq:ChiFinalRatio}) is given by Eqs.~(\ref{Eq:tauastBast}) as
\bse
\bea
B^\ast &=& -\frac{j_0 + 2j_1\cos^2(kd) +2\cos^2(2kd)}{j_0 + 2j_1\cos(kd) +2\cos(2kd)}\label{Eq:BStarb}\\*
&=&\frac{2-8\cos^2(kd)[1+\cos(kd)-\cos^2(kd)]+j_0}{2+4\cos^2(kd)-j_0}.\nonumber\\*
\label{Eq:B*J0J1zJ2z}
\eea
\ese
By eliminating $j_0$ in the simultaneous Eqs.~(\ref{Eq:ffromj0kz}) and~(\ref{Eq:B*J0J1zJ2z}), one obtains the very useful result
\be
B^\ast = 2(1-f)\cos(kd)[1+\cos(kd)] - f,
\label{Eq:BastFromfkz}
\ee
which only depends on the measurable parameters $kd$ and $f$.

\subsection{Reformulation of the In-Plane Magnetic Susceptibility in Terms of the $J_0$-$J_{z1}$-$J_{z2}$ Model}

Using Eqs.~(\ref{Eq:ChiT0Ratio}) and~(\ref{Eq:BastFromfkz}) we obtain the reduced  in-plane \mbox{$T=0$} susceptibility as
\be
\frac{\chi_{xy}(T=0)}{\chi(T_{\rm N})} = \frac{1}{2\big[1+2\cos(kd) + 2\cos^2(kd)\big]}.
\label{Eq:ChiT0TNkz}
\ee
This general result agrees with Yoshimori's pioneering calculation of $\chi_{xy}(T=0)/\chi(T_{\rm N})$ in his Eq.~(50) for the specific case of the \mbox{$c$-axis} helix in $\beta$-MnO$_2$ with the rutile structure, assuming a specific set of exchange constants,\cite{Yoshimori1959} and using the substitutions $\theta\to kd,\ \cos\theta\to-\cos\theta\ {\rm and}\ A_1/(4A_2)\to -\cos(kd)$ in his Eq.~(50).

Interestingly, the reduced $T=0$ in-plane susceptibility in Eq.~(\ref{Eq:ChiT0TNkz}) is expressed solely in terms of the turn angle $kd$ where $k$ is the magnitude of the helix or cycloid wave vector and $d$ is the distance between adjacent planes in the helix or cycloid.  A plot of this dependence is shown in Fig.~2(a) of Ref.~\onlinecite{Johnston2012}.  Lines of constant $kd$, and hence of constant normalized zero-temperature susceptibility, are shown above in Fig.~\ref{Fig:J0_Jz1_Jz2_Phase_Diagram}.  The behavior in Fig.~2(a) of Ref.~\onlinecite{Johnston2012} is unexpected for two reasons.  First, $\chi_{xy}(0)/\chi(T_{\rm N})$ varies nonmonotonically with $kd$.  Second, a peak appears in $\chi_{xy}(0)/\chi(T_{\rm N})$ at the unexpected wave vector $kd=2\pi/3$ for which $\chi_{xy}(0)/\chi(T_{\rm N}) = 1$.  The latter result $\chi_{xy}(0) = \chi(T_{\rm N})$ suggests that for this wave vector, $\chi_{xy}$ is independent of $T$ for $T\leq T_{\rm N}$, which is confirmed below.

When $\chi_{xy}(0)/\chi(T_{\rm N}) < 1/2$, Fig.~2(a) of Ref.~\onlinecite{Johnston2012} shows that the turn angle between layers of moments along the helix or cycloid axis is less than 90$^\circ$, which corresponds to a dominant FM interaction between a moment and the moments in an adjacent layer.  This is because a moment in one layer has a component in the same direction as the moment in an adjacent layer.  On the other hand, when $\chi_{xy}(0)/\chi(T_{\rm N}) > 1/2$, Fig.~2(a) of Ref.~\onlinecite{Johnston2012} shows that the turn angle between layers of moments along the helix or cycloid axis is greater than 90$^\circ$, which corresponds to a dominant AF interaction between a moment and the moments in an adjacent layer.  

Using Eq.~(\ref{Eq:BastFromfkz}), one can express $\chi_{xy}(T)/\chi(T_{\rm N})$ in Eq.~(\ref{Eq:ChiFinalRatio}) completely in terms of the measurable parameters $S$, $\bar{\mu}_0$, $t$, $f$ and now $kd$.  Plots of $\chi_{xy}(T)$ versus $T/T_{\rm N}$ obtained using Eqs.~(\ref{Eq:ChiFinalRatio}) and~(\ref{Eq:BastFromfkz}) for spins $S = 7/2$  (Ref.~\onlinecite{Johnston2012}) and~1/2 and various helix turn angles $kd$ and $f$ ratios are shown in Fig.~2(b) of Ref.~\onlinecite{Johnston2012}.  The maximum in $\chi_{xy}(T=0)$ versus $kd$ that appears in Fig.~2(a) of Ref.~\onlinecite{Johnston2012} is confirmed.  Furthermore, one sees that $\chi_{xy}$ is independent of $T$ for a turn angle $kd=2\pi/3$~rad as suspected above.

\begin{figure}
\includegraphics [width=3.in]{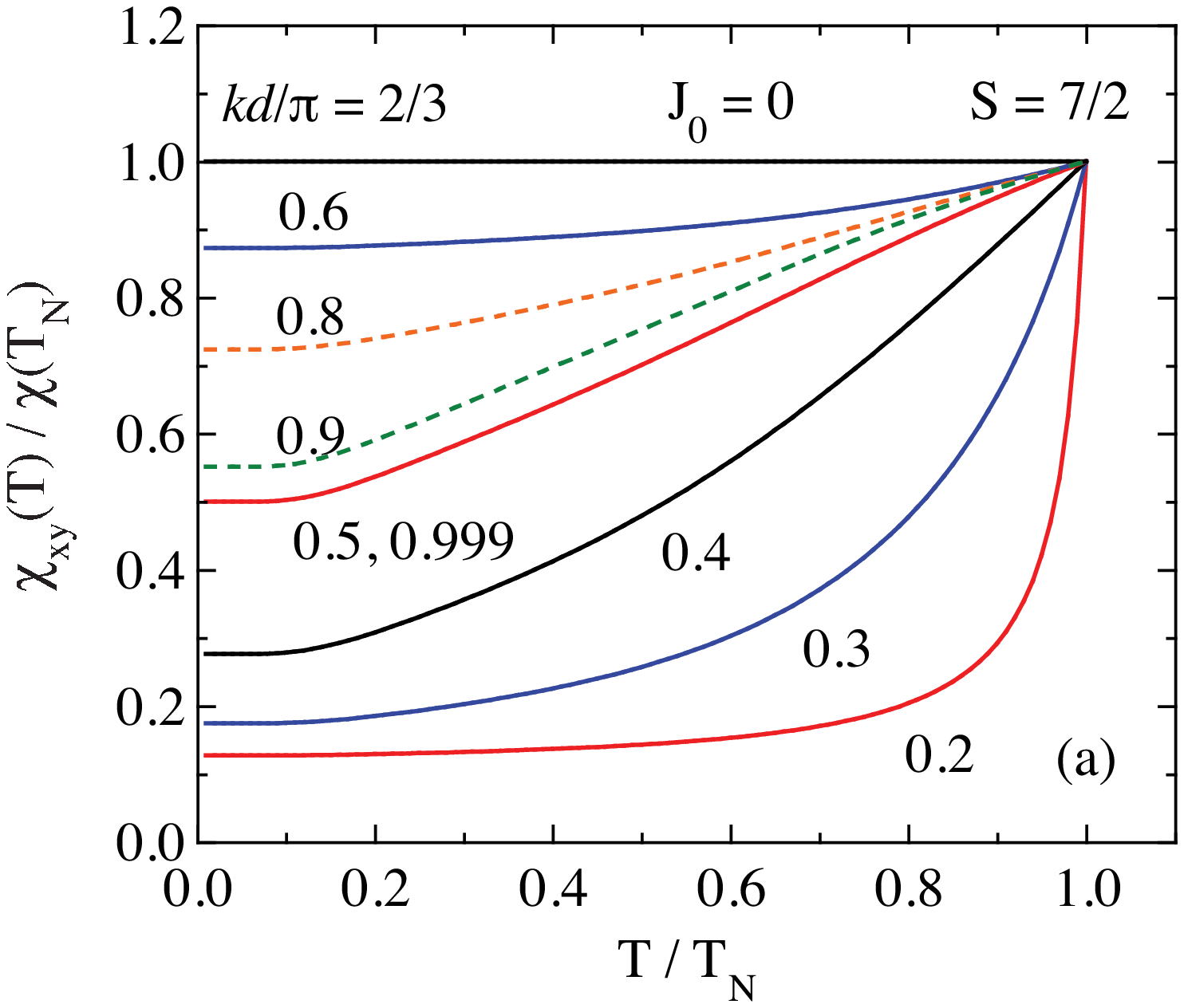}
\includegraphics [width=3.in]{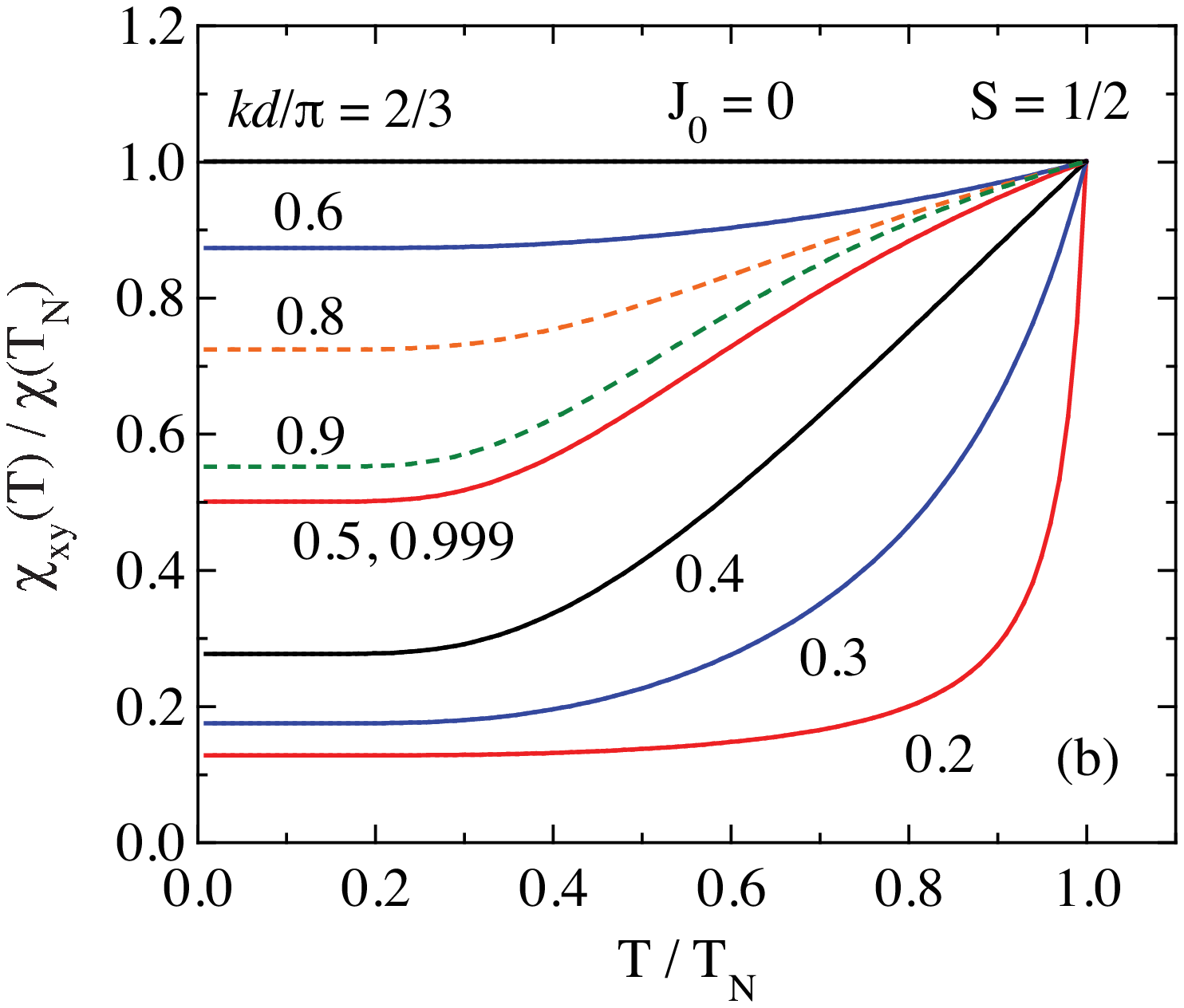}
\caption{(Color online) In-plane magnetic susceptibility $\chi_{xy}(T)$ versus temperature $T$ for the helical or cycloidal magnetic structure in the $J_0$-$J_{z1}$-$J_{z2}$ model with $J_0=0$ and the listed magnitudes of the helical or cycloidal turn angle $kd$ for spin (a) $S = 7/2$ and (b) $S = 1/2$.}
\label{Fig:J0Jz1Jz2ChiTonChiTN}
\end{figure}

Instead of using Eq.~(\ref{Eq:BastFromfkz}) for $B^\ast$, one can use Eq.~(\ref{Eq:BStarb}) and set $j_0=0$ to obtain an expression for $\chi_{xy}(T)/\chi(T_{\rm N})$ that only depends on the parameter $kd$ as in Eq.~(\ref{Eq:ChiT0TNkz}).  Plots of $\chi_{xy}(T)/\chi(T_{\rm N})$ versus $kd$ for $J_0=0$ are shown in Figs.~\ref{Fig:J0Jz1Jz2ChiTonChiTN}(a) and~\ref{Fig:J0Jz1Jz2ChiTonChiTN}(b) for $S = 7/2$ and~1/2, respectively.  These plots are useful for certain compounds such as noncollinear linear chain helical or cycloidal AFs where the interchain interactions are negligible compared to the intrachain ones, and also for higher-dimensional helical or cycloidal antiferromagnets such as $\beta$-MnO$_2$ with the rutile structure, where $J_3 = J_0/4 \approx 0$ for the $c$-axis helix has been estimated.\cite{Yoshimori1959}

\subsection{Noncollinear $120^\circ$ Helical or Cycloidal Antiferromagnets}

The turn angle $\phi_{ji} = kd=2\pi/3$~rad is special, since we found from the above results that
\be
\frac{\chi_{xy}(T\leq T_{\rm N})}{\chi(T_{\rm N})} =1 \qquad (kd=2\pi/3).
\label{Eq:kzd2pi3}
\ee
To check the generality of this important and unique result, we go back to the general expression for $B^\ast$ in Eq.~(\ref{Eq:BastFromfkz}) and substitute $\cos(kd=2\pi/3)=-1/2$, which gives
\be
B^\ast = -\frac{f+1}{2}.
\ee
Substituting this expression for $B^\ast$ into the general Eq.~(\ref{Eq:ChiFinalRatio}) for $\chi_{xy}(T)/\chi(T_{\rm N})$ and simplifying gives Eq.~(\ref{Eq:kzd2pi3}) identically, irrespective of the value of the spin~$S$.

The perpendicular susceptibility in Eq.~(\ref{Eq:ChiPerpSoln3}) also obeys Eq.~(\ref{Eq:kzd2pi3}).  Thus we predict that for AFs with a 120$^\circ$ helical or cycloidal magnetic structure, the $\chi(T\leq T_{\rm N})$ is isotropic and temperature-independent with the value at $T_{\rm N}$, irrespective of the value of $S$\@.  This prediction is strongly confirmed by experimental data on single crystals of a variety of 120$^\circ$ triangular-lattice AFs.\cite{Johnston2012}

For the special case of only the six nearest-neighbor interactions $J$ in a triangular lattice being nonzero, using $\phi_{ji} = kd = 120^\circ$ one obtains from Eqs.~(\ref{Eq:TmGeneral}) and~(\ref{Eq:WeissTemp}) 
\bea
T_{\rm N} &=& -\frac{S(S+1)}{3k_{\rm B}}\sum_j J_{ij}\cos\phi_{ji} = \frac{S(S+1)J}{k_{\rm B}},\nonumber\\*
\theta_{\rm p} &=& -\frac{S(S+1)}{3k_{\rm B}}\sum_j J_{ij} = -\frac{2S(S+1)J}{k_{\rm B}},\nonumber\\*
f &=& \frac{\theta_{\rm p}}{T_{\rm N}} =-2,\nonumber\\*
T_{\rm N} - \theta_{\rm p} &=& \frac{3S(S+1)J}{k_{\rm B}}.\label{Eq:TNmQPTri}
\eea
Thus from Eqs~(\ref{Eq:CurieConst2}), (\ref{Eq:CWLaw}) and~(\ref{Eq:TNmQPTri}) one obtains
\be
\chi_\perp = \chi(T_{\rm N}) =\frac{C_1}{T_{\rm N}-\theta_{\rm p}} = \frac{g^2\mu_{\rm B}^2}{9J},
\ee
which is independent of~$S$.

For the classical ($S\to\infty$) isolated triangular layer AF, one also obtains for the ground state at $T=0$ a nontrivial isotropy in $\chi(T=0)$ with the same value of $\chi(T=0)$ as we just obtained for finite spin by MFT.\cite{Kawamura1985, Chubukov1994}  In addition, classical Monte Carlo simulations for the single triangular layer indicated that $\chi$ is isotropic and nearly independent of $T$ at low $T$.\cite{Kawamura1984}  Our MFT results thus significantly extend the previous calculations for single classical triangular lattice layers to finite quantum spins $S$ and long-range AF ordering that occur in real systems.

\section{\label{Sec:HiFieldPerpM_hA1zero} Internal Energy, Magnetization, Phase Diagram and Heat Capacity of Collinear and Planar Noncollinear Antiferromagnets in a High Perpendicular Magnetic Field}

\begin{figure}[t]
\includegraphics [width=2.5in]{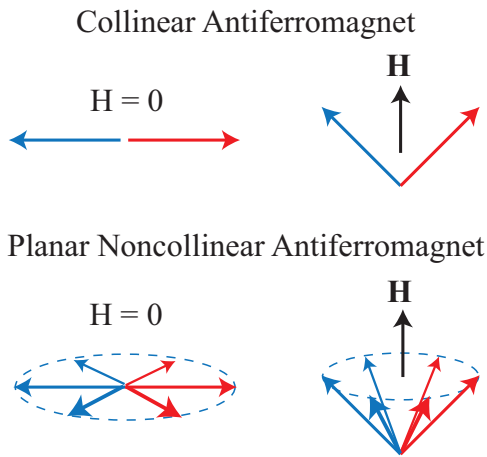}
\caption{(Color online) Influence on the generic magnetic structure due to a high magnetic field applied perpendicular to the ordering axis of a collinear antiferromagnet (AF) (top panel) and to the ordering plane of a planar noncollinear AF (bottom panel).  Hodographs of the zero-field magnetic moment vectors are shown on the left.  In high fields as shown on the right, the AF structures become canted towards the field.  The ordered moments of the collinear AF are now within a vertical plane, whereas those of the noncollinear AF now lie on the surface of a cone with the axis of the cone along the magnetic field axis as shown.  At a sufficiently high field $H=H_{\rm c\perp}$ given by Eq.~(\ref{Eq:HcPerpTgtr0}), the moments in either case become parallel to the applied field and each other and a second-order transition from the canted AF to the paramagnetic (PM) state occurs at that field.}
\label{Fig:High_Perp_Field_Structs}
\end{figure}

In this section a MFT calculation of the high-field magnetization and magnetic heat capacity with fields applied perpendicular to the zero-field ordered moments is carried out for generic collinear and planar noncollinear AFs containing identical magnetic moments interacting by Heisenberg exchange on the same footing.  These high-field calculations are included in the present paper because as in the previous sections we calculate the thermodynamics without the use of magnetic sublattices and express the results as laws of corresponding states in terms of measurable parameters.  

The influence of magnetocrystalline anisotropy on $\chi(T)$ for both $T>T_{\rm N}$ and $T<T_{\rm N}$, on $T_{\rm N}$ itself and on the high-field $M(H,T)$ behaviors and $H$-$T$ phase \mbox{diagrams} are discussed in  Refs.~\onlinecite{Johnston2014} and~\onlinecite{Anand2014b} for collinear and noncollinear AFs.  When a high field is applied parallel to the ordering axis of a collinear AF, where an anisotropy field is present that is sufficiently large to prevent a spin-flop transition from occurring, one must define separate up and down moment sublattices because within MFT the thermal-average magnitudes of the up and down moments are not the same.  In the present paper we only consider magnetic structures and behaviors where the concept of magnetic sublattices is not necessary and hence the discussion is limited to high perpendicular fields.

The generic responses of collinear and planar noncollinear AF structures to a high perpendicular magnetic field are illustrated in Fig.~\ref{Fig:High_Perp_Field_Structs}.  Whereas the tilted moments of a collinear structure due to the field reside within a vertical plane including the applied field, a hodograph of the tilted moments of a planar noncollinear structure lie on the surface of a cone with the magnetic field direction corresponding to the continuous rotational axis of the cone.  Both cases are treated here within the same formalism by the use of the spherical coordinates defined in Fig.~\ref{Fig:chiPerp3}, where the former AF structure is a special case of the latter.

\begin{figure}[t]
\includegraphics [width=1.5in]{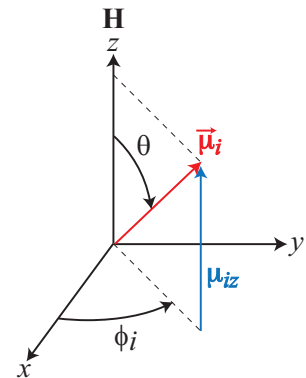}
\caption{(Color online) The rotation of a representative magnetic moment $\vec{\mu}_i$ out of the $xy$~plane towards the $z$~axis upon applying a magnetic field ${\bf H} = H\hat{\bf k}$.  The orientation of $\vec{\mu}_i$ for $H>0$ is described by spherical coordinates $\theta_i$ and $\phi_i$, where $\theta_i = \theta$ is the same for all moments.  Each moment was originally in the $xy$~plane for $H=0$ with coordinates $\theta = 90^\circ$ and $\phi = \phi_i$.   The azimuthal angle $\phi_i$ is in general different for different moments.}
\label{Fig:chiPerp3}
\end{figure}

\subsection{High-Field Magnetization Perpendicular to the Ordering Axis or Plane at $T = 0$}

The magnetic field is applied along the polar $z$ axis
\be
{\bf H} = H\hat{\bf k},
\label{Eq:HDef}
\ee
as shown in Fig.~\ref{Fig:chiPerp3}.  For $H=0$ the ordered magnetic moments lie in the $xy$-plane with polar angle $\theta=\pi/2$.  In the presence of a high perpendicular field, at $T = 0$ one has
\bea
\mu_{\rm sat} &=& gS\mu_{\rm B}\nonumber\\*
\phi_{ji} &=& \phi_j-\phi_i\nonumber\\*
\phi_j &=& \phi_i+\phi_{ji}\nonumber\\*
\vec{\mu}_i &=& \mu_{\rm sat}\big[\sin\theta(\cos\phi_i\,\hat{\bf i} + \sin\phi_i\,\hat{\bf j}) + \cos\theta\,\hat{\bf k}\big]\label{Eq:HiHMuTEqs}\\*
\vec{\mu}_j &=& \mu_{\rm sat}\big[\sin\theta(\cos\phi_j\,\hat{\bf i} + \sin\phi_j\,\hat{\bf j}) + \cos\theta\,\hat{\bf k}\big]\nonumber\\*
&=& \mu_{\rm sat}\Big\{\sin\theta\Big[(\cos\phi_i\cos\phi_{ji} - \sin\phi_i\sin\phi_{ji})\,\hat{\bf i}\nonumber \\*
&& +\ (\sin\phi_i\cos\phi_{ji} + \cos\phi_i\sin\phi_{ji})\,\hat{\bf j}\Big] + \cos\theta\,\hat{\bf k}\Big\},\nonumber
\eea
where $\phi_{i,j}$ are the azimuthal angles of ordered moments $\vec{\mu}_{i,j}$ with respect to the positive $x$-axis, $\phi_{ji} = \phi_j - \phi_i$ and at $T=0$ the magnitude of each magnetic moment is the saturation magnetic moment $\mu_{\rm sat} = gS\mu_{\rm B}$.  

The torque on a particular moment $\vec{\mu}_i$ due to the exchange field in Eq.~(\ref{Eq:HexchiDef}) is obtained using Eqs.~(\ref{Eq:HiHMuTEqs}) as
\be
\bs
\vec{\mu}_i\times {\bf H}_{{\rm exch}\,i} &= \frac{\mu_{\rm sat}^2}{g^2\mu_{\rm B}^2}\sin\theta\cos\theta (-\sin\phi_i\,\hat{\bf i} + \cos\phi_i\,\hat{\bf j})\\*
&\hspace{0.3in} \times \sum_j J_{ij}(1-\cos\phi_{ji}) 
\end{split}
\label{Eq:ExchTorque}
\ee
where we have only kept terms that do not contain $\sum_j J_{ij}\sin\phi_{ji}$ according to Eq.~(\ref{Eq:SumSinZero}).  The torque on $\vec{\mu}_i$ due to {\bf H} is
\be
\vec{\mu}_i\times {\bf H} = \mu_{\rm sat}H\sin\theta(\sin\phi_i\,\hat{\bf i}-\cos\phi_i\,\hat{\bf j}).
\label{muCrossHperp}
\ee
In equilibrium, the net torque is
\be
\vec{\tau} = \vec{\mu}_i \times {\bf H}_{{\rm exch}\,i} + \vec{\mu}_i \times {\bf H} = 0,
\label{Eq:tauDef}
\ee
which contains the two terms in Eqs.~(\ref{Eq:ExchTorque}) and~(\ref{muCrossHperp}).  Setting either the $x$ or $y$ component of the net torque equal to zero gives
\be
\cos\theta = \frac{g^2\mu_{\rm B}^2H}{\mu_{\rm sat}\sum_j J_{ij}(1-\cos\phi_{ji})}.
\label{Eq:costheta}
\ee
From Eqs.~(\ref{Eq:WeissTemp}) and~(\ref{Eq:TmGeneral}), one respectively obtains
\bse
\label{Eq:Sums}
\bea
\sum_j J_{ij} &=& -\frac{3k_{\rm B}}{S(S+1)}\theta_{\rm p},\label{Eq:Sum1}\\*
\sum_j J_{ij}\cos\phi_{ji} &=& -\frac{3k_{\rm B}}{S(S+1)}T_{\rm N}.
\label{Eq:Sum2}
\eea
\ese
Then using Eqs.~(\ref{Eq:CurieConst2}), (\ref{Eqs:CWTNs}) and~(\ref{Eq:Sums}), Eq.~(\ref{Eq:costheta}) can be written
\be
\cos\theta =\frac{\chi(T_{\rm N})H}{\mu_{\rm sat}}
\label{Eq:costheta2}
\ee

Referring to Fig.~\ref{Fig:chiPerp3}, the $z$-component $\mu_z$ of the induced magnetic moment of each spin is
\be
\mu_z = \mu_{\rm sat}\cos\theta.
\label{Eq:muzVsCosTheta}
\ee
Inserting Eq.~(\ref{Eq:costheta2}) into this expression gives the perpendicular susceptibility as
\be
\chi_\perp = \frac{\mu_z}{H} = \chi(T_{\rm N}) \qquad (T=0,\ H \leq H_{c\perp}).
\label{Eq:ChiPerpT0}
\ee
Thus the induced magnetic moment is proportional to $H$ until at a critical perpendicular field $H_{\rm c\perp}$ one obtains $\mu_z = \mu_{\rm sat}$.  This critical field occurs when $\theta=0$ ($\cos\theta=1$), which Eq.~(\ref{Eq:costheta2}) gives simply as
\be
H_{\rm c\perp} = \frac{\mu_{\rm sat}}{\chi(T_{\rm N})} \qquad(T=0).
\label{Eq:HcPerpT0}
\ee
At higher fields, $\mu_z$ cannot increase any further and is constant at the saturation value $\mu_z = \mu_{\rm sat} = gS\mu_{\rm B}$.  Thus a second-order phase transition occurs at $T=0$ with increasing $H$ at $H=H_{\rm c\perp}$ where there is a discontinuity in the slope of $\mu_z$ versus $H$ [see Fig.~\ref{Fig:MHs5fm1}(a) below].

\subsection{\label{Sec:MPerpVsHPerp} High-Field Magnetization Perpendicular to the Ordering Axis or Plane at $0 \leq T \leq T_{\rm N}$}

Because the calculation of the magnetization in a high perpendicular field at finite temperatures $0 \leq T \leq T_{\rm N}$ within MFT is more involved than the above calculation at $T=0$, we treat it separately in this section.  At each temperature and field, the magnitude $\mu(T)$ of each ordered moment is the same for all magnetic moments, because they are all equivalent with respect to the effect of the applied field.  Using Eqs.~(\ref{Eq:HexchiDef}) and~(\ref{Eq:HiHMuTEqs}), the component of the exchange field in the direction of the central magnetic moment $\vec{\mu}_i$ is
\be
\bs
H_{{\rm exch}\,i} &= -\frac{\bar{\mu}S}{g\mu_{\rm B}}\sum_j J_{ij}\hat{\mu}_i\cdot \hat{\mu}_j \\*
&=-\frac{\bar{\mu}S}{g\mu_{\rm B}}\bigg[\cos^2\theta\sum_j J_{ij} + \sin^2\theta\sum_j J_{ij}\cos\phi_{ji}\bigg],
\end{split}
\label{Eq:Hexchiperp}
\ee
where we recall that $\phi_i$, $\phi_j$ and hence $\phi_{ji} = \phi_j-\phi_i$ are independent of $H$, with only $\theta$ changing with $H$ (see Fig.~\ref{Fig:High_Perp_Field_Structs}).  Inserting Eqs.~(\ref{Eq:Sums}) into~(\ref{Eq:Hexchiperp}) gives
\bea
H_{{\rm exch}\,i} &=& \frac{3\bar{\mu}k_{\rm B}}{g\mu_{\rm B}(S+1)}(\theta_{\rm p}\cos^2\theta+T_{\rm N}\sin^2\theta)\hspace{0.4in}\nonumber\\*
&=&\frac{3\bar{\mu}k_{\rm B}T_{\rm N}}{g\mu_{\rm B}(S+1)}\big[1-(1-f)\cos^2\theta\big],\label{Eq:Hexchitheta}
\eea
where we have used $f \equiv \theta_{\rm p}/T_{\rm N}$ according to Eq.~(\ref{Eq:fRatioDef}).

We define the reduced magnitude of each ordered moment as
\be
\bar{\mu}(T) \equiv \frac{\mu(T)}{\mu_{\rm sat}},
\label{Eq:barmuz}
\ee
analogous to Eq.~(\ref{Eq:barmu0Def}) for $H=0$.  The value of $\bar{\mu}$ of each magnetic moment versus $H$ and $T$ is governed by the Brillouin function $B_S(y)$.  Substituting Eq.~(\ref{Eq:Hexchitheta}) into~(\ref{Eq:muimagnitude0}) gives
\bea
\bar{\mu} &=& B_S\left[\left(\frac{g\mu_{\rm B}}{k_{\rm B}T}\right)(H_{{\rm exch}\,i} + H\cos\theta)\right]\label{Eq:barmu}\\*
&=& B_S\left\{\frac{3\bar{\mu}}{(S+1)t}\big[1-(1-f)\cos^2\theta\big] + \frac{h\cos\theta}{t}\right\},\nonumber
\eea
where $H\cos\theta$ is the component of {\bf H} in the direction of each of the magnetic moments according to Fig.~\ref{Fig:chiPerp3}, the reduced field is $h\equiv g\mu_{\rm B}H/k_{\rm B}T_{\rm N}$ from Eq.~(\ref{Eq:barhDef}) and the reduced temperature is $t\equiv T/T_{\rm N}$ according to Eq.~(\ref{Eq:tDef}).

However, there are two unknowns, $\bar{\mu}$ and $\theta$, in Eq.~(\ref{Eq:barmu}), so we need another equation to solve for both.  For that, we set the net torque $\vec{\tau}$ on $\vec{\mu}_i$ to zero according to Eq.~(\ref{Eq:tauDef}).  The first term in Eq.~(\ref{Eq:tauDef}) is obtained from Eq.~(\ref{Eq:ExchTorque}) with the substitutions in Eqs.~(\ref{Eq:Sums}) and~(\ref{Eq:barmuz}), yielding
\be
\bs
\vec{\mu}_i \times {\bf H}_{{\rm exch}\,i} &= \frac{3\bar{\mu}^2Sk_{\rm B}}{S+1}\sin\theta\cos\theta(T_{\rm N}-\theta_{\rm p})\\*
& \hspace{0.5in}\times(-\sin\phi_i\,\hat{\bf i} + \cos\phi_i\,\hat{\bf j}).
\end{split}
\label{Eq:Term1}
\ee
The second term in Eq.~(\ref{Eq:tauDef}) is obtained from Eq.~(\ref{muCrossHperp}) with the substitution $\mu_{\rm sat}\to \mu = \bar{\mu}g\mu_{\rm B}S$, yielding
\be
\vec{\mu}_i \times {\bf H} = \bar{\mu}g\mu_{\rm B}S H\sin\theta(\sin\phi_i\,\hat{\bf i} - \cos\phi_i\,\hat{\bf j}).
\label{Eq:Term2}
\ee
Substituting Eqs.~(\ref{Eq:Term1}) and~(\ref{Eq:Term2}) into~(\ref{Eq:tauDef}) gives
\be
\frac{3\bar{\mu}k_{\rm B}}{S+1}(T_{\rm N}-\theta_{\rm p})\cos\theta = g\mu_{\rm B}H.
\label{Eq:tauequalszero}
\ee
Dividing each side by $k_{\rm B}T_{\rm N}$ gives
\be
\frac{3\bar{\mu}\cos^2\theta}{(S+1)t}\left(1-f\right) = \frac{h\cos\theta}{t}.\label{Eq:mubarbarh}
\ee
Substituting the left-hand side of Eq.~(\ref{Eq:mubarbarh}) for $h\cos\theta/t$ into Eq.~(\ref{Eq:barmu}) yields
\be
\bar{\mu}=B_S\left[\frac{3\bar{\mu}}{(S+1)t}\right].
\label{Eq:ordMoment}
\ee
This expression is identical to Eq.~(\ref{Eq:mubar0}) for determining $\bar{\mu}_0(t)$ for \mbox{$H=0$}.  In other words,  a perpendicular applied field has no influence on the magnitude of the $T$-dependent ordered moment, as long as the $z$-component of the moment is less than that magnitude at that $T$\@.  This general result from MFT is of course also valid for the special case of collinear AFs in a perpendicular magnetic field.

From Eq.~(\ref{Eq:tauequalszero}), one obtains
\be
\cos\theta = \frac{g^2\mu_{\rm B}^2S(S+1)H}{3\mu k_{\rm B}(T_{\rm N}-\theta_{\rm p})} = \frac{C_1H}{\mu(T_{\rm N}-\theta_{\rm p})},
\label{Eq:costheta3}
\ee
where we have used $C_1$ from Eq.~(\ref{Eq:CurieConst2}) and $\bar{\mu} = \mu/(g\mu_{\rm B}S)$ from Eq.~(\ref{Eq:barmuz}).  Then from Fig.~\ref{Fig:chiPerp3} one obtains
\be
\bs
\mu_z &= \mu \cos\theta = \frac{g^2\mu_{\rm B}^2S(S+1)H}{3 k_{\rm B}(T_{\rm N}-\theta_{\rm p})}\\*
&= \frac{C_1H}{T_{\rm N}-\theta_{\rm p}} = \chi(T_{\rm N})H,
\end{split}
\label{Eq:M(H)perp}
\ee
where we have used $\chi(T_{\rm N})$ from Eq.~(\ref{Eq:CWTN}).  Thus for $H\leq H_{\rm c\perp}(T)$ in MFT, the perpendicular susceptibility is
\be
\chi_\perp(T) = \frac{\mu_z}{H} = \chi(T_{\rm N}) \quad (T\leq T_{\rm N},\ H\leq H_{c\perp}),
\ee
analogous to Eq.~(\ref{Eq:ChiPerpT0}) for $T=0$.  The $\chi_\perp$ remains constant with increasing~$H$ at fixed~$T$ until the induced magnetic moment $\mu_z$ becomes equal to the ordered moment at the particular temperature at the perpendicular critical field $H_{c\perp}$, where
\be
H_{\rm c\perp}(T) = \frac{\mu_0(T)}{\chi(T_{\rm N})},
\label{Eq:HcPerpTgtr0}
\ee
which is analogous to the zero-temperature result in Eq.~(\ref{Eq:HcPerpT0}).  Above this field, the system is in the PM state with each induced moment aligned parallel to {\bf H}.  This generic behavior of the magnetization versus transverse magnetic field is of course also found for special cases such as for the simple N\'eel antiferromagnet with only nearest-neighbor interactions where $\theta_{\rm p} = -T_{\rm N}$ and $f=-1$.

\subsection{Magnetic Phase Diagram and Magnetization versus Field Isotherms for Magnetic Fields Applied Perpendicular to the Ordering Axis or Plane}

In the previous section we saw that a perpendicular field does not affect the magnitude of the reduced ordered moment $\bar{\mu}$ for $H\leq H_{\rm c\perp}$ and thus $\bar{\mu} = \bar{\mu}_0$ where the latter is the value in $H=0$ in Eq.~(\ref{Eq:mubar0}).  The critical field $H_{\rm c\perp}$ is the field at which the induced magnetic moment $\bar{\mu}_z$ equals $\bar{\mu}$ at that temperature.  At that field the ordered moment is pointing in the direction of ${\bf H}$.  From Eq.~(\ref{Eq:M(H)perp}), on the critical field curve with $\bar{\mu}_z = \bar{\mu}_0$ one obtains
\be
\bs
\bar{\mu}_0 &=  \frac{(S+1)}{3(T_{\rm N} - \theta_{\rm p})} \frac{g\mu_{\rm B}H_{\rm c\perp}}{k_{\rm B}}= \frac{(S+1)}{3(1 - f)} \frac{g\mu_{\rm B}H_{\rm c\perp}}{k_{\rm B}T_{\rm N}}\\*
&= \frac{(S+1)h_{\rm c\perp}}{3(1 - f)},
\end{split}
\label{Eq:muAtHcr}
\ee
where we have used the definition of $\bar{\mu}_0 = \mu_0/(g\mu_{\rm B}S)$ in Eq.~(\ref{Eq:barmu0Def}), of $f$ in Eq.~(\ref{Eq:fRatioDef}) and of $h$ in Eq.~(\ref{Eq:barhDef}).  Thus the reduced critical field is given from Eq.~(\ref{Eq:muAtHcr}) as

\be
h_{\rm c\perp}(t) \equiv \frac{g\mu_{\rm B}H_{\rm c\perp}(t)}{k_{\rm B}T_{\rm N}}= \frac{3(1-f)}{S+1}\bar{\mu}_0(t) ,
\label{Eq:BarHvsT}
\ee
which demonstrates the important property that $h_{\rm c\perp}(t) \propto \bar{\mu}_0(t)$.  Since $\bar{\mu}_0(t=0) = 1$, one obtains
\be
\frac{H_{\rm c\perp}(t)}{H_{\rm c\perp}(0)} = \bar{\mu}_0(t).
\label{Eq:HperpHperp0}
\ee
The critical field divides the $H$-$T$ plane into a (canted) AF state and the PM state, as shown in Fig.~\ref{Fig:mu0_vs_t} for $S=5$.  One can invert the axes in Fig.~\ref{Fig:mu0_vs_t} to obtain the field dependence of the N\'eel temperature.

\begin{figure}
\includegraphics [width=3.in]{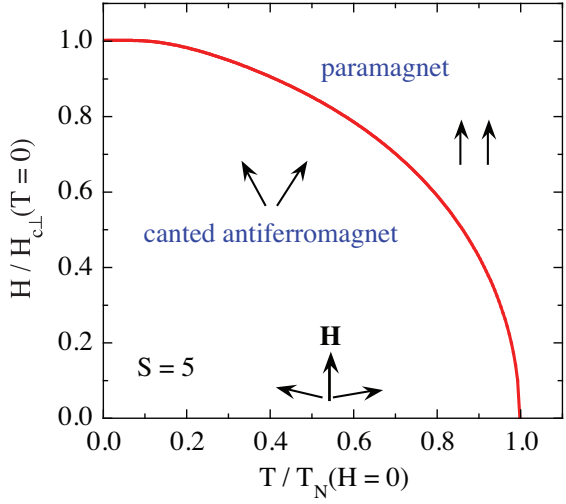}
\caption{(Color online) Phase diagram in the magnetic field-temperature $H$-$T$ plane for magnetic fields applied perpendicular to the ordering axis (collinear AF) or plane (planar noncollinear AF) of a Heisenberg AF with spin $S=5$.  The phase boundary between the AF and PM states is the critical field $H_{\rm c\perp}(T)$ that was calculated using Eqs.~(\ref{Eq:mubar0}) and~(\ref{Eq:HperpHperp0}).}
\label{Fig:mu0_vs_t}
\end{figure}

On the critical field curve with $h = h_{\rm c\perp}$, the ordered moment has the value $\bar{\mu}_z = \bar{\mu}_0$ given in the PM state by Eqs.~(\ref{Eq:muvsBrill3}) and~(\ref{Eq:BarHvsT})  as
\bse
\bea
\bar{\mu}_z &=& B_S\left[\frac{3\bar{\mu}_zf}{(S+1)t} + \frac{h_{\rm c\perp}}{t}\right]\\*
&=& B_S\left[\frac{3\bar{\mu}_z}{(S+1)t}\right]\quad(h = h_{\rm c\perp}).
\label{Eq:muzcrit}
\eea
\ese
A comparison of Eq.~(\ref{Eq:muzcrit}) with Eq.~(\ref{Eq:ordMoment}) shows explicitly that $\mu_z$ is continuous on crossing the critical line from the canted AF state into the PM state and hence the phase transition is second order.

To summarize, the reduced $z$-axis magnetic moment $\bar{\mu}_z$ versus reduced magnetic field $h$ in the $z$ direction is given for $h \leq h_{\rm c\perp}$ by Eq.~(\ref{Eq:muAtHcr}) with $\bar{\mu}_0$ replaced by $\bar{\mu}_z$ and $h_{c\perp}$ replaced by $h$, and for $h \geq h_{\rm c\perp}$ by Eq.~(\ref{Eq:muvsBrill3}), i.e., 
\bse
\label{Eqs:muVsHAFPMPerp}
\be
\bar{\mu}_z = \frac{S+1}{3(1-f)}\, h \hspace{0.85in}(h \leq h_{\rm c\perp}),
\label{Eq:muVsHAF}
\ee
\be
\bar{\mu}_z = B_S\left[\frac{3f\bar{\mu}_z}{(S+1)t} + \frac{h}{t}\right] \qquad(h \geq h_{\rm c\perp}),
\label{Eq:muVsHPM}
\ee
\ese
where $B_S(y)$ is given in Eq.~(\ref{Eq:BrillouinFunction}), $\bar{\mu}_z(t,h)$ is calculated from Eqs.~(\ref{Eqs:muVsHAFPMPerp}) in the relevant field range and $h_{c\perp}(t)$ is given in Eq.~(\ref{Eq:BarHvsT}).
 
The derivative $(d\bar{\mu}_z/dt)_h$ for $h\geq h_{c\perp}$ which we will need later is calculated by taking the total derivative of Eq.~(\ref{Eq:muVsHPM}) with respect to $t$ at fixed field and solving for $(d\bar{\mu}_z/dt)_h$, yielding
\bse
\label{Eqs:dmuzdh}
\be
\left(\frac{d\bar{\mu}_z}{dt}\right)_h = -\frac{\bar{\mu}_z + \frac{(S+1)h}{3f}}{t\Big[\frac{(S+1)t}{3fB_S^\prime(y)}-1\Big]} \qquad(h \geq h_{\rm c\perp}),
\label{Eq:dmuzdtPM}
\ee
where
\be
y = \frac{3f\bar{\mu}_z}{(S+1)t} + \frac{h}{t}
\label{Eq:yPM}
\ee
\ese
and $B_S^\prime(y)$ is given in Eq.~(\ref{Eq:dBSy0}).  

\begin{figure}
\includegraphics [width=3.3in]{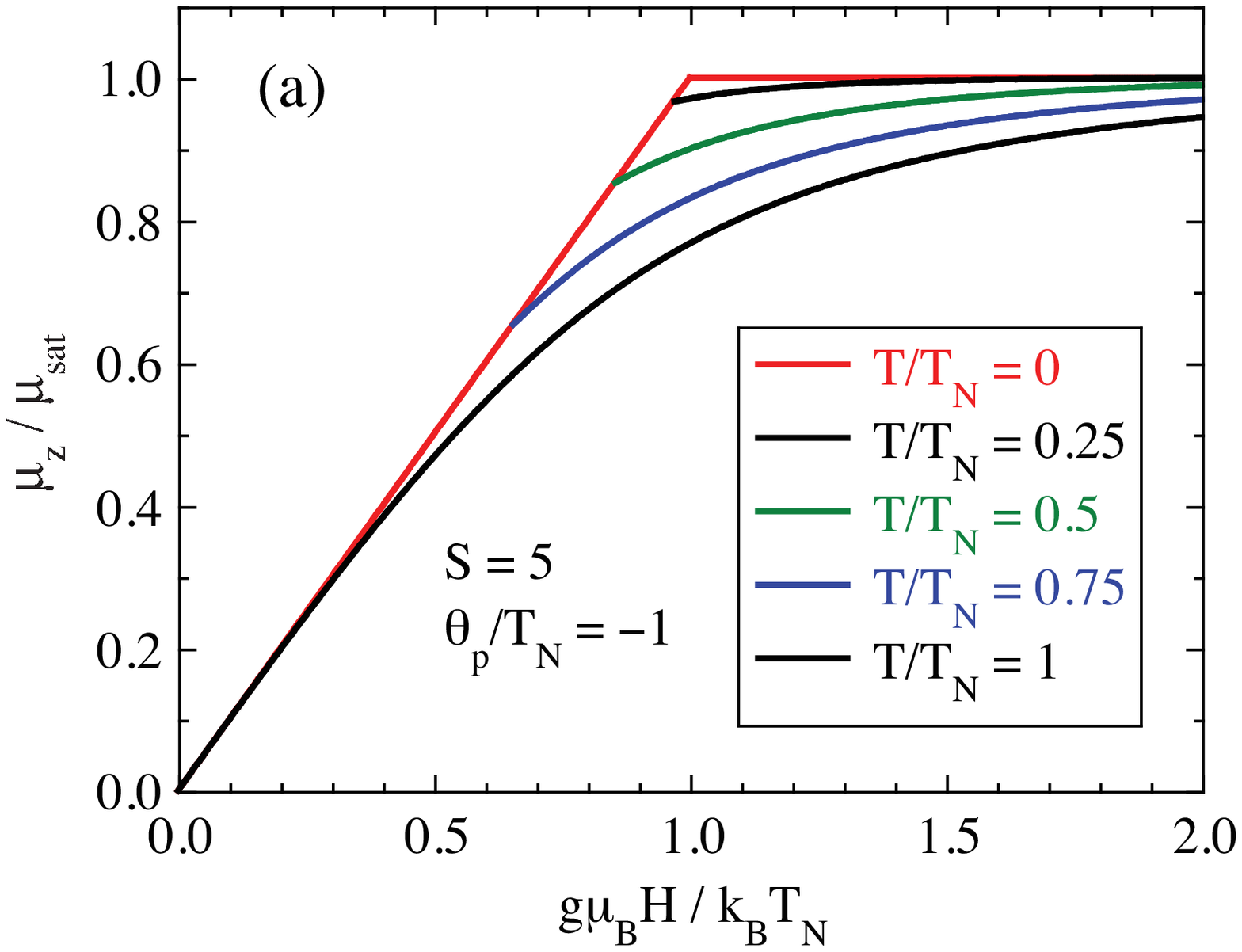}
\includegraphics [width=3.3in]{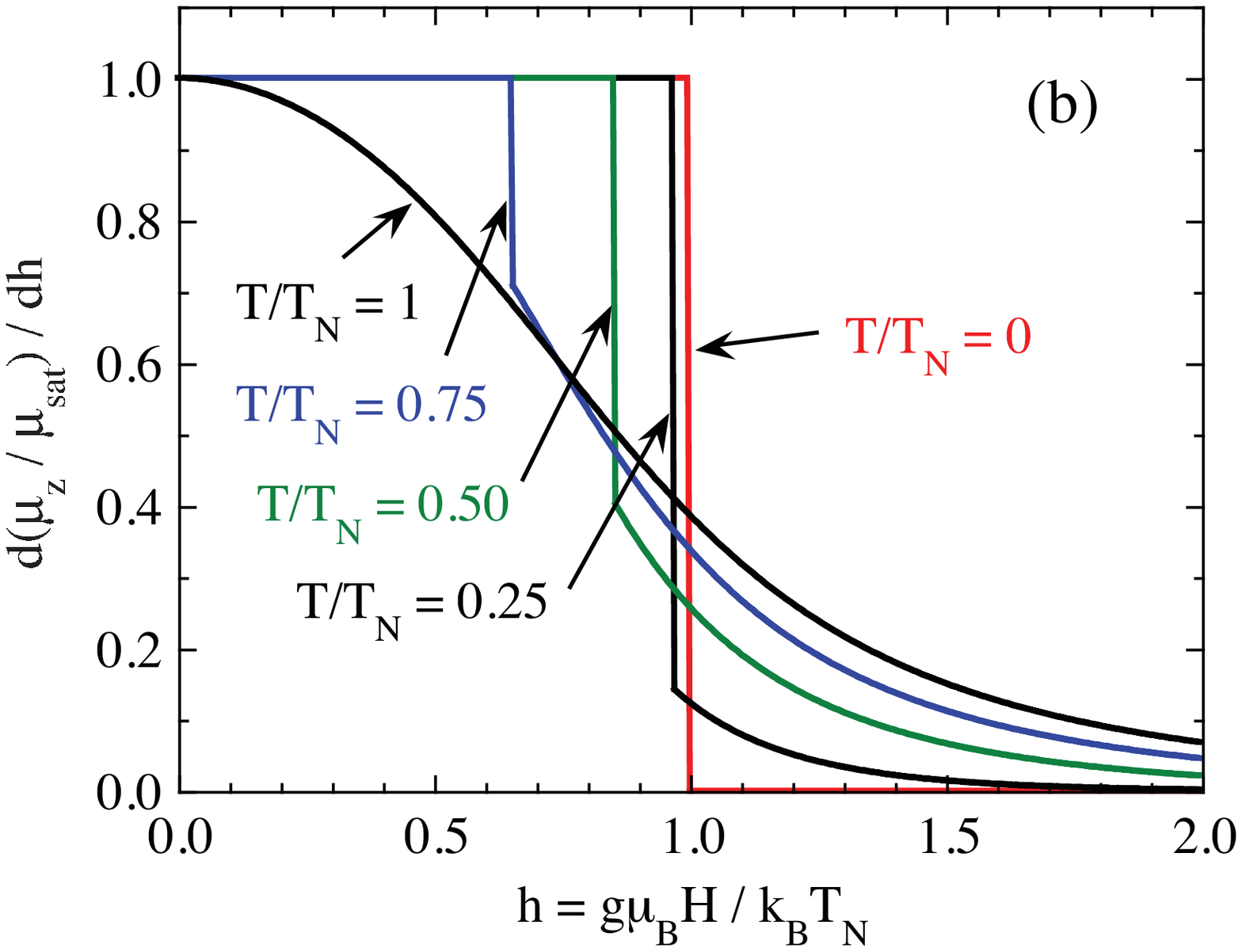}
\caption{(Color online) (a) Reduced induced perpendicular magnetic moment $\bar{\mu}_z = \mu_z/\mu_{\rm sat}$ versus reduced perpendicular magnetic field $h = g\mu_{\rm B}H/(k_{\rm B}T_{\rm N})$ at different reduced temperatures $t = T/T_{\rm N}$ as indicated, where $T_{\rm N}$ refers to $T_{\rm N}(H=0)$ and $\mu_{\rm sat} = gS\mu_{\rm B}$.  The ordered moments at $H=0$ are in the $xy$~plane and a perpendicular field $H$ is applied along the $z$~axis as shown in Fig.~\ref{Fig:High_Perp_Field_Structs}.  The spin is arbitrarily chosen to be $S=5$ and the ratio of the Weiss temperature to the N\'eel temperature is assumed to be $f\equiv \theta_{\rm p}/T_{\rm N} = -1$.  The curves were calculated using Eqs.~(\ref{Eqs:muVsHAFPMPerp}).  The field region preceding the sharp change in slope is the AF region, and the higher field region is the PM region at each $T$\@.  The second-order phase transition between these two regimes defines the N\'eel temperature $T_{\rm N}(H)$. (b)~Differential susceptibility $d\bar{\mu}_z/dh$ versus~$h$ calculated from Eqs.~(\ref{Eqs:muVsHAFPMPerp}) and~(\ref{Eqs:dmuzdh}).}
\label{Fig:MHs5fm1}
\end{figure}

Equation~(\ref{Eq:muVsHPM}) is applicable to the entire PM region of the $(h,t)$ phase diagram in Fig.~\ref{Fig:mu0_vs_t}, including the part where $T > T_{\rm N}\ (t > 1)$, where here $T_{\rm N}$ refers to $T_{\rm N}(H=0)$, and also the part where $h\geq h_{\rm c\perp}$ and $T < T_{\rm N}\ (t < 1)$.  Several $\bar{\mu}_z$ versus~$h$ isotherms calculated from Eqs.~(\ref{Eqs:muVsHAFPMPerp}) are plotted in Fig.~\ref{Fig:MHs5fm1}(a) for $0\leq t \leq 1$. The respective differential susceptibilities $d\bar{\mu}_z/dh$ are calculated from Eqs.~(\ref{Eqs:muVsHAFPMPerp}) and~(\ref{Eqs:dmuzdh}) and plotted versus~$h$ in Fig.~\ref{Fig:MHs5fm1}(b).  A discontinuous change in $d\bar{\mu}_z/dh$ versus~$h$ occurs on crossing the critical curve in Fig.~\ref{Fig:mu0_vs_t}, as emphasized in Fig.~\ref{Fig:MHs5fm1}(b), because $\bar{\mu}_z \propto h$ for $h < h_{\rm c\perp}$ but $\bar{\mu}_z(h)$ exhibits negative curvature for $h > h_{\rm c\perp}$ and hence $\bar{\mu}_z(H)$ is nonanalytic at $h = h_{\rm c\perp}$.  This discontinuity in slope is most apparent for $T \ll T_{\rm N}(H=0)$.  Theoretical curves similar to those in Fig.~\ref{Fig:MHs5fm1}(a) were plotted previously as derived from MFT,\cite{Kumar2012} although the equations used were not given.

\subsection{\label{Sec:CpHPerp} Magnetic Internal Energy and Heat Capacity in the PM Phase and in the AF Phase with Magnetic Fields Perpendicular to the Ordering Axis or Plane}

Here we calculate the magnetic heat capacity $C_{\rm mag}(T)$ for the perpendicular field orientation and study the evolution of $C_{\rm mag}(T)$ with increasing field.  We expect strong effects because the $T_{\rm N}$ can be driven to zero with sufficiently high fields as illustrated in Fig.~\ref{Fig:mu0_vs_t}.  From Figs.~\ref{Fig:mu0_vs_t} and~\ref{Fig:MHs5fm1}(b), the discontinuity in slope of $\mu_z$ versus $t$ decreases with increasing field, so we expect the discontinuity in $C_{\rm mag}$ at $T_{\rm N}(H)$ to also decrease with increasing field.  Moreover, the PM phase at $T > T_{\rm N}(H)$ must have a nonzero contribution to $C_{\rm mag}$ because the induced moment is nonzero for $T>T_{\rm N}(H)$, in contrast to the MFT prediction $C_{\rm mag}(T\geq T_{\rm N})=0$ for $H=0$ (zero induced moment) in Eq.~(\ref{Eq:Cmag}) and Fig.~11 of Ref.~\onlinecite{Johnston2011}.  In the following two Secs.~\ref{CmagAF} and~\ref{Sec:CmagPM} we derive the magnetic heat capacity in the AF and PM regimes separately, and then in Sec.~\ref{Sec:CmagAFPM} combine the results to obtain $C_{\rm mag}(T)$ at fixed $H$ including both the AF and PM regimes.

\subsubsection{\label{CmagAF} Magnetic Internal Energy and Heat Capacity of the AF-Ordered Phase}

The $C_{\rm mag}(H,T)$ in a perpendicular field is calculated in MFT from the internal energy per moment $E_i$, which is the same for each ordered and/or field-induced moment $\vec{\mu}_i$ for this field configuration because they are all equivalent with respect to the field as shown in Fig.~\ref{Fig:High_Perp_Field_Structs}.  In Sec.~\ref{Sec:MPerpVsHPerp} we determined that $\bar{\mu}_i$ is independent of field within the AF-ordered phase and is therefore equal to the zero-field value $\bar{\mu}_0$.  With the applied field given in Eq.~(\ref{Eq:HDef}) and the axis notation in Fig.~\ref{Fig:chiPerp3}, one obtains
\bse
\label{Eqs:ExchEnergyPerp}
\be
E_i = E_{{\rm exch}\,i} + E_{H},
\label{Eq:EiSum}
\ee
where
\bea
E_{{\rm exch}\,i} &=&  -\frac{1}{2}\mu_0 H_{{\rm exch}\,i} \nonumber\\*
&=& -\frac{1}{2}gS\mu_{\rm B} \bar{\mu}_0 H_{{\rm exch}\,i},\label{EexchiDefPerpH}\\*
E_{H} &=& -\mu_0 H\cos\theta = -\mu_z H,\label{Eq:E_H}
\eea
\ese
$\mu_0=gS\mu_{\rm B}\bar{\mu}_0$ from Eq.~(\ref{Eq:barmu0Def}), $\mu_z=\mu_0\cos\theta$, we use the fact that the magnitude $\mu_0$ of the ordered moment is the same for each $\vec{\mu}_i$, and have defined $H_{{\rm exch}\,i}$ as the {\it component} of ${\bf H}_{{\rm exch}\,i}$ in the direction of $\vec{\mu}_i$ as in Eq.~(\ref{Eq:HexchDef3}).  The factor of 1/2 in Eq.~(\ref{EexchiDefPerpH}) arises because the exchange energy is equally shared between each pair of interacting moments, whereas the exchange field seen by a given moment is assumed to be due only to the neighbors of the moment that interact with the moment with no contribution from the moment itself.

From Figs.~\ref{Fig:chiPerp} and~\ref{Fig:High_Perp_Field_Structs}, all ordered moments have the same angle $\theta$ with respect to the applied field, so for the general case of a planar noncollinear AF, which of course includes the collinear case, $H_{{\rm exch}\,i}$ is given by Eq.~(\ref{Eq:Hexchitheta}). 
Inserting Eq.~(\ref{Eq:Hexchitheta}) with $\bar{\mu}=\bar{\mu}_0$ into~(\ref{EexchiDefPerpH}) yields
\be
E_{{\rm exch}\,i} = -\frac{3S\bar{\mu}_0^2k_{\rm B}T_{\rm N}}{2(S+1)}\big[1-(1-f)\cos^2\theta\big].
\label{Eq:EexchDef6}
\ee
We normalize the energy by the thermal energy $k_{\rm B}T_{\rm N}$, yielding the reduced exchange energy
\be
\varepsilon_{{\rm exch}\,i} \equiv \frac{E_{{\rm exch}\,i}}{k_{\rm B}T_{\rm N}} = -\frac{3S\bar{\mu}_0^2}{2(S+1)}\big[1-(1-f)\cos^2\theta\big].
\label{Eq:EexchDef6B}
\ee
One can write Eq.~(\ref{Eq:costheta3}) for $\cos\theta$ with $\mu\to\mu_0$ as
\be
\cos\theta = \frac{(S+1)h}{3\bar{\mu}_0(1-f)},
\label{Eq:cosThetaPerp}
\ee
where $f=\theta_{\rm p}/T_{\rm N}$ and $h \equiv g\mu_{\rm B}H/(k_{\rm B}T_{\rm N})$ is the reduced magnetic field in Eq.~(\ref{Eq:barhDef}).  Substituting Eq.~(\ref{Eq:cosThetaPerp}) into~(\ref{Eq:EexchDef6B}) gives
\be
\varepsilon_{{\rm exch}\,i} = -\frac{3S\bar{\mu}_0^2}{2(S+1)} + \frac{S(S+1)h^2}{6(1-f)}.
\label{Eq:RedEexchDef}
\ee

Using Eq.~(\ref{Eq:HDef}) for {\bf H}, the expression for $\vec{\mu}_i$ in Eqs.~(\ref{Eq:HiHMuTEqs}) and the definition of $h$, the expression $\mu_z = \bar{\mu}_0 g\mu_{\rm B}S\cos\theta$ and Eq.~(\ref{Eq:cosThetaPerp}) for $\cos\theta$, the contribution of the external field to the internal energy per moment is 
\bse
\bea
E_{H} &=& -\mu_z H = -\frac{S(S+1)k_{\rm B}T_{\rm N}h^2}{3(1-f)}, \\*
\epsilon_H &\equiv& \frac{E_{H}}{k_{\rm B}T_{\rm N}} = -\frac{S(S+1)h^2}{3(1-f)}.\label{Eq:EHReduced}
\eea
\ese
The total reduced internal energy per moment in the AF state with a perpendicular magnetic field applied is obtained from Eqs.~(\ref{Eq:EiSum}), (\ref{Eq:RedEexchDef}) and~(\ref{Eq:EHReduced}) as
\be
\varepsilon_i \equiv \frac{E_i}{k_{\rm B}T_{\rm N}} = -\frac{3S\bar{\mu}_0^2}{2(S+1)} - \frac{S(S+1)h^2}{6(1-f)}\quad (h\leq h_{\rm c\perp}) ,
\label{Eq:EpsOfH}
\ee
where the reduced critical field $h_{\rm c\perp}$ is given in Eq.~(\ref{Eq:BarHvsT}), which defines the field boundary between the AF and PM phases.

The magnetic heat capacity per magnetic moment $C_{\rm mag}$ versus temperature at constant perpendicular field is obtained from Eq.~(\ref{Eq:EpsOfH}) using $t\equiv T/T_{\rm N}$ from Eq.~(\ref{Eq:tDef}) as
\bse
\be
\frac{C_{\rm mag}}{k_{\rm B}} = \left(\frac{d\varepsilon_i(t)}{dt}\right)_h = -\frac{3S}{(S+1)}\bar{\mu}_0(t)\frac{d\bar{\mu}_0(t)}{dt}\quad (h\leq h_{\rm c\perp}).
\label{Eq:CmaginH}
\ee
Substituting Eq.~(\ref{Eq:dbarmu0dt}) for $d\bar{\mu}_0(t)/dt$ into~(\ref{Eq:CmaginH}) gives  $C_{\rm mag}$ in the (canted) AF phase as
\be
\frac{C_{\rm mag}}{k_{\rm B}} = \frac{3S\bar{\mu}_0^2(t)}{(S+1)t\Big[\frac{(S+1)t}{3B_S^\prime(y_0)}-1 \Big]}\quad (h\leq h_{\rm c\perp}),
\label{Eq:CmaginH2}
\ee
\ese
Here $\bar{\mu}_0(t)$ is calculated by numerically solving Eq.~(\ref{Eq:mubar0}), the derivative $B_S^\prime(y)$ is given in Eq.~(\ref{Eq:dBSy0}) and $h_{\rm c\perp}$ is given in Eq.~(\ref{Eq:BarHvsT}).  Equation~(\ref{Eq:CmaginH2}) is identical to Eq.~(\ref{Eq:Cmag}) for $H=0$, except that we have now shown that it is also valid for perpendicular magnetic fields less than the $t$-dependent $h_{\rm c\perp}$.  Equation~(\ref{Eq:CmaginH2}) is valid in the magnetically-ordered state of any collinear or planar noncollinear Heisenberg AF containing identical crystallographically equivalent spins. At higher fields $h\geq h_{\rm c\perp}$, the $C_{\rm mag}$ in the PM state derived in the following section must be used in place of Eq.~(\ref{Eq:CmaginH2}).

\subsubsection{\label{Sec:CmagPM} Magnetic Internal Energy and Heat Capacity of the Paramagnetic Phase}

In the PM state all  magnetic moments $\mu_z$ are field-induced, have the same magnitude and are all in the same (perpendicular) direction of the applied field {\bf H}\@.  Equation~(\ref{Eq:Hexchipara}) gives the exchange field seen by each induced moment in the PM state as
\bse
\be
H_{{\rm exch}\,i} = -\frac{S\bar{\mu}_z(t)}{g\mu_{\rm B}}\sum_j J_{ij},
\label{Eq:HexchDef6}
\ee
where we used the definition $\bar{\mu}_z \equiv \mu_z/\mu_{\rm sat}=\mu_z/(gS\mu_{\rm B})$ as in Eq.~(\ref{Eq:barmu0Def}).  Inserting the expression for the sum given in Eq.~(\ref{Eq:Sum1}) yields
\be
H_{{\rm exch}\,i} = \frac{3\bar{\mu}_z(t)k_{\rm B}\theta_{\rm p}}{g\mu_{\rm B}(S+1)}.
\label{Eq:HexchDef7}
\ee
\ese
Then using Eq.~(\ref{EexchiDefPerpH}) with $\bar{\mu}_0\to\bar{\mu}_z$ one obtains the exchange energy as
\bse
\be
E_{{\rm exch}\,i} = -\frac{3S\bar{\mu}_z^2(t)k_{\rm B}\theta_{\rm p}}{2(S+1)}.
\label{Eq:EexchDef7}
\ee
From the definition of the reduced exchange energy as in Eq.~(\ref{Eq:EexchDef6B}) one obtains
\be
\varepsilon_{{\rm exch}\,i} = -\frac{3\bar{\mu}_z^2(t) f S}{2(S+1)},
\label{Eq:EpsHexch}
\ee
\ese
where we used the definition $f\equiv \theta_{\rm p}/T_{\rm N}$ from Eq.~(\ref{Eq:fRatioDef}).  The part of the internal magnetic energy per moment due to the applied magnetic field is given by Eq.~(\ref{Eq:E_H}), which we write in terms of reduced variables as
\be
\varepsilon_H = -Sh\bar{\mu}_z(t) .
\label{Eq:EpsH}
\ee

The total reduced internal magnetic energy per spin in the PM phase from Eqs.~(\ref{Eq:EpsHexch}) and~(\ref{Eq:EpsH}) is
\be
\varepsilon_i = -\frac{3fS}{S+1}\bigg[\frac{\bar{\mu}_z^2(t)}{2} + \frac{(S+1)h\bar{\mu}_z(t)}{3f}\bigg].
\label{Eq:EpsPM}
\ee
The $C_{\rm mag}$ per spin at fixed field is then given by the first equality in Eq.~(\ref{Eq:CmaginH}) as
\bse
\label{Eqs:CmagPMAll}
\be
\frac{C_{\rm mag}}{k_{\rm B}} = -\frac{3fS}{S+1}\bigg[\bar{\mu}_z(t) + \frac{(S+1)h}{3f}\bigg]\frac{d\bar{\mu}_z(t)}{dt}\Big|_h\quad (h\geq h_{\rm c\perp}).\label{Eq:CmagPM}
\ee
Substituting $d\bar{\mu}_z/dt$ from Eq.~(\ref{Eq:dmuzdtPM}) into~(\ref{Eq:CmagPM}) yields the $C_{\rm mag}$ per spin in the PM phase as
\be
\frac{C_{\rm mag}}{k_{\rm B}} = \frac{3fS\Big[\bar{\mu}_z(t) + \frac{(S+1)h}{3f}\Big]^2}{(S+1)t\Big[\frac{(S+1)t}{3fB_S^\prime(y)} - 1\Big]}  \quad(h\geq h_{\rm c\perp}),
\label{Eq:CmagPM2}
\ee
\ese
where $y$ is given in Eq.~(\ref{Eq:yPM}), $B_S^\prime(y)$ is given in Eq.~(\ref{Eq:dBSy0}), $\bar{\mu}_z(t)$ is obtained by numerically solving Eq.~(\ref{Eq:muvsBrill3}) and $h_{\rm c\perp}$ is given in Eq.~(\ref{Eq:BarHvsT}).

\subsubsection{\label{Sec:CmagAFPM} Magnetic Heat Capacity and Entropy of the Combined Antiferromagnetic and Paramagnetic Phases}

\begin{figure}
\includegraphics [width=2.8in]{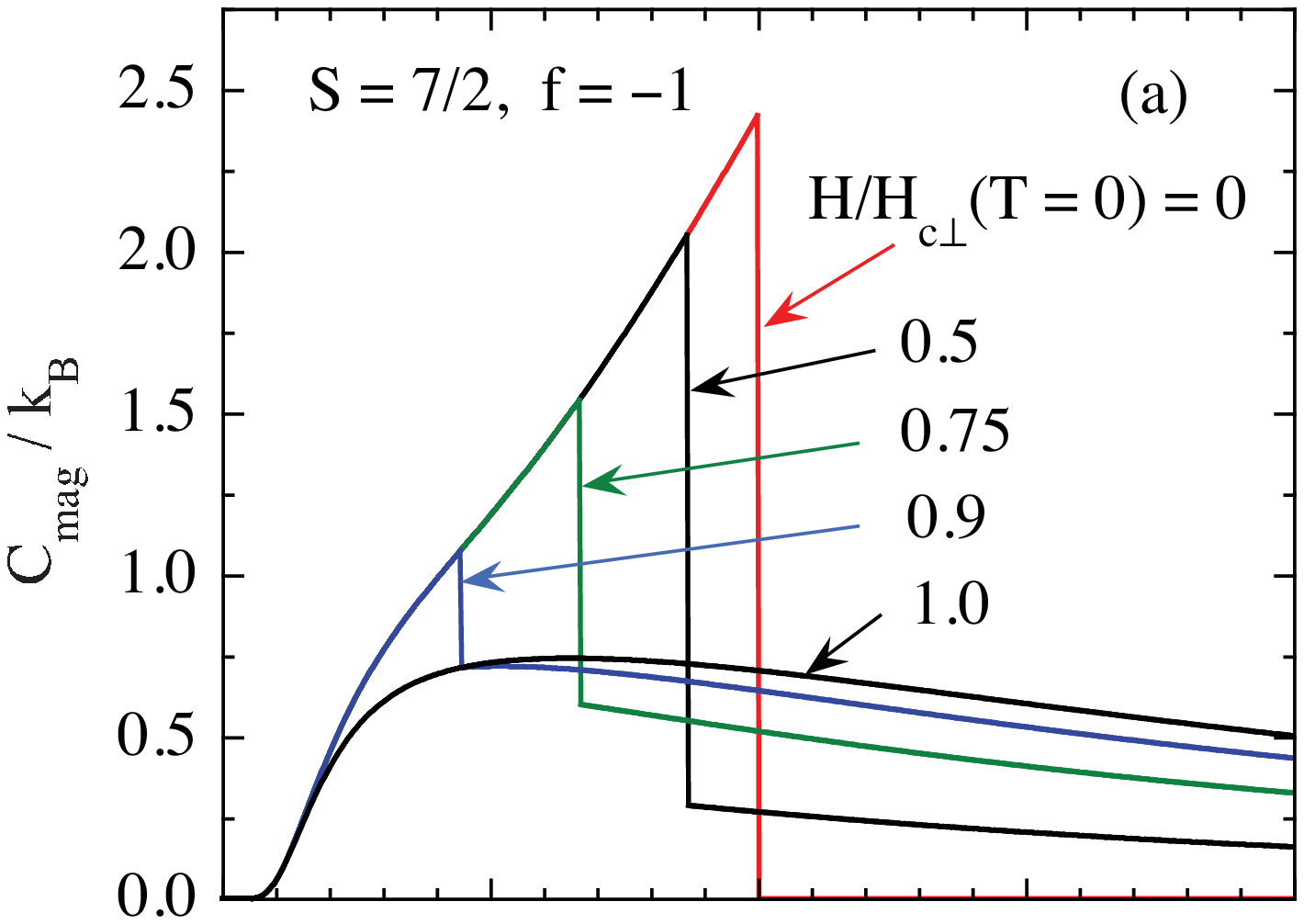}\vspace{-0.219in}
\includegraphics [width=2.8in]{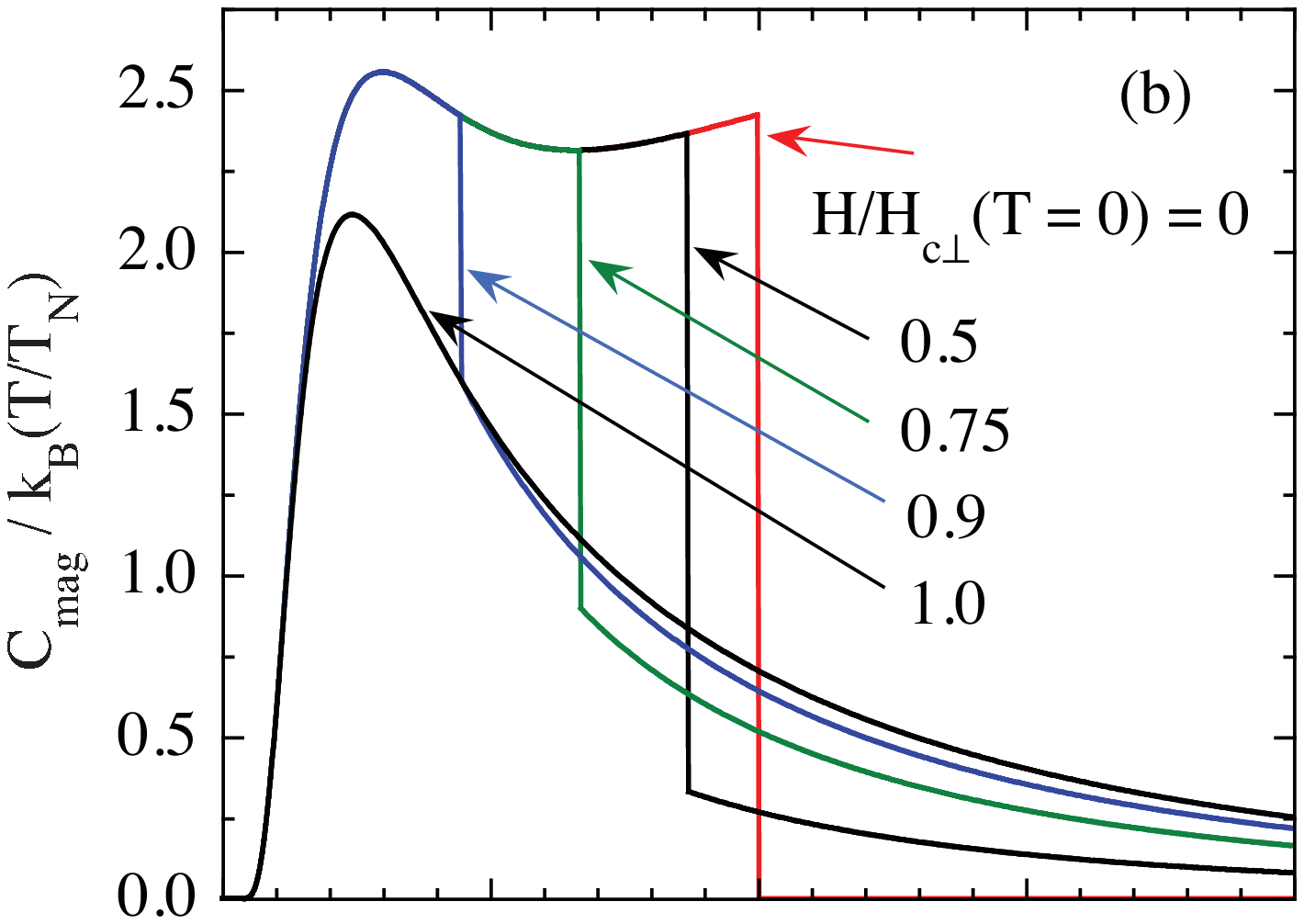}\vspace{-0.219in}
\includegraphics [width=2.8in]{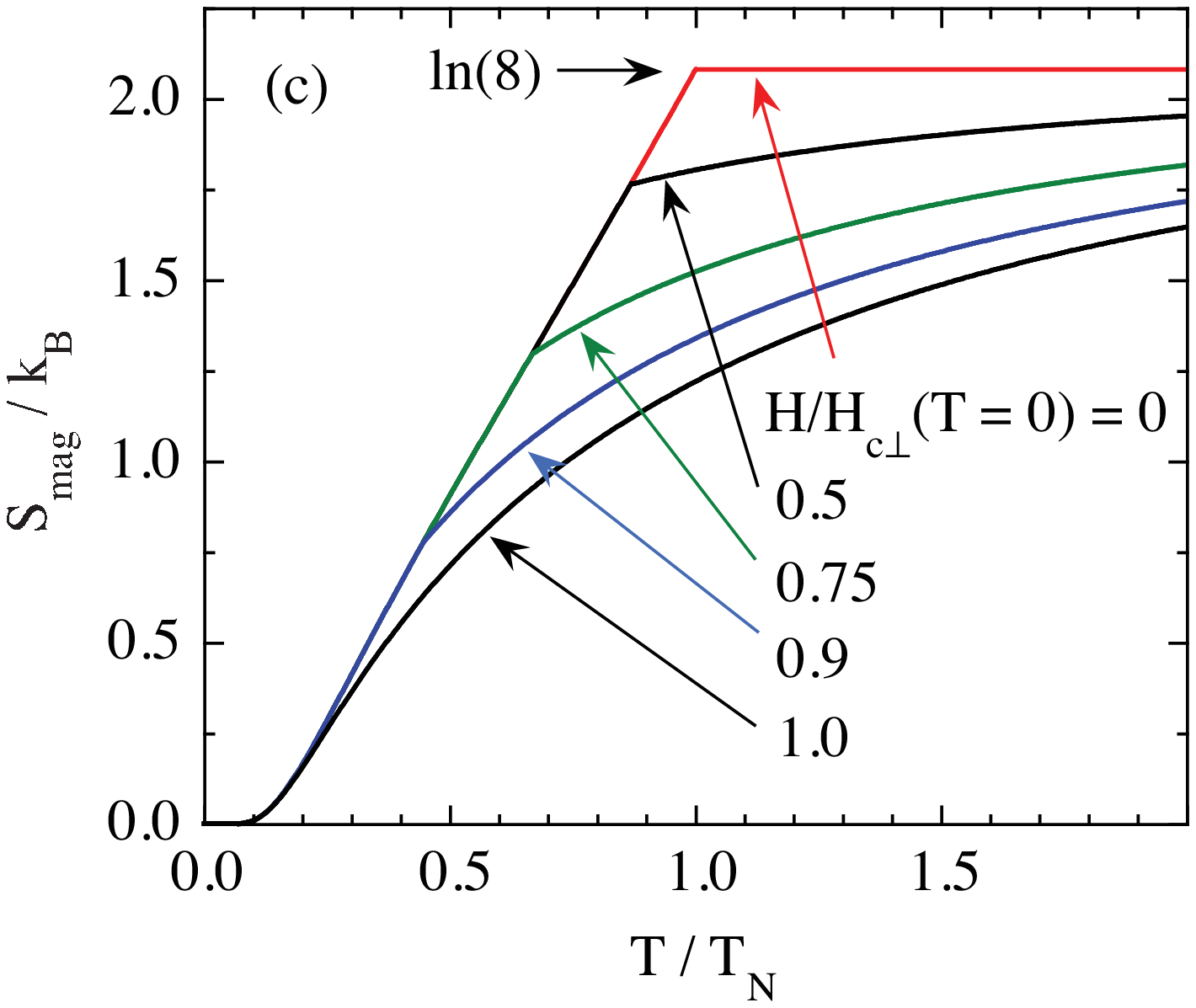}
\caption{(Color online) (a) Magnetic heat capacity $C_{\rm mag}$, (b) $C_{\rm mag}/T$ and (c) magnetic entropy $S_{\rm mag}$ versus temperature $T$ for a spin $S=7/2$ Heisenberg antiferromagnet with $f=-1$ for various magnetic fields $H$ normalized by the critical field $H_{\rm c\perp}(T=0)$.  The discontinuities in the respective figures versus temperature separate the low-$T$ and/or low-$H$ AF regime from the high-$T$ and/or high-$H$ PM regime.  The curves in (a) and~(b) were obtained using Eq.~(\ref{Eq:CmaginH2}) for the AF regime and Eq.~(\ref{Eq:CmagPM2}) for the PM regime and the curves in~(c) were obtained using Eq.~(\ref{Eq:Smag}). }
\label{Fig:cMags72fm1}
\end{figure}

Plots of $C_{\rm mag}$ versus $T$ for $f = -1$ and spin $S=7/2$ obtained for the AF and PM temperature and field regions using Eqs.~(\ref{Eq:CmaginH2}) and~(\ref{Eq:CmagPM2}), respectively, are shown in Fig.~\ref{Fig:cMags72fm1}(a) for values of $h$ given by $h/h_{\rm c\perp}(t=0) = 0$, 0.5, 0.75, 0.9 and 1, where 
\be
h_{\rm c\perp}(t=0) = \frac{3(1-f)}{S+1} 
\ee
using Eq.~(\ref{Eq:BarHvsT}) with $\bar{\mu}_0(t=0)=1$.  One sees that a jump in $C_{\rm mag}(t)$ is present at each $t_{\rm N}(h)$ as given above in Fig.~\ref{Fig:mu0_vs_t}, but the size of the jump decreases as $T_{\rm N}(H)/T_{\rm N}(H=0)$ decreases and disappears when $h = h_{\rm c\perp}(t=0)$.  This behavior of the heat capacity jump with field is reflected in the variation in the discontinuity in slope at $T=T_{\rm N}(H)$ in plots of $\bar{\mu}_z$ versus~$h$ for various~$t$ in Fig.~\ref{Fig:MHs5fm1}.  

Plots of $C_{\rm mag}/t$ versus $t$ obtained from the data in Fig.~\ref{Fig:cMags72fm1}(a) are shown in Fig.~\ref{Fig:cMags72fm1}(b).  The magnetic entropy $S_{\rm mag}(t)$ is obtained by integrating the data in Fig.~\ref{Fig:cMags72fm1}(b) versus $t$ according to
\be
\frac{S_{\rm mag}(t)}{k_{\rm B}} = \int_0^t \frac{C_{\rm mag}(t)}{k_{\rm B}t}\,dt,
\label{Eq:Smag}
\ee
and the results are shown in Fig.~\ref{Fig:cMags72fm1}(c).  Interestingly, the $S_{\rm mag}$ versus~$t$  plots at different $h$  are similar in shape to the $\mu_z$ versus~$h$ plots in Fig.~\ref{Fig:MHs5fm1}(a) at different $t$.  In the limit of high~$t$ the entropy for fixed spin~$S$ and all values of $h$ must be the same value $S_{\rm mag}(T\to\infty) = k_{\rm B}\ln(2S+1) = k_{\rm B}\ln(8)$, as indicated in Fig.~\ref{Fig:cMags72fm1}(c).  However, the approach to this asymptotic value with increasing $t$ is very slow in the PM phase, especially when $H$ is large, because the field tends to align the moments in the direction of {\bf H} that reduces the magnetic entropy, which competes with temperature-induced disorder.  For example, when $h$ is sufficiently high to destroy the AF state completely at $h = h_{\rm c\perp}(t=0)$ in Fig.~\ref{Fig:cMags72fm1}(c), the integral in Eq.~(\ref{Eq:Smag}) must be extended for spin $S = 7/2$ and $f=-1$ up to $t \equiv T/T_{\rm N} \approx20$ in order for $S_{\rm mag}$ to reach 99.5\% of its high-$T$ asymptotic value per moment of $k_{\rm B}\ln(8)$.  Because the PM part of $C_{\rm mag}$ grows strongly with increasing $h$ for temperatures $T>T_{\rm N}(H)$ and is distributed over a large $T$ range, it may be difficult to experimentally separate this PM contribution from the phonon contribution in heat capacity measurements of AF materials.

\section{\label{Sec:Discussion} Discussion}

In a system of \emph{noninteracting} spins-$S$ with \mbox{$z$-component} of the magnetic moment $\mu_z = -gm_S\mu_{\rm B}$, an applied magnetic field ${\bf H} = H\hat{\bf k}$ lifts the degeneracy of the $2S+1$ Zeeman levels labeled by the spin magnetic quantum number $m_S = -S,\ -S+1, \ldots,\ S$ and splits them in energy according to $E=-\mu_zH = gm_S\mu_{\rm B}H$, where the $m_S=-S$ state lies lowest in energy.  Within the Weiss MFT, the molecular field (exchange field ${\bf H}_{\rm exch}$) in the ordered state of a system of \emph{interacting} spins in zero applied field is assumed to act like a uniform  applied field in a FM or a staggered field in an AF\@.  This exchange field therefore results in the same splitting of the Zeeman levels of a magnetic moment in a FM or AF as happens due to a uniform field applied to a system of noninteracting spins.  Thus in the ordered state of a FM or AF there is an energy gap between the ground state and the first excited state given by $E_{\rm gap} = g\mu_{\rm B}H_{\rm exch}$ even in zero applied field.  According to Eq.~(\ref{Eq:HexchioTm}), the exchange field is proportional to the ordered moment, the $T$ dependence of which is shown in Fig.~\ref{Fig:OrderedMomentMFT3}.  This energy gap grows monotonically with decreasing $T$ and approaches a constant value for $T\to0$.  Thus all thermodynamic properties of the system approach their $T = 0$ values exponentially with decreasing temperature, including $C_{\rm mag}$ and $\chi$ along the easy axis (collinear AFs) or plane (planar noncollinear AFs).

However, many spin systems do not show such activated behaviors in the ordered state for $T\to0$, and this is a failure of the MFT\@.  Instead, FMs and AFs often show power-law behaviors in these properties at low $T$\@.  The reason for this failure is that MFT does not take into account magnetic excitations associated with tilting of the individual magnetic moment directions.  These excitations are propagating spin waves with a linear dispersion relation $\omega = vk$ for AFs where $v$ is the spin-wave velocity, $\omega$ is the spin-wave angular frequency and $k$ is the magnitude of the wave vector, or $\omega = Ak^2$ for FMs where $A$ is a constant.  These dispersion relations give rise to $T^{3/2}$ and $T^3$ contributions to $C_{\rm mag}$ at temperatures low compared to the transition temperature of three-dimensional (3D) ferromagnets and antiferromagnets, respectively.\cite{Keffer1966, Keffer1952, Majlis2007, JohnstonRef}  On the other hand, MFT can predict $C_{\rm mag}$ over the entire $T$ range below the magnetic ordering temperature, in contrast to spin-wave theory that is useful only at temperatures much lower than the ordering temperature.

Whereas spin-wave theory can produce more accurate predictions for the magnetic and thermal properties of Heisenberg spin systems than MFT for temperatures much lower than the magnetic ordering temperature, quantum mechanical high-temperature series expansions (HTSEs) of $\chi$ and $C_{\rm mag}$ of Heisenberg AFs in powers of $1/T$ yield predictions more accurate than MFT in the high-$T$ region above the magnetic ordering temperature.  For example, the first two terms in the HTSE for $\chi$ give the Curie-Weiss law.  Subsequent terms give corrections to this behavior that become more important as $T$ decreases.  The minimum $T$ at which accurate descriptions of the magnetic and thermal properties are obtained using HTSEs decreases with increasing number of terms in the HTSE\@.  Depending on the spin lattice, the spin-lattice dimensionality and the value of $S$, such HTSEs typically contain $\sim 10$--20 terms. 

The MFT prediction in Eq.~(\ref{Eq:TmGeneral}) for the magnetic transition temperature does not take into account quantum fluctuations associated with a low dimensionality of the spin lattice, because the same formula applies to all spin lattices irrespective of their dimensionality.  The Mermin-Wagner theorem states that long-range magnetic order cannot occur at a {\it finite} temperature in 1D or 2D Heisenberg spin lattices.\cite{Mermin1966} Perhaps surprisingly, this theorem does not rule out long-range AF order at $T=0$ in 2D, which is actually predicted to occur in the 2D $S=1/2$ square lattice Heisenberg AF.\cite{Johnston1997}  Of course, from theory and experiment such long-range ordering does occur in 3D spin lattices.  This suppression of magnetic ordering in low-dimensional spin lattices arising from quantum fluctuations is related to the reduction in the number of nearest neighbors of a given spin as the dimensionality of the spin lattice decreases.

The Weiss MFT predicts that the ordered moment at $T=0$ in the magnetically ordered state of a FM or AF in $H=0$ is equal to the saturation moment: $\mu_0(T=0) = gS\mu_{\rm B}$.  Another manifestation of quantum fluctuations is a reduction in this $T=0$ ordered moment that becomes increasingly pronounced as the effective dimensionality of the spin lattice and/or the spin~$S$ decrease.  For example, in ${\rm La_2CuO_4}$ containing spins-1/2 on a square lattice, the ordered moment at $T\to0$ is found experimentally and theoretically to be suppressed by about 30\% compared to the MFT prediction $\mu_0(T=0) = 1~\mu_{\rm B}$/Cu assuming a spectroscopic splitting factor $g=2$.\cite{Johnston1997}

\acknowledgments

The author is grateful to V. K. Anand, R. J. Goetsch, A.~Honecker and M.~E.~Zhitomirsky for helpful discussions.  This work was partially supported by the U.S. Department of Energy, Office of Basic Energy Sciences, Division of Materials Sciences and Engineering.  Ames Laboratory is operated for the U.S. Department of Energy by Iowa State University under Contract No.~DE-AC02-07CH11358.\\


\end{document}